\documentclass[a4paper,12pt]{article}
\usepackage[utf8]{inputenc}
\pdfoutput=1

\usepackage{jheppub}
\usepackage{graphicx}
\usepackage{psfrag}
\usepackage{color}
\usepackage{amssymb}
\usepackage{amsmath}
\usepackage{amsbsy}
\usepackage{braket}
\usepackage{epstopdf}
\usepackage{rotating}
\usepackage{subfigure}

\begin{document}

\allowdisplaybreaks[2]

\title{Light-front holographic $\rho$-meson distributions in the momentum space}

\author[a]{Satvir Kaur}
\author[b,c,d]{Chandan Mondal}
\author[a]{Harleen Dahiya}

\affiliation[a]{Department of Physics, Dr. B. R. Ambedkar National Institute of Technology Jalandhar, Jalandhar-144011, India}
\affiliation[b]{Institue for Modern Physics, Chinese Academy of Sciences, Lanzhou-730000, China}
\affiliation[c]{School of Nuclear Science and Technology, University of Chinese Academy of Sciences, Beijing 100049, China}
\affiliation[d]{CAS Key Laboratory of High Precision Nuclear Spectroscopy, Institute of Modern Physics, Chinese
Academy of Sciences, Lanzhou 730000, China}
                                         
\emailAdd{satvirkaur578@gmail.com}
\emailAdd{mondal@impcas.ac.cn}
\emailAdd{dahiyah@nitj.ac.in}
%
%
%
%

\abstract{
We present the leading-twist quark transverse momentum-dependent parton distribution functions (TMDs) for the spin-1 target, such as the $\rho$-meson, in the light-front framework. Specifically, we predict the TMDs in the light-front holographic model and compare with the light-front quark model predictions. We obtain the TMDs using the overlap of the light-front wave functions. We evaluate the ${\bf k}_\perp$ moments upto second order and compare with the available theoretical predictions. Further, we analyze the leading-twist parton distribution functions (PDFs) of the $\rho$-meson in the light-front holographic model which are found to be in accord with the Nambu-Jona-Lasinio (NJL) model and the light-front quark model predictions. We further study the QCD evolution of the PDFs. The positivity bounds on the TMDs and the PDFs are also discussed. We also present the quark spin densities in the transverse momentum plane for different polarization configurations of the quark and the $\rho$-meson target. 
 
}

\maketitle
\section{Introduction}

Different aspects of hadron structure are described by the various partonic distributions. At partonic level, one of the most intimated function to reveal the structure of hadron is the parton distribution function (PDF) \cite{Collins:1981uw, Martin:1998sq, Gluck:1994uf, Gluck:1998xa}. Being function of longitudinal momentum fraction ($x$) only, the PDFs do not provide any knowledge about the spatial location and the transverse motion of partons inside the hadron. However, the modern distributions, i.e. the generalized parton distributions (GPDs) \cite{Diehl:2003ny, Garcon:2002jb, Belitsky:2005qn} and the transverse momentum-dependent parton distributions (TMDs) \cite{Diehl:2015uka, Angeles-Martinez:2015sea, Boer:1997nt, Boer:1999mm}, have been widely investigated in both experimentally and theoretically to perceive the combined hadronic structural information. The modern tomography is able to explain the three-dimensional structural information of the hadron. 
Basically, the three-dimensional TMDs are the extended version of collinear PDFs, predicting the information of the hadronic consituents within the transverse momentum space. These distributions also help to gain the knowledge about the correlation between spins of the hadron and the parton.  
The bunch of hidden information inside the hadron can be retrieved with the selection of high energy scattering processes. The complimentary method to acquire the TMDs are Drell-Yan processes \cite{ Ralston:1979ys, Donohue:1980tn, Tangerman:1994eh, Zhou:2009jm,Collins:2002kn} and $Z^0/W^{\pm}$ production \cite{Collins:1984kg, Catani:2015vma, Scimemi:2017etj}. The conventionally used process for measuring TMDs is semi-inclusive deep inelastic scattering (SIDIS) \cite{ Brodsky:2002cx, Ji:2004wu, Bacchetta:2017gcc}. 

 Our aim of this paper is to predict the holographic quark distributions of $\rho$-meson in the momentum space using the light-front (LF) dynamics. By illustrating the relativistic importance, the light-front dynamics having remarkable accomplishments provide a suitable framework to study the hadron structure \cite{Dirac:1949cp, Brodsky:1997de, Harindranath:1996hq}. To study the leading twist $\rho$-meson TMDs, we take the minimal Fock-state expansion into account, i.e. $|M \rangle = \sum | q \bar{q} \rangle \psi_{q \bar{q}}$.
 In the case of $\rho$-meson, there are total nine T-even TMDs at leading twist. These functions arise from the different matrix elements emerging from the bilocal operator. However, three TMDs are special distributions for $\rho$-meson: $f_{1LL}(x, {\bf k}^2_\perp), f_{1LT}(x, {\bf k}^2_\perp)$,and $f_{1TT}(x, {\bf k}^2_\perp)$, which are absent for spin-$0/\frac{1}{2}$ targets. The variables $x$ and ${\bf k}_\perp$ are the longitudinal momentum fraction and transverse momentum carried by the active quark. By applying certain integrations on TMDs, one can get the four collinear PDFs in $\rho$-meson. In other words, if we do not consider any perturbative effects, spin-1 hadron  can produce four PDFs, one of which is tensor-polarized corresponding to the unpolarized quark. 
   
   The tensor polarized PDF $b_1(x)$ being sensitive to the parton's orbital angular momentum is of great importance theoretically as well as experimentally \cite{Hoodbhoy:1988am, Close:1990zw, Umnikov:1996qv,Detmold:2005iz,Kumano:2010vz,Islam:2012zua, Miller:2013hla, Kumano:2016ude, Cosyn:2017fbo}. The utmost existing experimental data of structure function $b_1$ is available for the deutron state, formed by the two weakly bounded spin-1/2 hadrons, only~\cite{Airapetian:2005cb}. On the other hand, experimental data is not yet available for the $\rho$-meson, whereas on the theoretical front, the PDF $b_1(x)$ has been studied using simple relativistic model in Ref. \cite{Mankiewicz:1988dk}. A detailed investigation on $\rho$-meson TMDs has been done in Nambu-Jona--Lasinio (NJL) model focusing on the covariant approach~\cite{Ninomiya:2017ggn}. An important discussion on deep inelastic inclusive processes for spin-1, such as $\rho$-meson, has been reported  in Ref.~\cite{Bacchetta:2000jk}. 
Furthermore, the distributions containing the quark transverse momentum and the corresponding fragmentation functions have been explained in Ref.~\cite{Bacchetta:2001rb}. Hino and Kumano discussed the polarized Drell-Yan processes to study the structure functions of spin-1 hadrons \cite{ Hino:1998ww,Hino:1999qi}.

The LF holographic model for the $\rho$-meson \cite{Forshaw:2012im} is widely implemented to successfully study the various decays \cite{Ahmady:2012dy, Ahmady:2013cga}. It is interesting to extend the investigation to study the leading twist quark TMDs in $\rho$-meson using this model. Also, it would be interesting to compare the TMDs obtained in the LF holographic model with one of another successful model, the LF quark model \cite{Yu:2007hp, Qian:2008px}. The overlap representation of light-front wave functions (LFWFs) approach is used to reveal the quark TMDs of $\rho$-meson in both the models.  On account of the light-front helicities of partons inside the hadron, this approach allows us to retrieve the understanding of encoded spin-spin and spin-orbit correlations in the TMDs explicitly \cite{Bacchetta:2015qka}. 


The work is arranged as follows. In  section~\ref{model}, the essential details on basic formalism of the LF holographic model are presented. Section~\ref{tmd_defi} contains the general relations between the various leading twist spin-1 hadron TMDs and the correlation functions. We evaluate the T-even TMDs in terms of the overlaps of the LFWFs via different light-front amplitudes and discuss the numerical results and their positivity constrains in this section. 
In section~\ref{pdfs_rho}, we provide the detailed study of the PDFs of $\rho$-meson and also discuss the scale evolution of the PDFs. The spin densities generated from the different polarization configurations of the quark and the $\rho$-meson are explained in section~\ref{density}. The summary is given in section~\ref{summary}. The details about the general formalism of LF quark model and explicit expressions of the TMDs are presented in Appendix \ref{LFmodel}. The required description of density matrix of spin-1 is given in Appendix \ref{densitymartix}.
\section{Light-front holographic $\rho$-meson wave functions}\label{model}
Let us begin with the holographic Schr\"odinger equation, which is derived in the semiclassical approximation to QCD in the light-front and is assured by the dynamical part of the holographic wave function. The holographic Schr\"odinger equation is derived as:
\begin{eqnarray}
\left(-\frac{{\rm d}^2}{{\rm d}\zeta^2}-\frac{1-4L^2}{4\zeta^2}+U_{\rm eff}(\zeta)\right)\Phi(\zeta)=M^2\Phi(\zeta),
\label{schrodinger_equation}
\end{eqnarray}
with $\zeta^2=x(1-x){\bf b}^2_\perp$, where ${\bf b}_\perp$ defines the transverse separation of quark and antiquark in the hadron. 
For deriving Eq. (\ref{schrodinger_equation}), it is assumed that there is neither any quark mass nor any quantum loop \cite{Brodsky:2006uqa, deTeramond:2005su, Brodsky:2014yha, deTeramond:2008ht}. The light-front holography provides a link between the weakly coupled string modes propagate in the higher dimensional AdS space-time and the QCD Hamiltonian formulated in the physical spacetime at the fixed light-front time. 
The comparison of the wave equation for the amplitude of spin-$J$ string modes propagating in a modified AdS$_5$ space-time, with the light-front wave equation described in Eq.~(\ref{schrodinger_equation}), leads to the identification of the fifth dimension $z$ in AdS$_5$ with the LF variable $\zeta$,  where $(2-J)^2=L^2 -(\mu R)^2$ with $R$ and $\mu$ being the radius of curvature of  AdS$_5$ space-time and the 5-d mass of the string modes, respectively \cite{Brodsky:2014yha}. The consequence of mapping $\zeta \leftrightarrow z$ allows us to solve the single variable LF Schr\"odinger equation with known effective confining potential $U_{\rm eff}(z)$. The dilation field $\varphi(z)$, which distorts the pure AdS$_5$ geometry, is used to derive the confining potential $U_{\rm eff}(\zeta)$:
\begin{eqnarray}
U_{\rm eff}(\zeta,J)= \frac{1}{2}\varphi^{\prime \prime}{(\zeta)}+\frac{1}{4}\varphi^\prime(\zeta)^2 +\frac{2J-3}{2\zeta}\varphi^\prime(\zeta).
\label{confining-potential}
\end{eqnarray}
The choice of dilation field conditions the underlying action leading to the holographic Schr\"odinger equation, Eq. (\ref{schrodinger_equation}), being conformally invariant. The quadratic confinement potential has tendency to do so, $U(\zeta)=\kappa^4 \zeta^2$ \cite{Brodsky:2013ar}, which requires the choice of dilation field to be $\varphi=\kappa^2 z^2$. Therefore, from Eq. (\ref{confining-potential}):
\begin{eqnarray}
U_{\rm eff}=\kappa^4 \zeta^2+2 \kappa^2 (J-1),
\label{potential}
\end{eqnarray}
where $\kappa$ is known as the mass scale parameter, which determines the strength of the dilation field in AdS spacetime. The value of $\kappa=523 \pm 24$ MeV, which is fixed by the fit to the Regge trajectories of light mesons \cite{Brodsky:2016rvj}. Solving Eq. (\ref{schrodinger_equation}) by substituting the confining potential, one can get the  eigenvalue as the meson mass spectra:
\begin{eqnarray}
M^2=(4n+2L+2)\kappa^2+2\kappa^2(J-1)
\end{eqnarray}
and 
\begin{eqnarray}
\Phi_{nL}(\zeta)=\kappa^{1+L}\sqrt{\frac{2n!}{(n+L)!}}\zeta^{1/2+L}\, {\rm exp}\left(-\frac{\kappa^2 \zeta^2}{2}\right)L_n^L(\kappa^2 \zeta^2),
\label{transverse_part}
\end{eqnarray}
as the dynamical part of the holographic wave function.

The complete holographic wave function is written as \cite{Brodsky:2014yha, Forshaw:2012im}:
\begin{eqnarray}
\psi(x,\zeta,\theta)=e^{\iota L \theta} X(x)\frac{ \Phi(\zeta)}{\sqrt{2\pi \zeta}}\,,
\label{holographic_wavefunction}
\end{eqnarray}
where $X(x)$ corresponds to the longitudinal part of the wave function, which is fixed by mapping the spacelike electromagnetic form factor calculated in AdS/QCD \cite{Brodsky:2008pf} and the light-front formalism \cite{Drell:1969km, West:1970av}. In AdS$_5$, the form factor is evaluated by an overlap integral of the incoming and outgoing hadronic modes convoluted with the bulk-to-boundary propagator
which maps onto the free electromagnetic current in 
physical spacetime. In physical spacetime, the form factor is expressed by the integral overlap of the meson LFWFs. This matching procedure yields $X(x)=\sqrt{x(1-x)}$ \cite{Brodsky:2014yha}.
Meanwhile, mapping of the gravitational form factor  
in the AdS$_5$ and the physical spacetime also provides an identical result~\cite{Brodsky:2008pf}.
The ground state holographic wave function is then expressed in the transverse impact-parameter space as:
\begin{eqnarray}
\psi(x,\zeta^2)=\frac{\kappa}{\sqrt{\pi}}\sqrt{x(1-x)}\,{\rm exp}\left(\frac{-\kappa^2 \zeta^2}{2}\right)\,,
\end{eqnarray}
and in the transverse momentum space as:
\begin{eqnarray}
\psi(x,{\bf k}^2_\perp) \propto \frac{1}{\sqrt{x(1-x)}}\,{\rm exp}\left(-\frac{\mathcal{M}^2}{2 \kappa^2 }\right)\,,\label{m0}
\end{eqnarray}
where the invariant mass of the quark-antiquark pair is determined by $\mathcal{M}^2={\bf k}^2_\perp/\{x(1-x)\}$. The Fourier transform of the transverse separation between quark and antiquark ${\bf b}_\perp$ leads to provide the quark transverse momentum ${\bf k}_\perp$. 
The above equation of the holographic wave function is true for the massless quarks. 

For massive quarks, one needs to assume something beyond the AdS$_5$ correspondence. Since, the quark masses are associated with the longitudinal part of the kinetic energy, not with the transverse, the introduction of massive quarks requires specification of either the longitudinal
wave function or the dynamical equation for this. Brodsky and de Téramond have proposed an ansatz for
the former~\cite{Brodsky:2008pg}, while the latter has been considered in Ref.~\cite{Chabysheva:2012fe}, which provide a light-front equation for the longitudinal part that includes quark masses and the model (t'Hooft) potential.

 Here, we adopt the Brodsky–de Téramond ansatz that extends the transverse momentum dependence to include the full invariant mass. A simple generalization of the LFWF Eq.~(\ref{m0}) for massive quarks follows from the
assumption that the momentum space LFWF is a function of the invariant off-shell energy. For massive quarks, this invariant mass should be 
\begin{equation}
\mathcal{M}_{f\bar{f}^\prime}^2= \frac{{\bf k}^2_\perp}{x(1-x)} + \frac{m_f^2}{x} + \frac{m^2_{\bar{f}^\prime}}{1-x}~,\label{modified_M}
\end{equation}
with $f ~(\bar{f}^\prime)$ being the flavor of the quark (antiquark). The ansatz replaces $\mathcal{M}^2$ with $\mathcal{M}_{f\bar{f}^\prime}^2$ and then, the form of the holographic wave function for the meson bound state with massive quarks becomes
\begin{eqnarray}
\psi(x,{\bf k}^2_\perp)\propto \frac{1}{\sqrt{x(1-x)}}\,\exp\left(-\frac{ {\bf k}^2_\perp}{2 \kappa^2 x(1-x)}\right)\,  \exp\left(-\frac{\mu_{12}^2}{2 \kappa^2}\right)\, ,
\label{holographic-wf-momentum}
\end{eqnarray} 
where $\mu_{12}^2=\frac{m_f^2}{x}+\frac{m_{\bar{f}^\prime}^2}{1-x}$. We emphasize that Eq.~(\ref{holographic-wf-momentum}) is an educated guess and adding $\mu_{12}^2$
to Eq.~(\ref{schrodinger_equation}) does not lead to  Eq.~(\ref{holographic-wf-momentum}) as the solution. However, this treatment is also noted and adopted in the literature~\cite{Vega:2009zb, Branz:2010ub, Swarnkar:2015osa}, which is misleading. Meanwhile, the holographic wave function, Eq.~(\ref{holographic-wf-momentum}), has been successfully used in the description of diffractive $\rho$ meson electroproduction at HERA~\cite{Forshaw:2012im}, in $B\to \rho \gamma$~\cite{Ahmady:2012dy} and $B\to K^* \gamma$~\cite{Ahmady:2013cva} decays as well as in the
prediction of $B\to \rho$~\cite{Ahmady:2013cga}, $B\to K^*$ form factors~\cite{Ahmady:2013cva} and $B\to K^*\mu^-\mu^+$ decays~\cite{Ahmady:2014cpa}. This LFWF has also been used to
investigate the spectrum~\cite{Branz:2010ub} and the distribution amplitudes~\cite{Hwang:2012xf} of light and heavy mesons. For the excited meson states one can follow the same procedure by replacing the key invariant mass variable in the polynomials in the LFWF using~Eq.~(\ref{modified_M})~\cite{Brodsky:2014yha}.

 The massive quarks also lead to the essential modification in the hadronic mass, i.e., a shift in the predicted meson masses,
\begin{equation}
	M^2 = M_0^{2} + \Delta M^2,
\end{equation}
where $M_0^{2}$ stands for the meson mass squared in the limit of massless quarks and the mass shift is given by \cite{Brodsky:2008pg, Brodsky:2014yha}
\begin{equation}
	\Delta M^2 = \left\langle \Psi \left|\frac{m^2_f}{x} + \frac{m^2_{\bar{f}^\prime}}{1-x}\right|\Psi \right \rangle \;.
\label{mass-shift}
\end{equation}
The modified holographic wavefunction, given by Eq. \eqref{holographic-wf-momentum}, is used to compute $\Delta M^2$.

Since, we are dealing with the $\rho$-meson in this work, which follows: $m_f=m_{\bar{f'}}=m_q$, where $m_q$ represents the mass of the light quarks ($u $ and $d$). Therefore, the holographic wave function, in Eq. (\ref{holographic-wf-momentum}), becomes:
\begin{eqnarray}
\psi(x,{\bf k}^2_\perp)\propto \frac{1}{\sqrt{x(1-x)}}\,{\rm exp}\left(-\frac{ {\bf k}^2_\perp+m^2_q}{2 \kappa^2 x(1-x)}\right)\,.
\label{holographic-wf}
\end{eqnarray}  

Till now, the helicities of the quark and the antiquark are not included. To take these into account, one can express the wave functions as \cite{Ahmady:2020mht}:
\begin{eqnarray}
\Psi^\Lambda_{h_q, h_{\bar{q}}}(x,{\bf k}_\perp)= \chi^{\Lambda}_{h_q, h_{\bar{q}}}\psi(x,{\bf k}^2_\perp)\,,
\end{eqnarray}
 with $\Lambda=L(T)$  as the longitudinal~(transverse) spin projection of the $\rho$-meson and
\begin{eqnarray}\label{nond}
\chi^L_{h_q, h_{\bar{q}}}=\frac{1}{\sqrt{2}}\delta_{h_q, -h_{\bar{q}}} \ \ \ ; \ \ \ \chi^{T(\pm)}_{h_q, h_{\bar{q}}}=\frac{1}{\sqrt{2}}\delta_{h_q \pm , h_{\bar{q}}\pm}.
\end{eqnarray}
Here, $h_q (h_{\bar{q}})$ are defined as the quark (antiquark) helicity. The spin structures in Eq.~(\ref{nond}) correspond to a nondynamical spin wavefunction.
Note that since the holographic wave function defined in Eq.~(\ref{holographic-wf-momentum}) does not depend on the spin, there is no distinction between the light pseudoscalar and the light vector mesons. Thus, with a universal AdS/QCD scale, this would yield the degenerate decay constants for the pseudoscalar and the vector mesons, in contradiction with the experiment~\cite{Tanabashi:2018oca}. On the other hand, this would also lead to the same decay constants for the longitudinally and transversely polarized vector mesons, in contradiction with lattice QCD~\cite{Becirevic:2003pn, Braun:2003jg}.
The mentioned shortcomings can be addressed by taking the dynamical spin effects into account. Considering dynamical spin effects, the vector meson wave functions can then be written as \cite{Forshaw:2012im, Ahmady:2020mht}:
\begin{eqnarray}
\Psi^\Lambda_{h_q, h_{\bar{q}}}(x,{\bf k}_\perp)= \chi^{\Lambda}_{h_q, h_{\bar{q}}}(x,{\bf k}_\perp)\psi(x,{\bf k}^2_\perp),
\end{eqnarray} 
where the Lorentz invariant spin structure for the vector meson is expressed by accounting the photon-quark-antiquark vertex: 
\begin{eqnarray}
\chi^{L(T)}_{h_q, h_{\bar{q}}}(x,{\bf k}_\perp)= \frac{\bar{u}_{h_q}(k^+,{\bf k}_\perp)}{\sqrt{x}} \, \epsilon_\Lambda \cdot \gamma \, \frac{v_{h_{\bar{q}}}(k^{\prime +},{\bf k}^{\prime}_\perp)}{\sqrt{1-x}}\,,
\label{spin-structure}
\end{eqnarray} 
where $k$ and $k^\prime$ denote the 4-momenta of the quark and the antiquark respectively. The longitudinal momentum fraction carried by the quark and the antiquark are defined as $x =\frac{k^+}{P^+}$ and $(1-x)=\frac{k^{\prime +}}{P^+}$, respectively.
The polarizations vectors, $\epsilon_\Lambda$, for the longitudinally polarized and the transversely polarized $\rho$-meson are given by
\begin{eqnarray}
\epsilon_L=\left(\frac{P^+}{M_\rho},-\frac{M_\rho}{P^+},0,0\right)\ \ \ ; \ \ \ \epsilon^{\pm}_T=\mp\frac{1}{\sqrt{2}}\left(0,0,1,\pm \iota\right).
\end{eqnarray}
This leads to the spin improved LFWFs for the longitudinally polarized $\rho$-meson at the scale $\mu_{\rm LFH}^2=0.20$ GeV$^2$ \cite{Forshaw:2012im, Ahmady:2020mht}:
\begin{equation}
\Psi_{h_q,h_{\bar{q}}}^L(x,{\bf k}_\perp)=\mathcal{N}_L \, \delta_{h_q,-h_{\bar{q}}} \left(M_\rho^2 x(1-x) + m_q^2+{\bf k}_\perp^2\right)\, \frac{\psi(x,{\bf k}^2_\perp)}{x(1-x)}\,,
\label{longitudinal-component}
\end{equation}
and for the transversely polarized $\rho$-meson as:
\begin{align}
\Psi_{h_q,h_{\bar{q}}}^{T(+)}(x,{\bf k}_\perp)=\mathcal{N}_T \,\big( k_\perp e^{ \iota \theta_{k_\perp}}(x \delta_{h_q+,h_{\bar{q}}-}&-(1-x)\delta_{h_q-,h_{\bar{q}}+})\nonumber\\
&+m_q \delta_{h_q+,h_{\bar{q}}+}\big) \, \frac{\psi(x,{\bf k}^2_\perp)}{x(1-x)}\,,
\label{transverse-component}
\end{align}
\begin{align}
\Psi_{h_q,h_{\bar{q}}}^{T(-)}(x,{\bf k}_\perp)=\mathcal{N}_T \,\big( k_\perp e^{- \iota \theta_{k_\perp}}((1-x)\delta_{h_q+,h_{\bar{q}}-}&-x \delta_{h_q-,h_{\bar{q}}+})\nonumber\\
&+m_q \delta_{h_q-,h_{\bar{q}}-}\big) \, \frac{\psi(x,{\bf k}^2_\perp)}{x(1-x)}\,,
\label{transverse-component}
\end{align}
where the normalization constants $\mathcal{N}_{L(T)}$ are determined by
\begin{eqnarray}
 \sum_{h_q,h_{\bar{q}}}\int  \frac{{\rm d} x \, {\rm d}^2{\bf k}_\perp  }{2(2 \pi)^3}\, |\Psi^\Lambda_{h_q,h_{\bar{q}}}(x,{\bf k}_\perp)|^2=1,
\end{eqnarray}
depending upon the polarization of the $\rho$-meson. 

 We state that the spin-improved holographic light-front wavefunctions are distinct from the boosted wave functions in the quark model obtained by boosting the non-relativistic Schr\"odinger wave function in the meson's rest frame to the light-front. This is usually performed using the Brodsky-Huang-Lepage prescription \cite{Brodsky:1981jv}, together with the Melosh-Wigner rotation \cite{Melosh:1974cu} for the spin structure.  On the other hand, the spin structures in holographic model are fixed by the rules of light-front field theory for coupling a quark and an antiquark into $\rho$ meson (point-like), while nonperturbative bound state effects are captured by the holographic wavefunction given by Eq.~(\ref{holographic-wf}). The holographic wavefunctions are directly formulated on the light-front and thus, they are frame-independent and avoid the ambiguities associated with a boosting prescription.  

Having said that, it is worth noting that the boosting of a harmonic oscillator rest frame Schr\"odinger wavefunction results in the boosted Gaussian wavefunction, Eq.~(\ref{momentum-space-wf}) in Appendix A, which seems similar to the holographic Gaussian wave function given by Eq.~(\ref{holographic-wf}). However, we must highlight three essential differences between these two wave functions: first, the harmonic confining potential in the holographic Schr\"odinger equation, Eq.~(\ref{schrodinger_equation}), is uniquely fixed by a specific mechanism of conformal symmetry breaking in semiclassical light-front QCD  unlike the assumed harmonic potential in the ordinary Schr\"odinger equation. Second the AdS/QCD scale parameter, $\kappa$, is extracted from the mass spectroscopic data and it fixes the width of the holographic Gaussian, where as the width of the boosted Gaussian is a free parameter, which has to be fixed by some constraint on the wave function.  Thirdly, the two wave functions differ by an overall notable factor: $1/\sqrt{x(1-x)}$  which makes their end-point behaviours different.

Though these wave functions are not derived from QCD first principle, so far these phenomenological models are proven to reproduce many interesting properties of $\rho$ meson. These wave functions have been successfully applied to describe the data on diffractive $\rho$ meson electroproduction~\cite{Forshaw:2012im, Ahmady:2016ujw}, decay constant~\cite{Ahmady:2016ujw, Qian:2008px}, distribution amplitude~\cite{Ahmady:2012dy}, electromagnetic and vector to pseudoscalar transition form factors~\cite{Choi:1996mq, Yu:2007hp, Ahmady:2020mht} etc.. It has been observed that the diffractive $\rho$-meson electroproduction data differentiate the two wave functions and found to be in agreement with the holographic Gaussian~\cite{Forshaw:2012im}. For vector mesons like $\rho$, there exist no experimental data for TMDs or PDFs, however, it is nevertheless important to investigate these in various nonperturbative approaches or phenomenological models of QCD. Additionally, 
the TMD fragmentation functions~\cite{Metz:2016swz, Accardi:2012qut} for an elementary quark fragmentation process to vector mesons can be obtained from the TMDs and may be measurable in the production of vector mesons~\cite{Ji:1993vw}.
We thus adopt two different approaches/models for illuminating $\rho$ meson TMDs.
On the other hand, till date, the $\rho$-meson TMDs and PDFs have been studied only in the framework provided by the NJL model~\cite{Nambu:1961tp, Nambu:1961fr}, incorporating important aspects of quark confinement via the infrared cutoff in the proper-time regularization scheme~\cite{Ebert:1996vx, Hellstern:1997nv, Bentz:2001vc}, where the chiral symmetry is dynamically broken. 
It has also been stated that the TMDs do not exhibit the familiar Gaussian behavior in the transverse momentum~\cite{Ninomiya:2017ggn}. It is therefore interesting to compare our predictions with the results of the NJL model.

\section{Transverse momentum-dependent parton distributions}\label{tmd_defi}
The quark TMDs of the hadron are defined through the transverse momentum-dependent quark correlation function. For the spin-1 target having ${\bf k}_\perp$ as its active constituent transverse momentum, the quark correlator function is given by \cite{Tangerman:1994eh, Bacchetta:2001rb,Bacchetta:2000jk, Ninomiya:2017ggn, Hino:1999qi, Pasquini:2014ppa, Meissner:2007rx}
\begin{eqnarray}
\Theta_{i j}^{(\Lambda)_{\bf \mathcal{S}}}(x, {\bf k}_\perp) &=&
 \int \frac{{\rm d}z^- \, {\rm d}^2 {\bf z}_\perp}{(2\pi)^3} \, 
e^{\iota k \cdot z} \ {}_{\bf {\mathcal{S}}}\langle P,\Lambda | \bar{\vartheta}_{j}(0) \mathcal{L}^\dagger(0,{\bf 0}_\perp|n)\mathcal{L}(z^-,{\bf z}_\perp|n)
\vartheta_{i}(z^-,{\bf z}_\perp) | P, \Lambda \rangle_{\bf {\mathcal{S}}}, \nonumber \\[0.2em]
&\equiv& \epsilon^*_{\Lambda (\mu)}(P) \ \Theta_{ij}^{\mu\nu}(x,{\bf k}_\perp) \ \epsilon_{\Lambda (\nu)}(P),
\label{phi1}
\end{eqnarray}
with the gauge link $\mathcal{L}$, defined as \cite{Pasquini:2014ppa,Meissner:2007rx}:
\begin{eqnarray}
\mathcal{L}(z^-,{\bf z}_\perp|n)&=&\mathcal{P} {\ \rm exp}\left(-\iota g \int_{z^-}^{n\cdot\infty} {\rm d}\eta^- \cdot A^+(\eta^- ,{\bf z}_\perp)\right) \nonumber\\
&&\times \mathcal{P} {\ \rm exp}\left(-\iota g \int_{{\bf z}_\perp}^{\infty} {\rm d}{\boldsymbol{\eta}}_\perp \cdot {\bf A}_\perp(z^-=n \cdot \infty,{\bf \eta}_\perp)\right).
\end{eqnarray}
The gauge link $\mathcal{L}$ guarantees the gauge invariance of the non-local operator in Eq. (\ref{phi1}).
For simplicity, in this work, we assume the gauge link to be unity, which leads us to determine the T-even TMDs only. 
%
%
In Eq. (\ref{phi1}), the state $|P, \Lambda \rangle_{\mathcal{S}}$ indicates that the projection of the target's spin on the direction $\mathcal{S}$ is equal to $\Lambda=\pm 1, 0$. $\vartheta$ represents the flavor SU(2) quark field operator. The quark correlation matrix is expressed as the contraction of  the polarization-independent Lorentz tensor matrix $\Theta_{ij}^{\mu\nu}$ with the polarization 4-vector $\epsilon_{\mu(\nu)}$.
We define the kinematical variables of the target's state in light-front frame as
\begin{eqnarray}
P=\left(P^+,P^-,{\bf P}_\perp \right)=\left(P^+, \frac{M_\rho^2}{P^+},{\bf 0}_\perp\right).
\end{eqnarray}

At the leading-twist, there are nine T-even TMDs for the $\rho$-meson, which are related to the quark-quark correlators depending on the different spin projections $\Lambda=0,\pm 1$ and, the longitudinal and transverse polarizations of the target as:
\begin{align}\label{form1} 
\langle \gamma^+ \rangle_{\boldsymbol{\mathcal{S}}}^{(\Lambda)} (x, {\bf k}_\perp) 
& = 
f_1(x, {\bf k}_\perp^2) + \mathcal{S}_{LL} \,
f_{1LL}(x, {\bf k}_\perp^2)\nonumber\\
& + \frac{\boldsymbol{ \mathcal{S}}_{LT} \cdot {\bf k}_\perp}{M_\rho}\, f_{1LT}(x, {\bf k}_\perp^2) 
+\frac{{\bf k}_\perp \cdot \boldsymbol{\mathcal{S}}_{TT} \cdot {\bf k}_\perp}{M_\rho^2} \, f_{1TT}(x, {\bf k}_\perp^2)\, , \\
%
%
\langle\gamma^{+} \gamma_{5}\rangle^{(\Lambda)}_{\boldsymbol{\mathcal{S}}} (x,{\bf k}_\perp)
&= \mathcal{S}_L\,g_{1L}(x,{\bf k}_\perp^2)+ \frac{{\bf k}_\perp \cdot \boldsymbol{\mathcal{S}}_\perp}{M_\rho} g_{1T}(x,{\bf k}_\perp^2)\, , \\
%
\langle\gamma^+\gamma^i\gamma_{5}\rangle^{(\Lambda)}_{\boldsymbol{\mathcal{S}}} (x,{\bf k}_\perp)
& =\mathcal{S}_\perp^ih_1(x,{\bf k}_\perp^2) 
+ \mathcal{S}_L\frac{k_\perp^i}{M_\rho}h_{1L}^\perp(x,{\bf k}_\perp^2) \nonumber\\
&+
\frac{1}{2\,M_\rho^2}
\left(2\,k_\perp^i {\bf k}_\perp \cdot \boldsymbol{ \mathcal{S}}_\perp - \mathcal{S}_\perp^i~{\bf k}_\perp^2\right) h_{1T}^{\perp}(x,{\bf k}_\perp^2)\, ,
\label{form3}
\end{align}
The correlations in Eqs. (\ref{form1})-(\ref{form3}) are defined by
\begin{eqnarray}
\langle \Gamma \rangle_{\bf \mathcal{S}}^{(\Lambda)}(x,{\bf k}_\perp)&=&\frac{1}{2} {\rm Tr}_{D}\left(\Gamma \Theta^{(\Lambda)_{\bf \mathcal{S}}}(x, {\bf k}_\perp)\right)\equiv  \epsilon^{*}_{\Lambda(\mu)} (P)~\langle \Gamma\rangle^{\mu \nu}(x,{\bf k}_\perp)~\epsilon_{\Lambda(\nu)}(P)
\label{ap}
\end{eqnarray}
where the Dirac matrices $\Gamma$ are $\gamma^+$, $\gamma^+ \gamma_5$ or $\gamma^+ \gamma^i \gamma_5$ with $i=1,2$, and we introduce the following quantities with implicit ${\bf \mathcal{S}}$ and  $\Lambda$ dependence:
 \begin{eqnarray}
\mathcal{S}_{LL}&=&\left(3 \Lambda^2-2 \right)\left(\frac{1}{6}-\frac{1}{2} \mathcal{S}_L^2\right),\\
\mathcal{S}^i_{LT}&=&\left(3 \Lambda^2-2 \right)\mathcal{S}_L \mathcal{S}^i_\perp,\\
\mathcal{S}_{TT}^{ij}&=&\left(3 \Lambda^2-2 \right)(\mathcal{S}_\perp^i \mathcal{S}_\perp^j-\frac{1}{2}\mathcal{S}_\perp^2~ \delta^{ij})\,.
\end{eqnarray}
Here, $\mathcal{S}_\perp^{i(j)}$ symbolizes the transverse polarization of the target meson in the directions $i(j)=x \ {\rm or} \ y$, while $\mathcal{S}_L$ is the longitudinal polarization of the target. The correlator equates the  respective spin-1 meson TMDs corresponding to the unpolarized quark, the longitudinally polarized quark and the transversely polarized quark identified with the several alphabets $f$, $g$ and $h$. The subscript $1$ in the various TMDs refers to the twist-2 or leading-twist. We emphasize that for $\rho$-meson to be longitudinally polarized means $| \mathcal{S}_L |=1$ and $| \mathcal{S}_\perp |=0$, which corresponds to the spin projections $\Lambda=0,\pm 1$ parallel to the direction of quark momentum. However, when the $\rho$-meson is transversely polarized, the condition converses, i.e. $| \mathcal{S}_L |=0$ and $| \mathcal{S}_\perp |=1$, which describes the spin projections $\Lambda=0, \pm 1$ perpendicular to the direction of the quark momentum.


\subsection{Overlap formalism}\label{overlap}
An equivalent way to derive the TMDs explicitly is to represent the correlator in the basis where
one considers the light-front helicities of both, the target and the active parton~\cite{Bacchetta:2015qka}.
The light-front helicity amplitudes with $h_q(h^\prime_q)$ and $\Lambda(\Lambda^\prime)$, which define the initial(final) state helicities of the active quark and the target, respectively, can be expressed as: 
\begin{eqnarray}
A_{h^\prime_q \Lambda^\prime, h_q \Lambda}(x,{\bf k}_\perp)&=&\frac{1}{(2 \pi)^3} \sum_{ h_{\bar{q}}} \Psi^{\Lambda^\prime *}_{h^\prime_q, h_{\bar{q}}}(x,{\bf k}_\perp)\,\Psi^\Lambda_{h_q, h_{\bar{q}}}(x,{\bf k}_\perp)\,,
\end{eqnarray}
where $\Psi^\Lambda_{h_q, h_{\bar{q}}}$ are the light-front wave functions.
By symmetry, we choose the row entries as $(h^\prime_q\, \Lambda^\prime)=(+ \, +), \,(+\,0),\,(+\,-),\,(-\,+),\,(-\,0),\,(-\,-)$, and the column entries as $(h_q\, \Lambda)=(+\,+),\,(+\,0),\,(+\,-),\,(-\,+),\,(-\,0),\,(-\,-)$. We can therefore express the light-front helicity amplitude matrix for spin-1 hadron as:
\begin{eqnarray}
\Phi= \left(
\begin{array}{cccccc}
A_{++,++} ~&~ A_{++,+0} ~&~ A_{++,+-} ~&~ A_{++,-+} ~&~ A_{++,-0} ~&~ A_{++,--}\\
A_{+0,++}& A_{+0,+0}& A_{+0,+-} & A_{+0,-+}& A_{+0,-0} & A_{+0,--}\\
A_{+-,++}& A_{+-,+0} & A_{+-,+-} & A_{+-,-+} & A_{+-,-0} & A_{+-,--}\\
A_{-+,++} & A_{-+,+0} & A_{-+,+-} & A_{-+,-+} & A_{-+,-0} & A_{-+,--}\\
A_{-0,++} & A_{-0,+0} & A_{-0,+-} & A_{-0,-+} & A_{-0,-0} & A_{-0,--}\\
A_{--,++} & A_{--,+0} & A_{--,+-} & A_{--,-+} & A_{--,-0} & A_{--,--}
\end{array}
\right)\,.
\label{matrix-helicity}
\end{eqnarray}
The light-front helicity amplitudes can be parametrized by the following combinations of $\rho$-meson TMDs~\cite{Ninomiya:2017ggn}:
\begin{eqnarray}
\Phi=\left(
\begin{array}{cccccc}
f^+ &  \frac{k_L}{\sqrt{2}\,M_\rho} g_{1T}^{(+)}
& \frac{k_L^2}{M_\rho^2}f_{1TT}    & \frac{k_R}{M_\rho} h_{1L}^{\perp}  & \sqrt{2} \, h_1  &  0\\  \\
\frac{k_R}{\sqrt{2}\,M_\rho}  g_{1T}^{(+)} & f^0  &
\frac{k_L}{\sqrt{2}\,M_\rho}  g_{1T}^{(-)} & \frac{k_R^2}{\sqrt{2}\,M_\rho^2} h_{1T}^{\perp}  &  0  &  \sqrt{2} \, h_1  \\  \\
 \frac{k_R^2}{M_\rho^2}  f_{1TT} 
& \frac{k_R}{\sqrt{2}\,M_\rho} g_{1T}^{(-)} & f^- & 0  &  \frac{k_R^2}{\sqrt{2}\,M_\rho^2}  h_{1T}^{\perp}  &  
- \frac{k_R}{M_\rho}   h_{1L}^{\perp}   \\   \\ 
\frac{k_L}{M_\rho}  h_{1L}^{\perp}  &  
\frac{k_L^2}{\sqrt{2}\,M_\rho^2}  h_{1T}^{\perp} & 0  & f^- &  -\frac{k_L}{\sqrt{2}\, M_\rho} g_{1T}^{(-)}
&  \frac{k_L^2}{M_\rho^2}  f_{1TT} \\  \\
\sqrt{2} \, h_1  &   0 &  \frac{k_L^2}{\sqrt{2}\,M_\rho^2}  h_{1T}^{\perp}  & -\frac{k_R}{\sqrt{2}\, M_\rho} g_{1T}^{(-)} & f^0  &
-\frac{k_L}{\sqrt{2}\, M_\rho} g_{1T}^{(+)} \\ \\
0  &  \sqrt{2} \, h_1  &  -\frac{k_L}{M_\rho}   h_{1L}^{\perp}  &  \frac{k^2_R}{M_\rho^2} f_{1TT} 
& -\frac{k_R}{\sqrt{2}\, M_\rho}  g_{1T}^{(+)} & f^+     
\label{matrix}
\end{array}
\right)\,,
\end{eqnarray}
where 
\begin{eqnarray}
f^+ &=& f_1 - \frac{1}{3}f_{1LL} + g_{1L},\\
f^0 &=& f_1 + \frac{2}{3} f_{1LL},\\
f^- &=& f_1 - \frac{1}{3} f_{1LL} - g_{1L},\\
g_{1T}^{(\pm)} &=& g_{1T} \pm f_{1LT},
\end{eqnarray}
and 
\begin{equation}
k_{R(L)}=k_x\pm \iota k_y.
\end{equation}
By comparing the two matrices given in Eqs.~(\ref{matrix-helicity}) and (\ref{matrix}), we obtain the $\rho$-meson TMDs in terms of the overlaps  of the LFWFs as:
\begin{eqnarray}
f_1(x,{\bf k}^2_\perp)&=&\frac{1}{6(2\pi)^3} \sum_{h_q,h_{\bar{q}}}  \Big(
|\Psi^{0}_{h_q,h_{\bar{q}}}(x,{\bf k}_\perp)|^2\nonumber\\
&&+|\Psi^{+1}_{h_q,h_{\bar{q}}}(x,{\bf k}_\perp)|^2+|\Psi^{-1}_{h_q,h_{\bar{q}}}(x,{\bf k}_\perp)|^2\Big)\,,
\label{f_1}
\end{eqnarray}
\begin{eqnarray}
g_{1L}(x,{\bf k}^2_\perp)&=&\frac{1}{4(2\pi)^3}\sum_{h_{\bar{q}}}  \Big(|\Psi_{+,h_{\bar{q}}}^{+1}(x,{\bf k}_\perp)|^2-|\Psi_{-,h_{\bar{q}}}^{+1}(x,{\bf k}_\perp)|^2
\nonumber\\
&&-|\Psi_{+,h_{\bar{q}}}^{-1}(x,{\bf k}_\perp)|^2 
+|\Psi_{-,h_{\bar{q}}}^{-1}(x,{\bf k}_\perp)|^2
\Big)\,,
\label{g_1l}
\end{eqnarray}
\begin{eqnarray}
g_{1T}(x,{\bf k}^2_\perp)&=&\frac{M_\rho}{4\sqrt{2}\,(2 \pi)^3{\bf k}_\perp^2} \sum_{h_{\bar{q}}}  \bigg(k_R \Big(\Psi^{+1^*}_{+,h_{\bar{q}}}(x,{\bf k}_\perp)\,\Psi^{0}_{+,h_{\bar{q}}}(x,{\bf k}_\perp)\nonumber\\
&&-\Psi^{+1^*}_{-,h_{\bar{q}}}(x,{\bf k}_\perp)\,\Psi^{0}_{-,h_{\bar{q}}}(x,{\bf k}_\perp)+\Psi^{0^*}_{+,h_{\bar{q}}}(x,{\bf k}_\perp)\,\Psi^{-1}_{+,h_{\bar{q}}}(x,{\bf k}_\perp)\nonumber\\
&&-\Psi^{0^*}_{-,h_{\bar{q}}}(x,{\bf k}_\perp)\,\Psi^{-1}_{-,h_{\bar{q}}}(x,{\bf k}_\perp)\Big) + k_L \Big(\Psi^{0^*}_{+,h_{\bar{q}}}(x,{\bf k}_\perp)\,\Psi^{+1}_{+,h_{\bar{q}}}(x,{\bf k}_\perp)\nonumber\\
&& -\Psi^{0^*}_{-,h_{\bar{q}}}(x,{\bf k}_\perp)\,\Psi^{+1}_{-,h_{\bar{q}}}(x,{\bf k}_\perp)+ \Psi^{-1^*}_{+,h_{\bar{q}}}(x,{\bf k}_\perp)\,\Psi^{0}_{+,h_{\bar{q}}}(x,{\bf k}_\perp) \nonumber\\
&&- \Psi^{-1^*}_{-,h_{\bar{q}}}(x,{\bf k}_\perp)\,\Psi^{0}_{-,h_{\bar{q}}}(x,{\bf k}_\perp)\Big)\bigg)\,,
\label{g_1t}
\end{eqnarray}
\begin{eqnarray}
h_1(x,{\bf k}^2_\perp)&=&\frac{1}{4\sqrt{2}(2\pi)^3}\sum_{h_{\bar{q}}} \Big(\Psi_{+,h_{\bar{q}}}^{+1^*}(x,{\bf k}_\perp)\, \Psi_{-,h_{\bar{q}}}^{0}(x,{\bf k}_\perp) \nonumber\\
&&+\Psi_{-,h_{\bar{q}}}^{0^*}(x,{\bf k}_\perp)\, \Psi_{+,h_{\bar{q}}}^{+1}(x,{\bf k}_\perp)+\Psi_{+,h_{\bar{q}}}^{0^*}(x,{\bf k}_\perp)\, \Psi_{-,h_{\bar{q}}}^{-1}(x,{\bf k}_\perp)\nonumber\\
&&+\Psi_{-,h_{\bar{q}}}^{-1^*}(x,{\bf k}_\perp)\, \Psi_{+,h_{\bar{q}}}^{0}(x,{\bf k}_\perp) \Big)\,,
\label{h_1}
\end{eqnarray}
\begin{eqnarray}
h_{1L}^\perp(x,{\bf k}^2_\perp)&=&\frac{M_\rho}{4 (2\pi)^3{\bf k}_\perp^2}\sum_{h_{\bar{q}}}  \bigg(k_R \Big(\Psi_{-,h_{\bar{q}}}^{+1^*}(x,{\bf k}_\perp)\, \Psi_{+,h_{\bar{q}}}^{+1}(x,{\bf k}_\perp)\nonumber\\
&&-\Psi_{-,h_{\bar{q}}}^{-1^*}(x,{\bf k}_\perp)\, \Psi_{+,h_{\bar{q}}}^{-1}(x,{\bf k}_\perp) \Big) + k_L \Big(\Psi_{+,h_{\bar{q}}}^{+1^*}(x,{\bf k}_\perp)\, \Psi_{-,h_{\bar{q}}}^{+1}(x,{\bf k}_\perp)\nonumber\\
&&-\Psi_{+,h_{\bar{q}}}^{-1^*}(x,{\bf k}_\perp)\, \Psi_{-,h_{\bar{q}}}^{-1}(x,{\bf k}_\perp)\Big)\bigg)\,,
\label{h_1Lp}
\end{eqnarray}
\begin{eqnarray}
h_{1T}^\perp(x,{\bf k}^2_\perp)&=&\frac{M_\rho^2}{2\sqrt{2} (2\pi)^3}\frac{1}{{\bf k}_\perp^4}\sum_{h_{\bar{q}}}  \bigg(k_R^2 \Big( \Psi_{-,h_{\bar{q}}}^{+1^*}(x,{\bf k}_\perp)\,\Psi_{+,h_{\bar{q}}}^{0}(x,{\bf k}_\perp)\nonumber\\
&&+ \Psi_{-,h_{\bar{q}}}^{0^*}(x,{\bf k}_\perp)\,\Psi_{+,h_{\bar{q}}}^{-1}(x,{\bf k}_\perp) \Big) +k_L^2 \Big( \Psi_{+,h_{\bar{q}}}^{0^*}(x,{\bf k}_\perp)\,\Psi_{-,h_{\bar{q}}}^{+1}(x,{\bf k}_\perp)\nonumber\\
&&+\Psi_{+,h_{\bar{q}}}^{-1^*}(x,{\bf k}_\perp)\,\Psi_{-,h_{\bar{q}}}^{0}(x,{\bf k}_\perp)\Big)\bigg)\,,
\label{h_1t}
\end{eqnarray}
\begin{eqnarray}
f_{1LL}(x,{\bf k}^2_\perp)&=&\frac{1}{2(2\pi)^3}\sum_{h_q,h_{\bar{q}}} \bigg(
|\Psi^{0}_{h_q,h_{\bar{q}}}(x,{\bf k}_\perp)|^2
\nonumber\\
&&-\frac{1}{2}\left(|\Psi^{+1}_{h_q,h_{\bar{q}}}(x,{\bf k}_\perp)|^2 +|\Psi^{-1}_{h_q,h_{\bar{q}}}(x,{\bf k}_\perp)|^2\right)\bigg)\,,
\label{f_1LL}
\end{eqnarray}
\begin{eqnarray}
f_{1LT}(x,{\bf k}^2_\perp)&=&\frac{M_\rho}{4\sqrt{2}(2\pi)^3{\bf k}_\perp^2} \sum_{h_{\bar{q}}} \bigg(k_R \Big( \Psi^{+1^*}_{+,h_{\bar{q}}}(x,{\bf k}_\perp)\,\Psi^{0}_{+,h_{\bar{q}}}(x,{\bf k}_\perp)\nonumber\\
&&+\Psi^{+1^*}_{-,h_{\bar{q}}}(x,{\bf k}_\perp)\,\Psi^{0}_{-,h_{\bar{q}}}(x,{\bf k}_\perp)-\Psi^{0^*}_{+,h_{\bar{q}}}(x,{\bf k}_\perp)\,\Psi^{-1}_{+,h_{\bar{q}}}(x,{\bf k}_\perp)\nonumber\\
&&-\Psi^{0^*}_{-,h_{\bar{q}}}(x,{\bf k}_\perp)\,\Psi^{-1}_{-,h_{\bar{q}}}(x,{\bf k}_\perp)\Big) + k_L \Big(\Psi^{0^*}_{+,h_{\bar{q}}}(x,{\bf k}_\perp)\,\Psi^{+1}_{+,h_{\bar{q}}}(x,{\bf k}_\perp)\nonumber\\
&& +\Psi^{0^*}_{-,h_{\bar{q}}}(x,{\bf k}_\perp)\,\Psi^{+1}_{-,h_{\bar{q}}}(x,{\bf k}_\perp)-\Psi^{-1^*}_{+,h_{\bar{q}}}(x,{\bf k}_\perp)\,\Psi^{0}_{+,h_{\bar{q}}}(x,{\bf k}_\perp)\nonumber\\
&&-\Psi^{-1^*}_{-,h_{\bar{q}}}(x,{\bf k}_\perp)\,\Psi^{0}_{-,h_{\bar{q}}}(x,{\bf k}_\perp)\Big)\bigg)\,,
\label{f_1lt}
\end{eqnarray}
\begin{eqnarray}
f_{1TT}(x,{\bf k}^2_\perp)&=&\frac{M^2_\rho}{4(2\pi)^3{\bf k}_\perp^4}\sum_{h_{\bar{q}}}\bigg(k_R^2\Big(\Psi_{+,h_{\bar{q}}}^{+1^*}(x,{\bf k}_\perp)\,\Psi_{+,h_{\bar{q}}}^{-1}(x,{\bf k}_\perp)\nonumber\\
&&+\Psi_{-,h_{\bar{q}}}^{+1^*}(x,{\bf k}_\perp)\,\Psi_{-,h_{\bar{q}}}^{-1}(x,{\bf k}_\perp)\Big)+k_L^2\Big( \Psi_{+,h_{\bar{q}}}^{-1^*}(x,{\bf k}_\perp)\,\Psi_{+,h_{\bar{q}}}^{+1}(x,{\bf k}_\perp)\nonumber\\
&&+ \Psi_{-,h_{\bar{q}}}^{-1^*}(x,{\bf k}_\perp)\,\Psi_{-,h_{\bar{q}}}^{+1}(x,{\bf k}_\perp)\Big)\bigg)\,.
\label{f_1tt}
\end{eqnarray}
%
Using the holographic LFWFs, Eqs.~(\ref{longitudinal-component})-(\ref{transverse-component}),
in the overlap representations, Eqs.~(\ref{f_1})-(\ref{f_1tt}), we extract the explicit expressions for the leading-twist T-even TMDs for the $\rho$-meson in the LF holographic model as:
\begin{eqnarray}
f_1(x,{\bf k}^2_\perp)&=& \frac{1}{3(2 \pi)^3}\bigg(\mathcal{N}^2_L\left( M^2_\rho~ x(1-x)+m_q^2+{\bf k}_\perp^2 \right)^2 \frac{|\psi(x,{\bf k}^2_\perp)| ^2 }{{x^2 (1-x)^2}}\nonumber\\
&+& \mathcal{N}^2_T\left(m_q^2 + {\bf k}^2_\perp \left(2 x^2-2 x+1\right)\right)\frac{|\psi(x,{\bf k}^2_\perp)|^2}{{x^2 (1-x)^2}} \bigg)\,,\label{f1-LFH} 
\end{eqnarray}
\begin{eqnarray}
g_{1L}(x,{\bf k}^2_\perp)&=& \frac{\mathcal{N}^2_T}{2(2 \pi)^3}\left(m_q^2+{\bf k}^2_\perp\left(2 x-1\right) \right)\frac{|\psi(x,{\bf k}^2_\perp)| ^2}{{x^2 (1-x)^2}}\,,\label{g1l-LFH}
\end{eqnarray}
\begin{eqnarray}
g_{1T}(x,{\bf k}^2_\perp)&=&\mathcal{N}_L \mathcal{N}_T\frac{M_\rho}{\sqrt{2}(2 \pi)^3}\left(M^2_\rho~ x(1-x)+m_q^2 +{\bf k}^2_\perp \right)\frac{|\psi(x,{\bf k}^2_\perp)|^2}{{x^2 (1-x)^2}}\,,\label{g1t-LFH}\nonumber\\
\end{eqnarray}
\begin{eqnarray}
h_1(x,{\bf k}^2_\perp)&=& \mathcal{N}_L \mathcal{N}_T\frac{m_q}{\sqrt{2}(2 \pi)^3}\left(M^2_\rho~ x(1-x)+m_q^2 +{\bf k}^2_\perp \right)\frac{|\psi(x,{\bf k}^2_\perp)|^2}{{x^2 (1-x)^2}}\,, \label{h1-LFH}\nonumber\\
\end{eqnarray}
\begin{eqnarray}
h^\perp_{1L}(x,{\bf k}^2_\perp)&=& -\mathcal{N}^2_T\frac{m_q \, M_\rho}{(2 \pi)^3}\frac{|\psi(x,{\bf k}^2_\perp)| ^2}{x^2 (1-x)}\,, \label{hlp-LFH}
\end{eqnarray}
\begin{eqnarray}
h^\perp_{1T}(x,{\bf k}^2_\perp)&=&0\,,\label{htp-LFH}
\end{eqnarray}
\begin{eqnarray}
f_{1LL}(x,{\bf k}^2_\perp)&=& \frac{1}{(2 \pi)^3}\bigg(\mathcal{N}^2_L\left( M^2_\rho~ x(1-x)+m_q^2+{\bf k}_\perp^2 \right)^2 \frac{|\psi(x,{\bf k}^2_\perp)| ^2 }{x^2 (1-x)^2}\nonumber\\
&&-\mathcal{N}^2_T\left(m_q^2 + {\bf k}^2_\perp \left(2 x^2-2 x+1\right)\right)\frac{|\psi(x,{\bf k}^2_\perp)|^2}{{2 x^2 (1-x)^2}}\bigg)\,,\label{f1ll-LFH}
\end{eqnarray}
\begin{eqnarray}
f_{1LT}(x,{\bf k}^2_\perp)&=&\mathcal{N}_L \mathcal{N}_T \frac{M_\rho}{ \sqrt{2}(2 \pi)^3}\left(2x-1 \right)\left( M^2_\rho~ x(1-x)+m_q^2+{\bf k}_\perp^2\right)\nonumber\\
&& \times \frac{|\psi(x,{\bf k}^2_\perp)|^2}{{x^2 (1-x)^2}}\,,\label{f1lt-LFH}
\end{eqnarray}
\begin{eqnarray}
f_{1TT}(x,{\bf k}^2_\perp)&=& \mathcal{N}^2_T\frac{M^2_\rho}{(2 \pi)^3} \frac{|\psi(x,{\bf k}^2_\perp)| ^2}{x (1-x)}\,,\label{f1tt-LFH}
\end{eqnarray}
where
$\psi(x,{\bf k}^2_\perp)$ is given in Eq. (\ref{holographic-wf}).
Meanwhile, all the T-even TMDs in the LF quark model are evaluated in Appendix~\ref{LFmodel}.
%
\begin{table}
\scriptsize
\centering
\begin{tabular}{c c c c c}
\hline \hline
\\
OAM $L_z$ & & Configurations: $\Ket{\Lambda} \rightarrow \Ket{h_q + h_{\bar{q}}}$ & & \\ \\ \hline
& & & & \\
$-2$ & $\Ket{-1} \rightarrow \Ket{+\frac{1}{2}+\frac{1}{2}}$ & & & \\
\\
$-1$ & $\Ket{0} \rightarrow \Ket{+\frac{1}{2}+\frac{1}{2}}$ & $\Ket{-1} \rightarrow \Ket{+\frac{1}{2}-\frac{1}{2}}$ & $\Ket{-1} \rightarrow \Ket{-\frac{1}{2}+\frac{1}{2}}$ \\ 
\\
$0$ & $\Ket{+1} \rightarrow \Ket{+\frac{1}{2}+\frac{1}{2}}$ & $\Ket{0} \rightarrow \Ket{+\frac{1}{2}-\frac{1}{2}}$ & $\Ket{0} \rightarrow \Ket{-\frac{1}{2}+\frac{1}{2}}$ & $\Ket{-1} \rightarrow \Ket{-\frac{1}{2}-\frac{1}{2}}$ \\
\\
$+1$ & $\Ket{+1} \rightarrow \Ket{+\frac{1}{2}-\frac{1}{2}}$ & $\Ket{+1} \rightarrow \Ket{-\frac{1}{2}+\frac{1}{2}}$ & $\Ket{0} \rightarrow \Ket{-\frac{1}{2}-\frac{1}{2}}$ \\
\\
$+2$ & $\Ket{+1} \rightarrow \Ket{-\frac{1}{2}-\frac{1}{2}}$ \\
\\
\hline
\end{tabular}
\caption{The possible orbital angular momentum $L_z$ contributions for $\rho$-meson light-front wave functions based on the different configurations of spin projections of valence quarks $h_q, h_{\bar{q}}$ and hadron spin $\Lambda$.}
\label{table}
\end{table}
%

The LFWFs satisfy the angular momentum conservation projected along $z$-axis, i.e. $J_z=\sum_{i=1}^n s_z^i +\sum^{n-1}_{j=1} L_z^j$, where $n=2$ in this particular case. The intrinsic spin contribution is denoted by $s^1_z+s^2_z$, while the relative orbital angular momentum (OAM) is $L_z$ for each configuration of LFWF. Here, $s_z^1$ and $s_z^2$ represent $h_q$ and $h_{\bar{q}}$, respectively and $L_z=\Lambda-(h_q + h_{\bar{q}})$. 
 For the $\rho$-meson, the different configurations of the LFWFs with the OAMs $L_z=0, \pm 1$ and $ \pm 2$, which correspond to the S, P and D wave compositions respectively, are listed in Table \ref{table}. 
 
 We observe that $f_1$, $g_{1L}$, $h_1$ and $f_{1LL}$ are all diagonal in OAM in the overlap representation. In other words, there is zero OAM transfer from the initial to the final state of the hadron. Meanwhile, the overlap configurations of the other TMDs show interference between several wave compositions, which may refer to the non-zero OAM transfer from the initial to the final state of the $\rho$-meson.
\begin{figure}[hbt]
\centering
\begin{minipage}[c]{1\textwidth}\begin{center}
(a)\includegraphics[width=.4\textwidth]{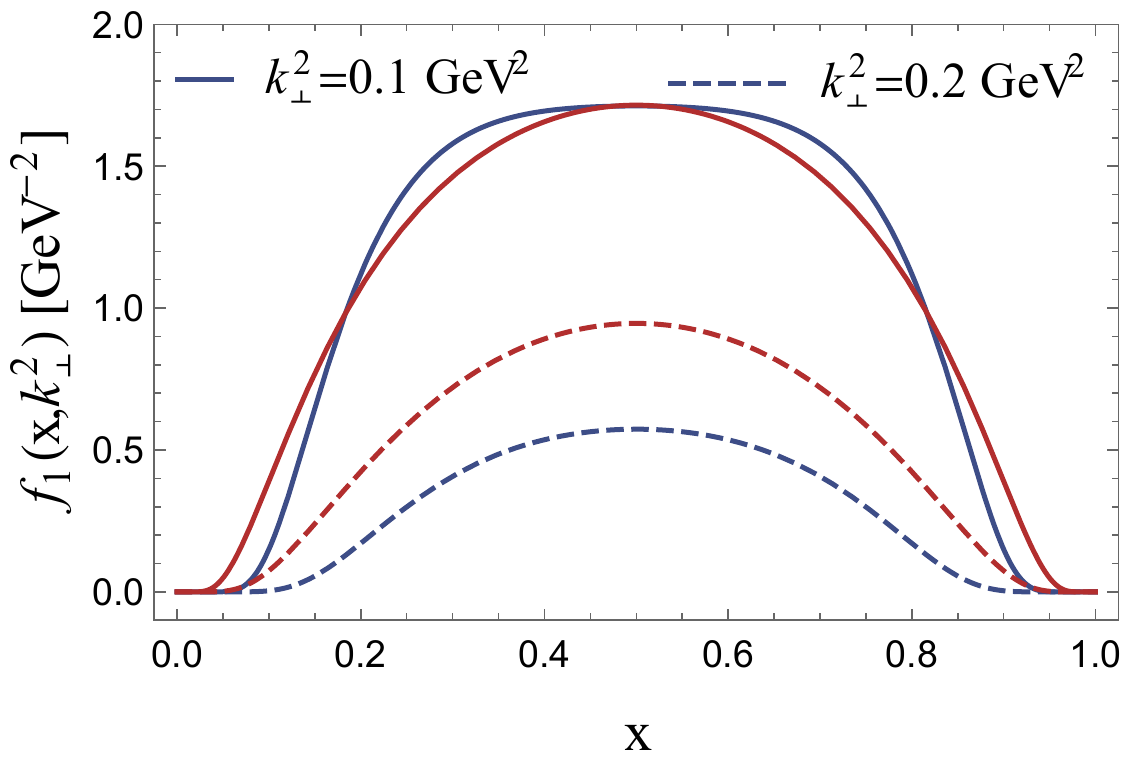}
(b)\includegraphics[width=.38\textwidth]{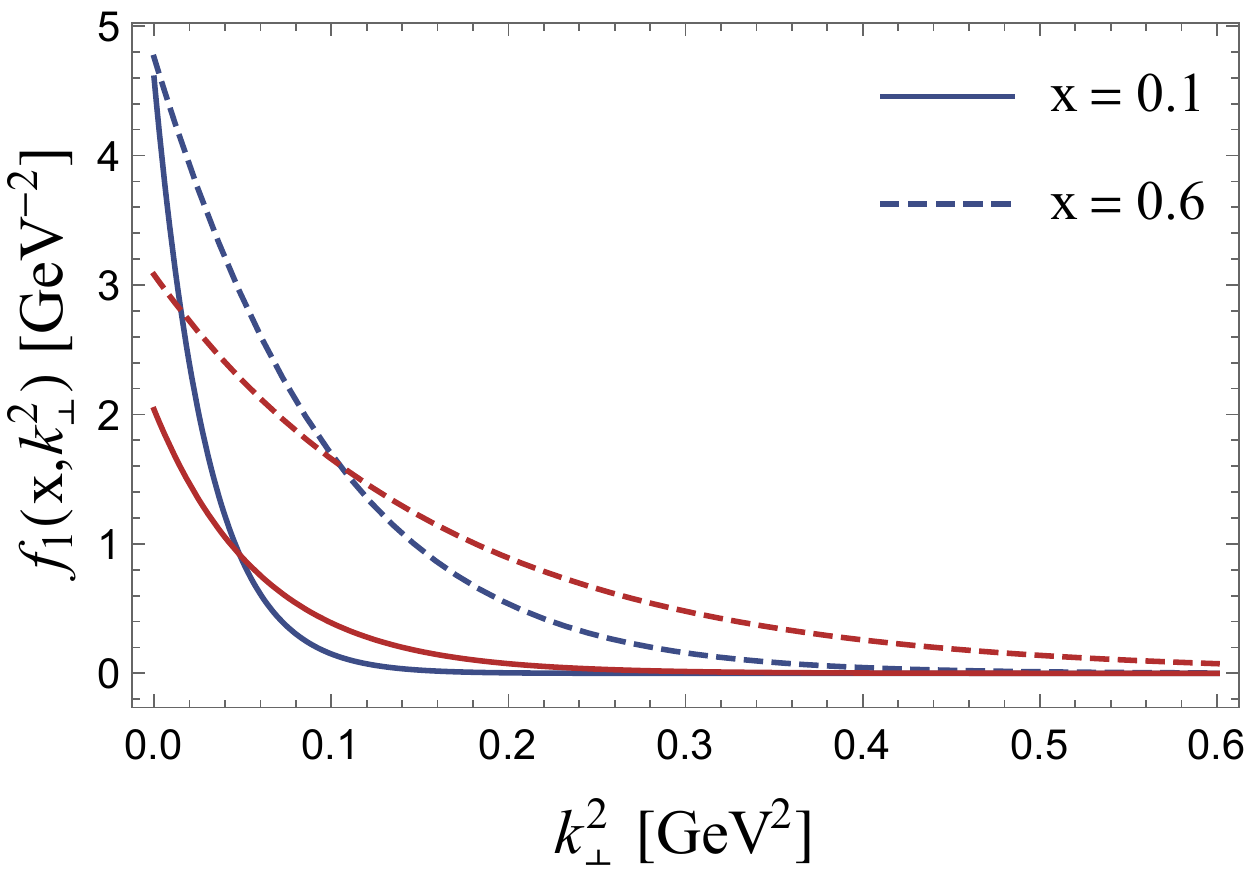}\end{center}
\end{minipage}
\begin{minipage}[c]{1\textwidth}\begin{center}
(c)\includegraphics[width=.38\textwidth]{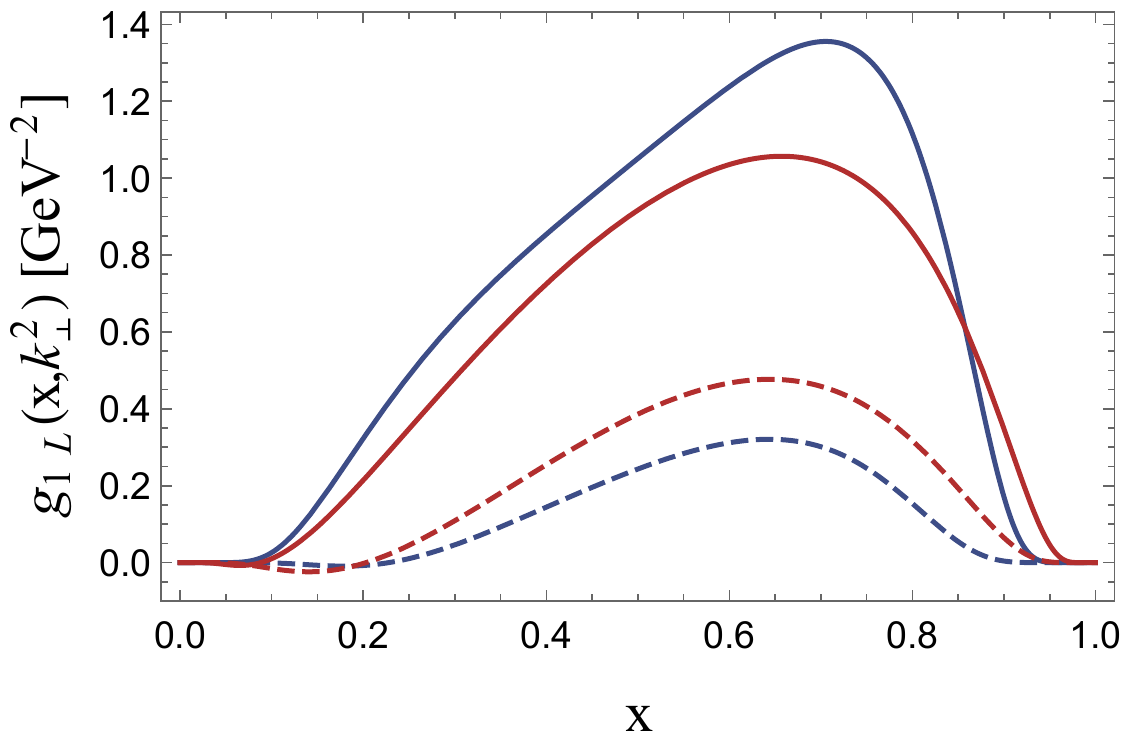}
(d)\includegraphics[width=.38\textwidth]{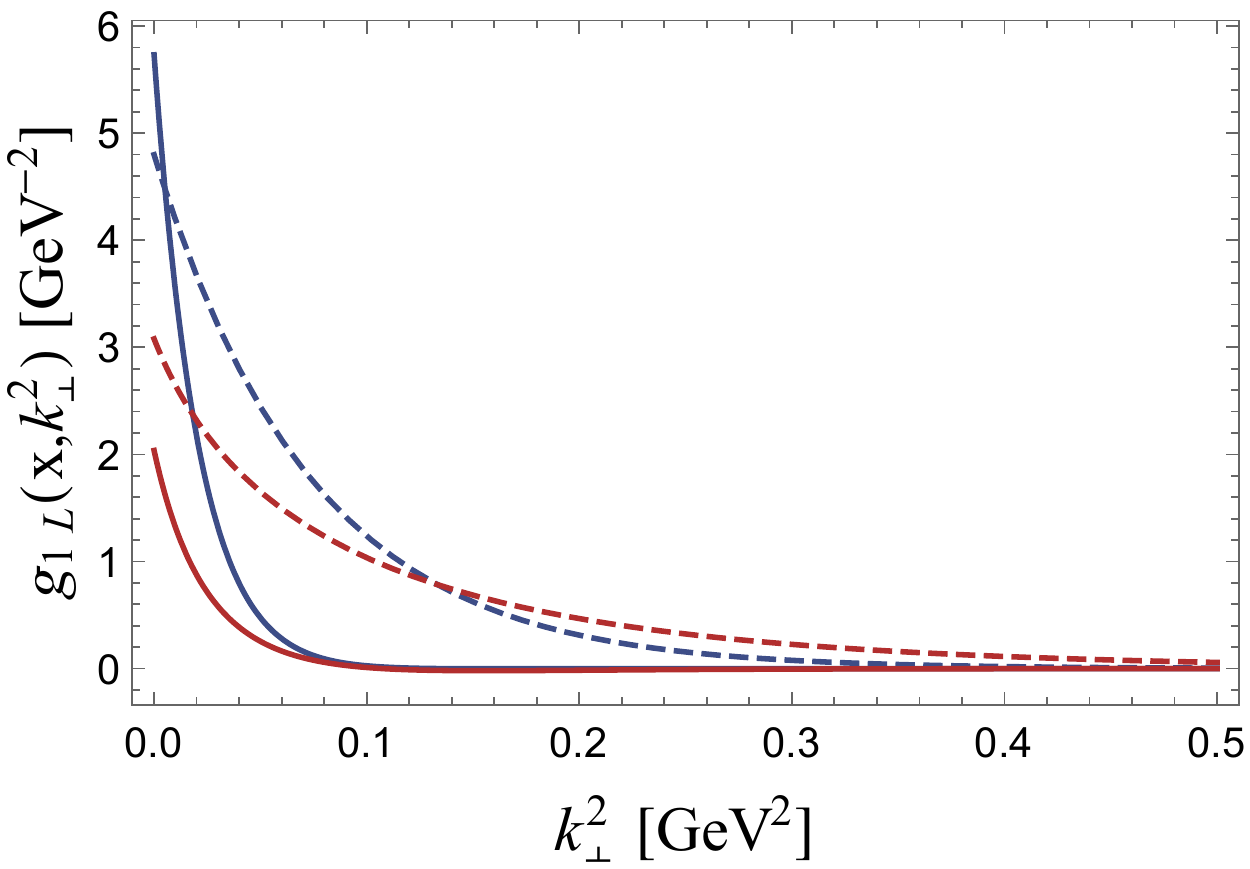}\end{center}
\end{minipage}
\begin{minipage}[c]{1\textwidth}\begin{center}
(e)\includegraphics[width=.38\textwidth]{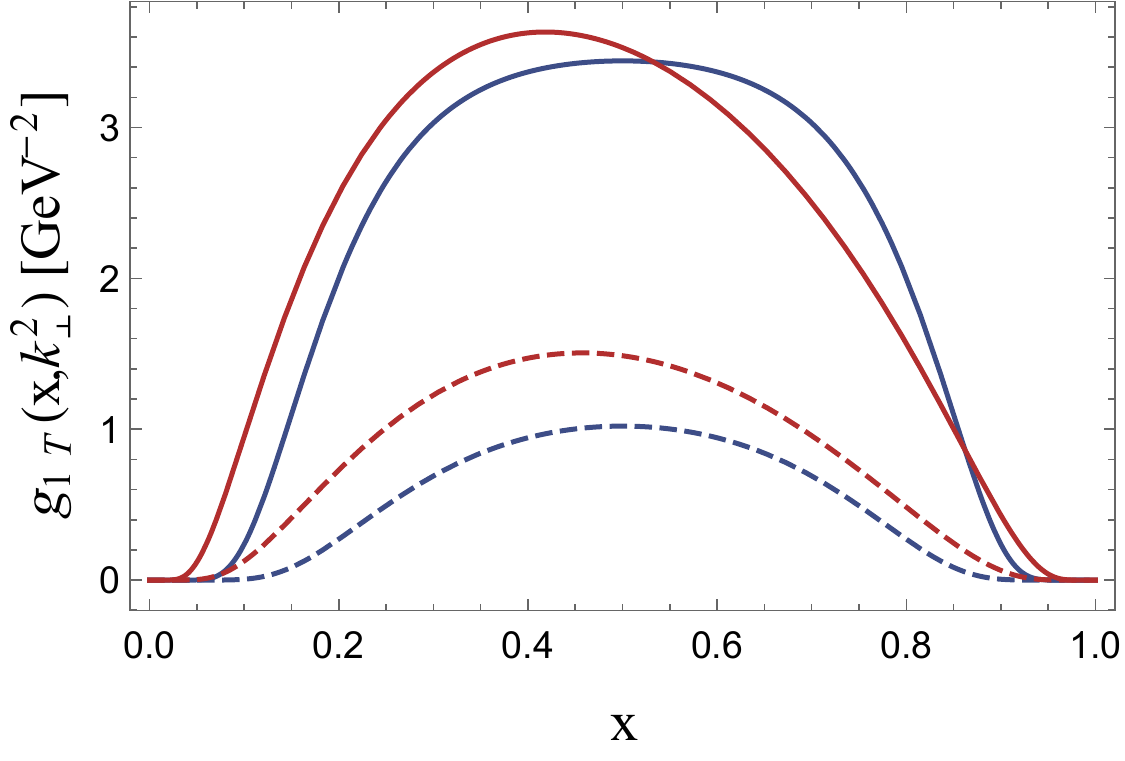}
(f)\includegraphics[width=.38\textwidth]{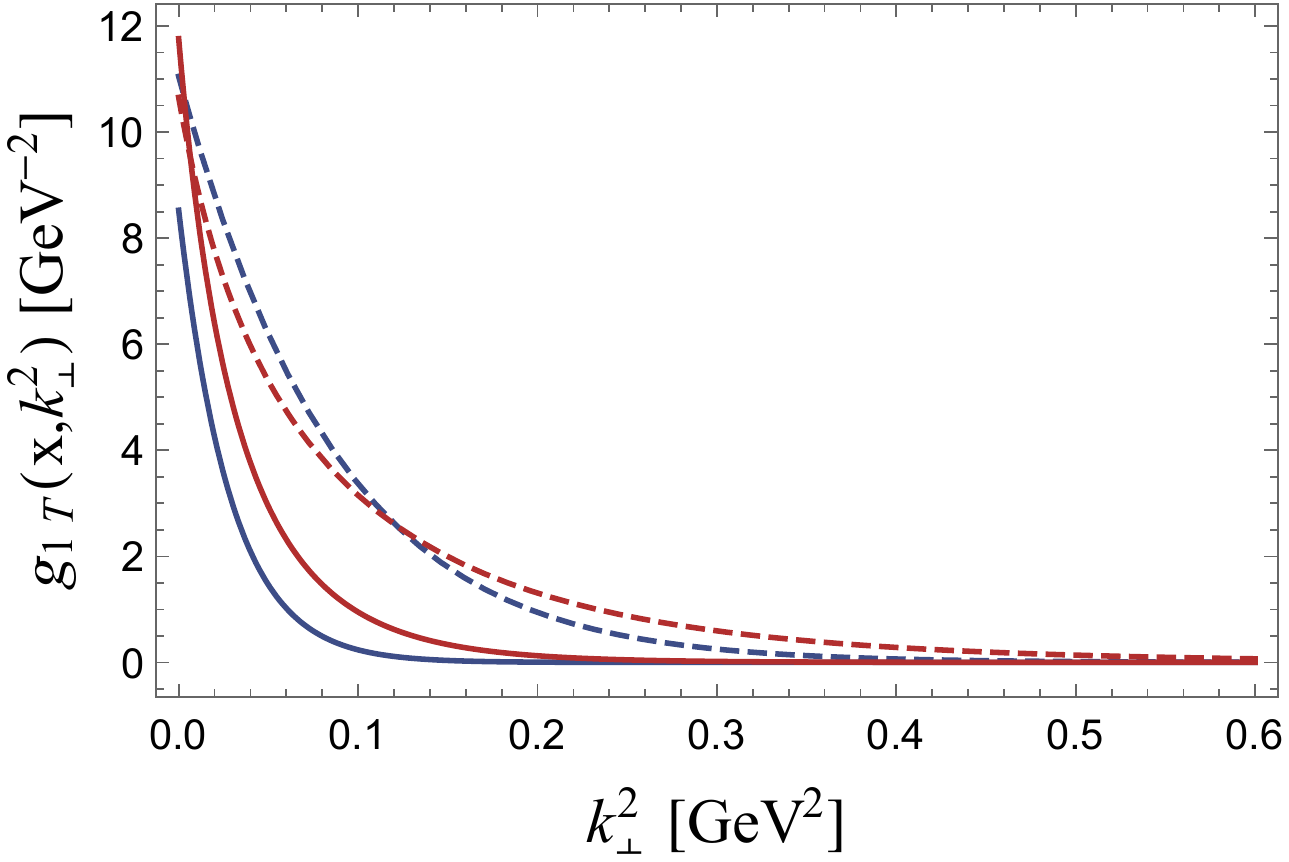}\end{center}
\end{minipage}
\caption{(Color online) The unpolarized TMD $f_1(x,{\bf k}^2_\perp)$ and the longitudinally polarized quark TMDs $g_{1L}(x,{\bf k}^2_\perp)$ and  $g_{1T}(x,{\bf k}^2_\perp)$ with respect to $x$ at different value of ${\bf k}^2_\perp$ (left panel), i.e., ${\bf k}^2_\perp=0.1 {\ \rm GeV}^2$ (solid curves) and ${\bf k}^2_\perp=0.2 {\ \rm GeV}^2$ (dashed curves). On the right panel, these TMDs are shown with respect to ${\bf k}^2_\perp$ at different values of $x$, i.e., $x=0.3$ (solid curves) and $x=0.6$ (dashed curves). The blue and red curves correspond to the light-front holographic model at the model scale $\mu_{\rm LFH}^2=0.20$ GeV$^2$ and the light-front quark model predictions at the model scale $\mu_{\rm LFQM}^2=0.19$ GeV$^2$, respectively.} 
\label{tmds1}
\end{figure}
\begin{figure}[hbt]
\begin{minipage}[c]{1\textwidth}\begin{center}
(a)\includegraphics[width=.38\textwidth]{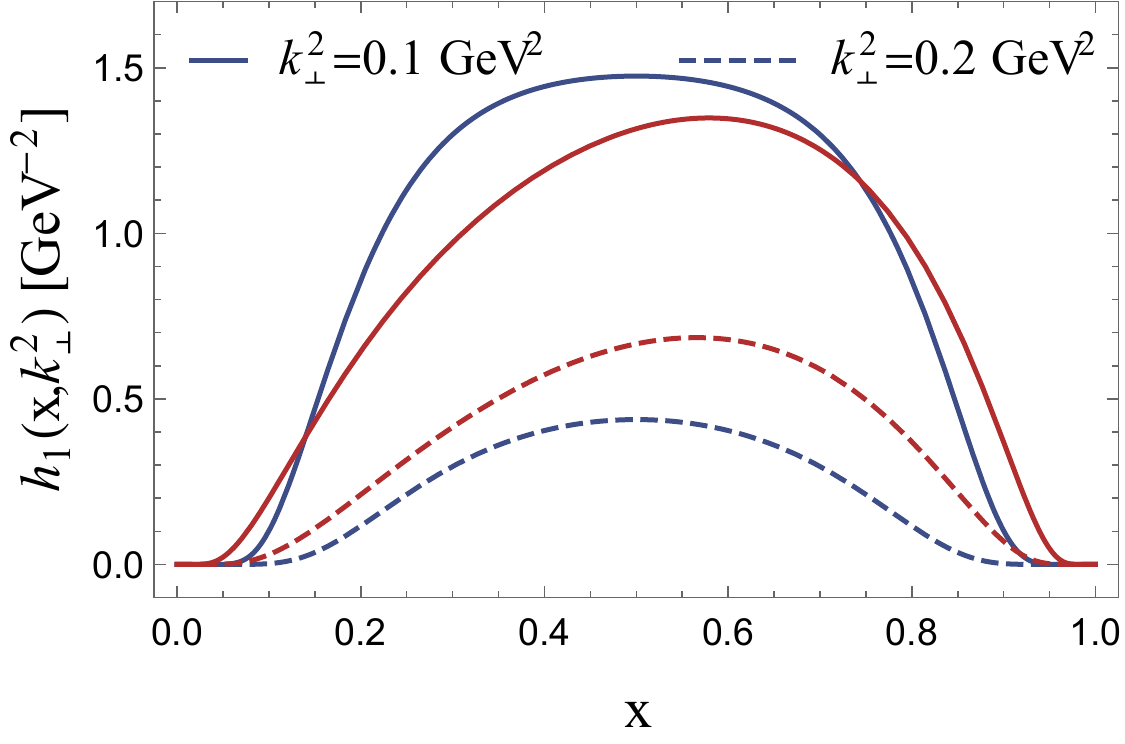}
(b)\includegraphics[width=.38\textwidth]{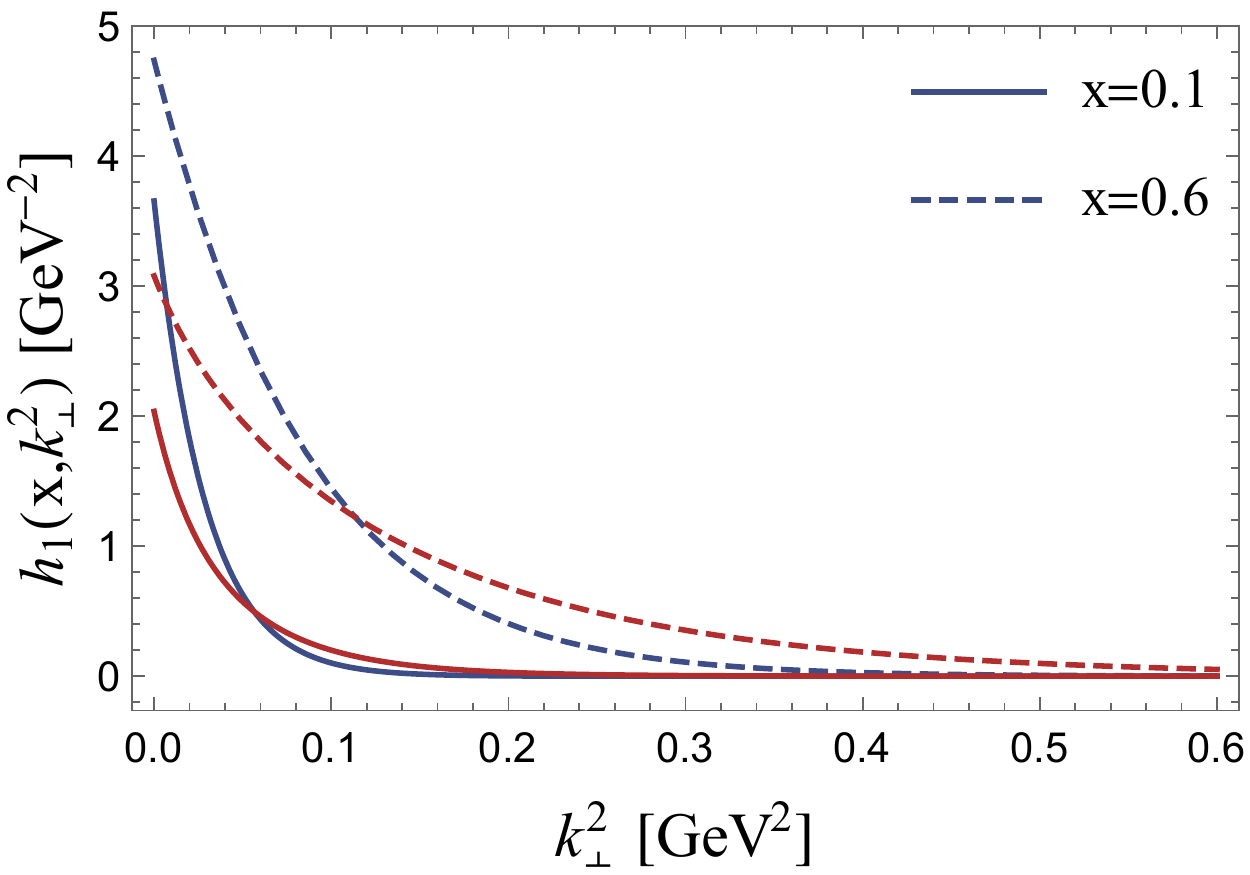}\end{center}
\end{minipage}
\begin{minipage}[c]{1\textwidth}\begin{center}
(c)\includegraphics[width=.38\textwidth]{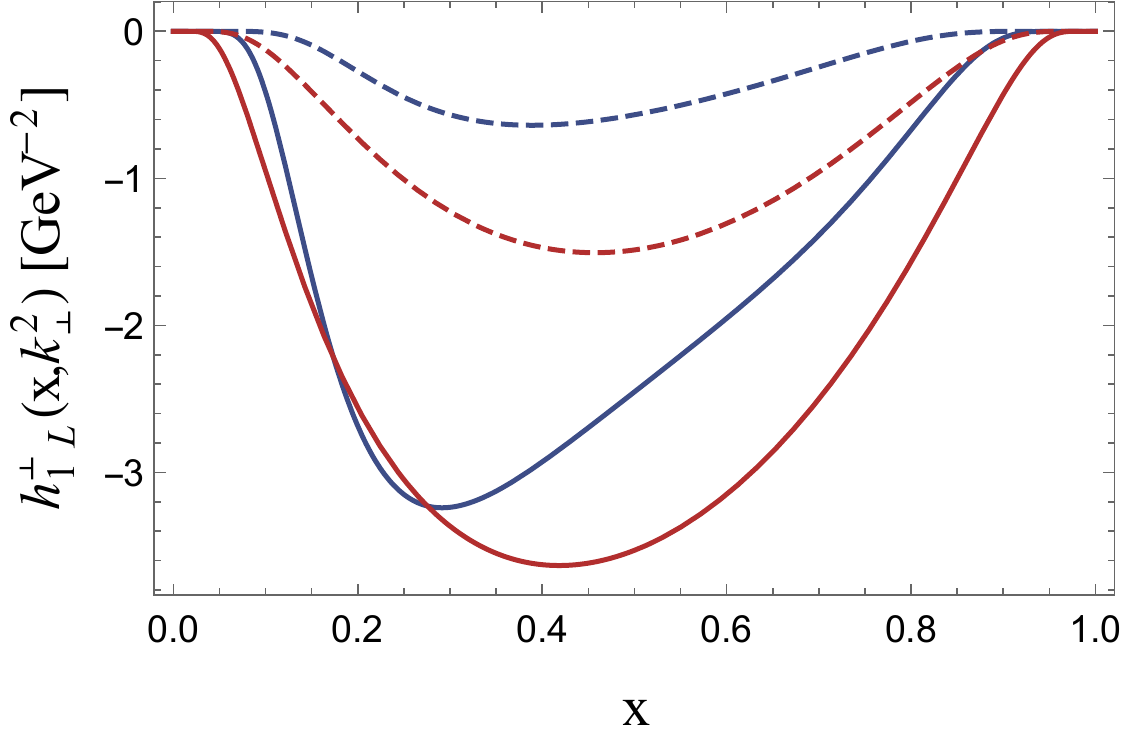}
(d)\includegraphics[width=.38\textwidth]{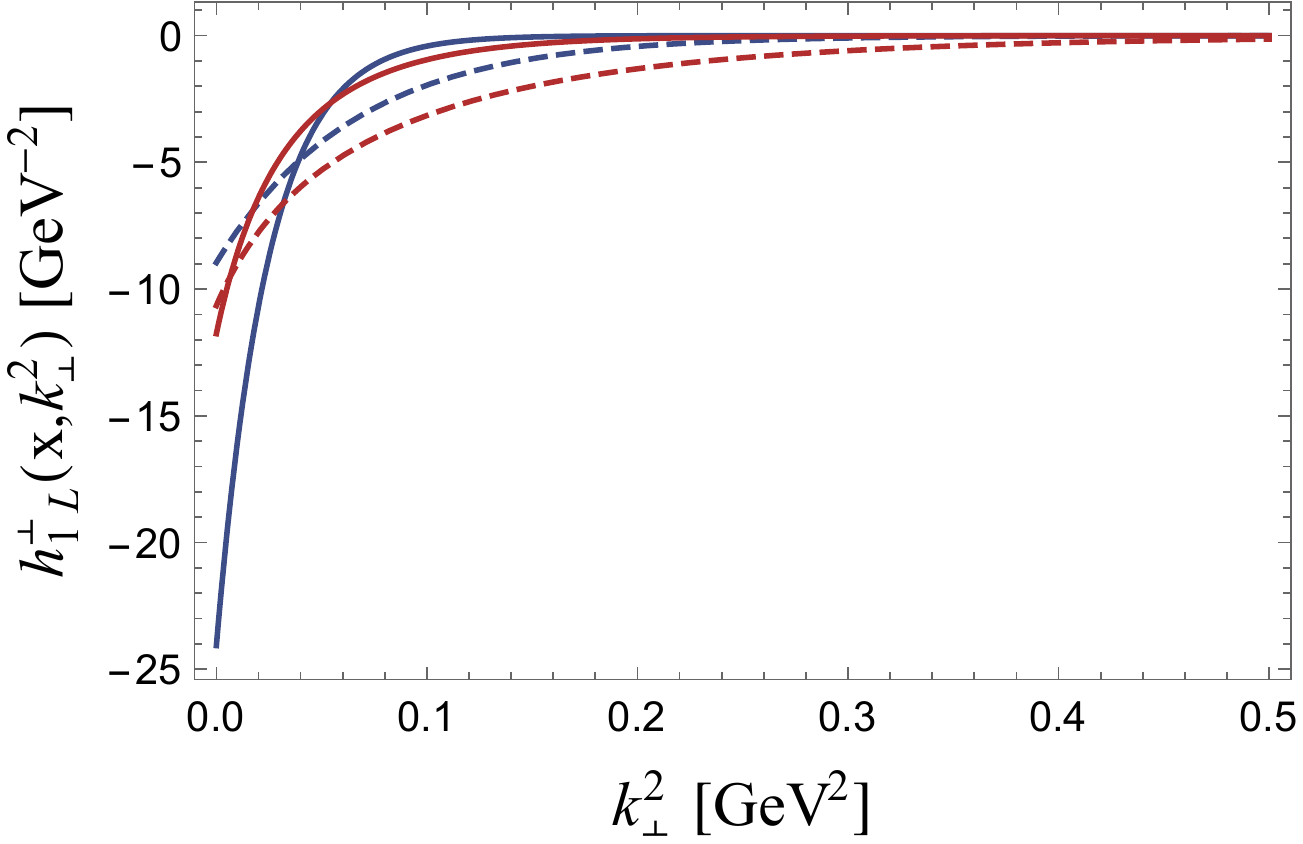}\end{center}
\end{minipage}
\begin{minipage}[c]{1\textwidth}\begin{center}
(e)\includegraphics[width=.38\textwidth]{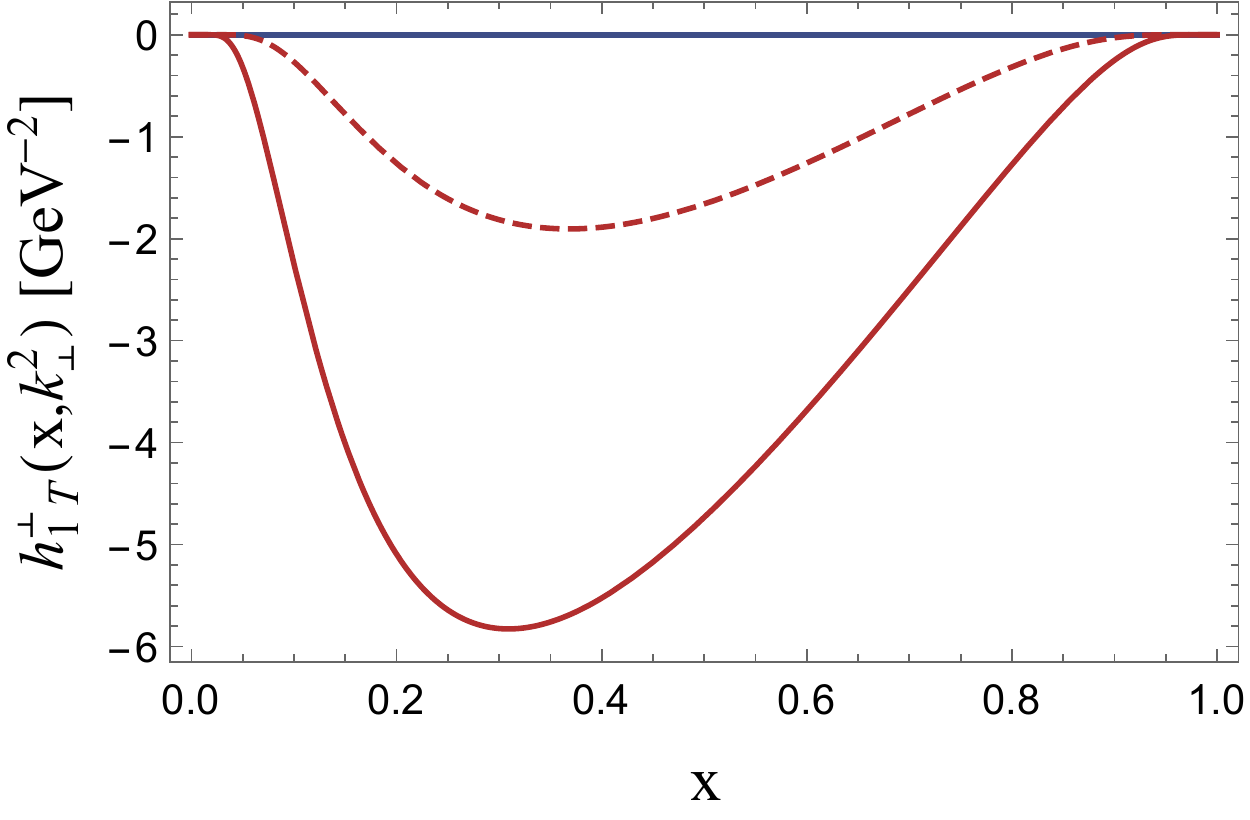}
(f)\includegraphics[width=.38\textwidth]{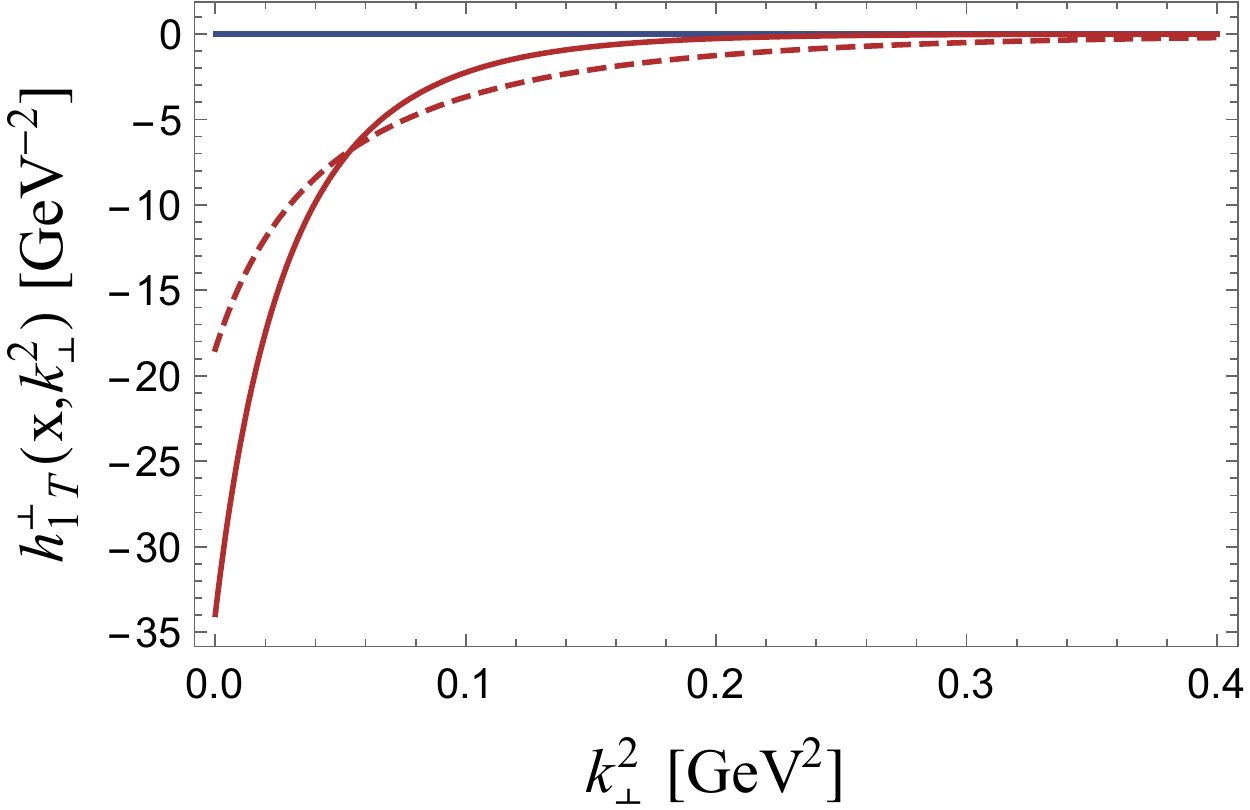}\end{center}
\end{minipage}
\caption{(Color online) The transversely polarized quark TMDs $h_1(x,{\bf k}^2_\perp)$, $h^\perp_{1L}(x,{\bf k}^2_\perp)$ and $h^\perp_{1T}(x,{\bf k}^2_\perp)$ with respect to $x$ at different value of ${\bf k}^2_\perp$ (left panel), i.e., ${\bf k}^2_\perp=0.1 {\ \rm GeV}^2$ (solid curves) and ${\bf k}^2_\perp=0.2 {\ \rm GeV}^2$ (dashed curves). On the right panel, these TMDs are shown with respect to ${\bf k}^2_\perp$ at different values of $x$, i.e., $x=0.3$ (solid curves) and $x=0.6$ (dashed curves). The blue and red curves correspond to the light-front holographic model at the model scale $\mu_{\rm LFH}^2=0.20$ GeV$^2$ and the light-front quark model predictions at the model scale $\mu_{\rm LFQM}^2=0.19$ GeV$^2$, respectively.} 
\label{tmds2}
\end{figure}
\begin{figure}[hbt]
\begin{minipage}[c]{1\textwidth}\begin{center}
(a)\includegraphics[width=.4\textwidth]{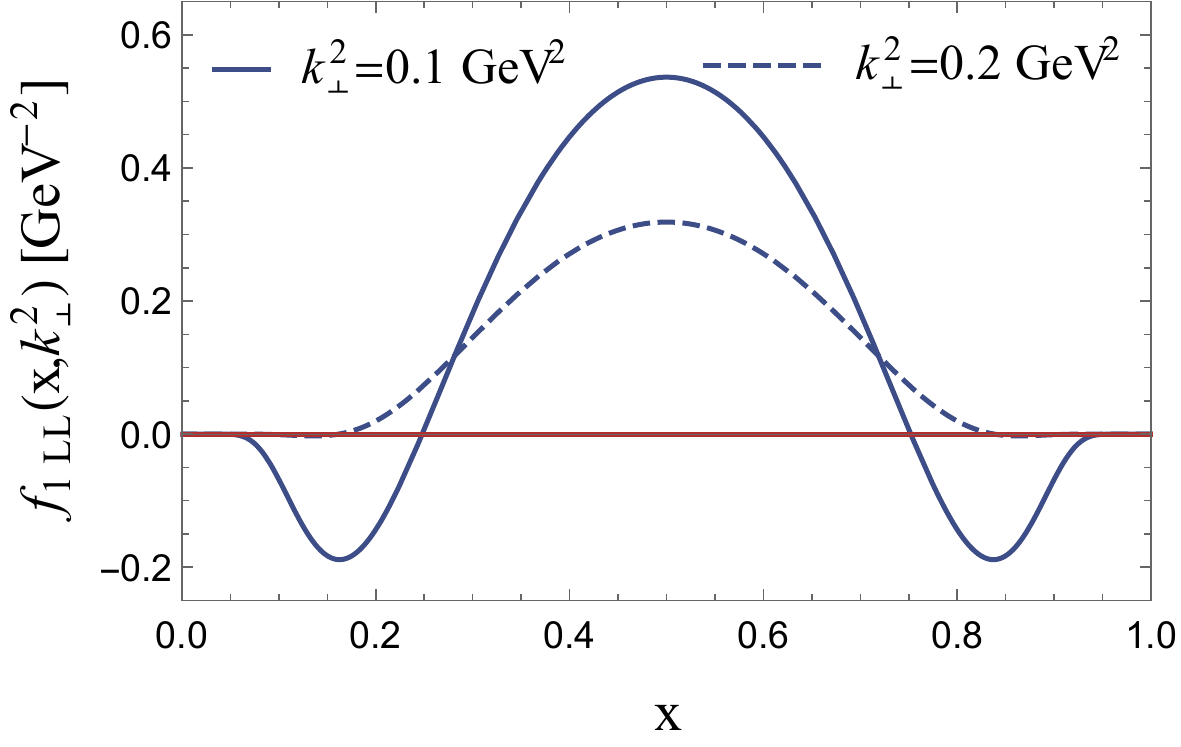}
(b)\includegraphics[width=.38\textwidth]{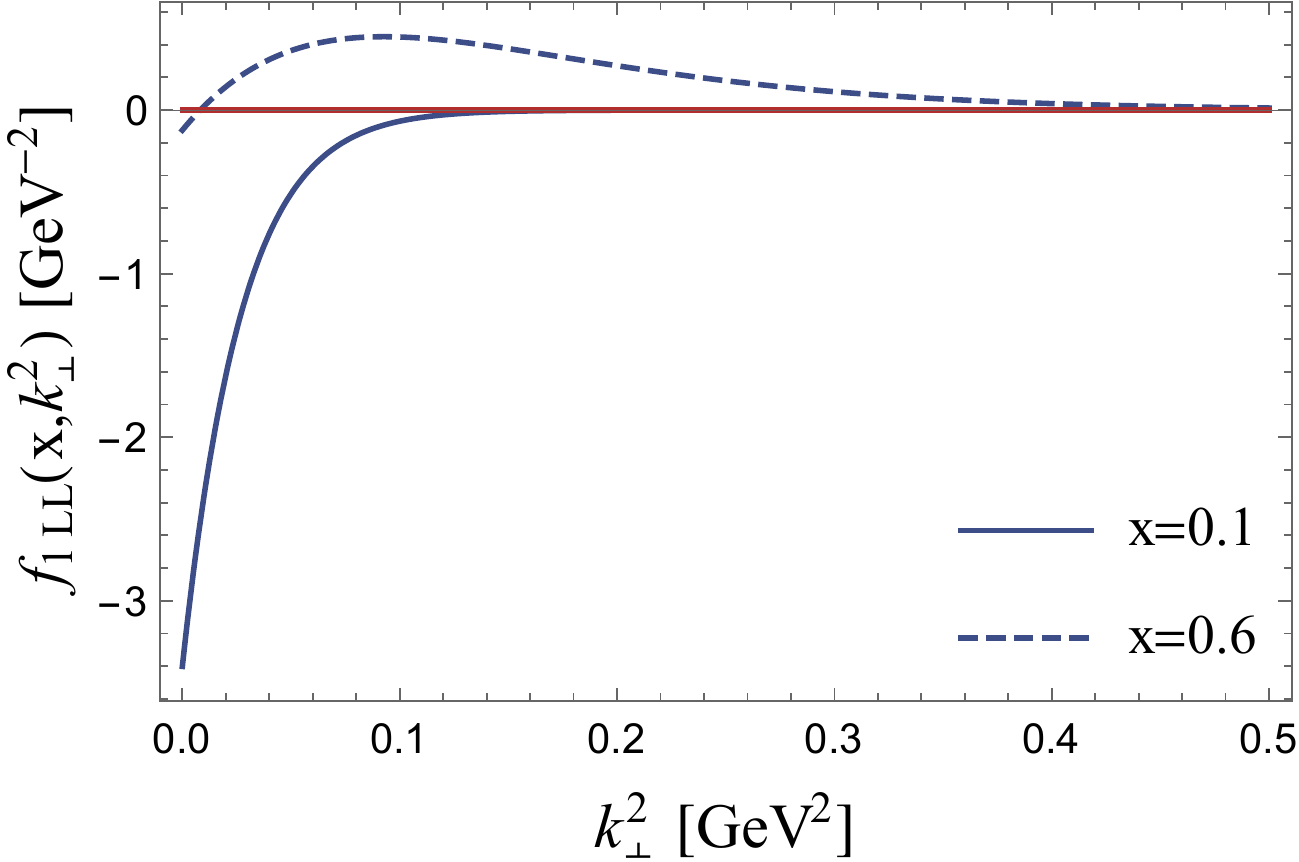}\end{center}
\end{minipage}
\begin{minipage}[c]{1\textwidth}\begin{center}
(c)\includegraphics[width=.38\textwidth]{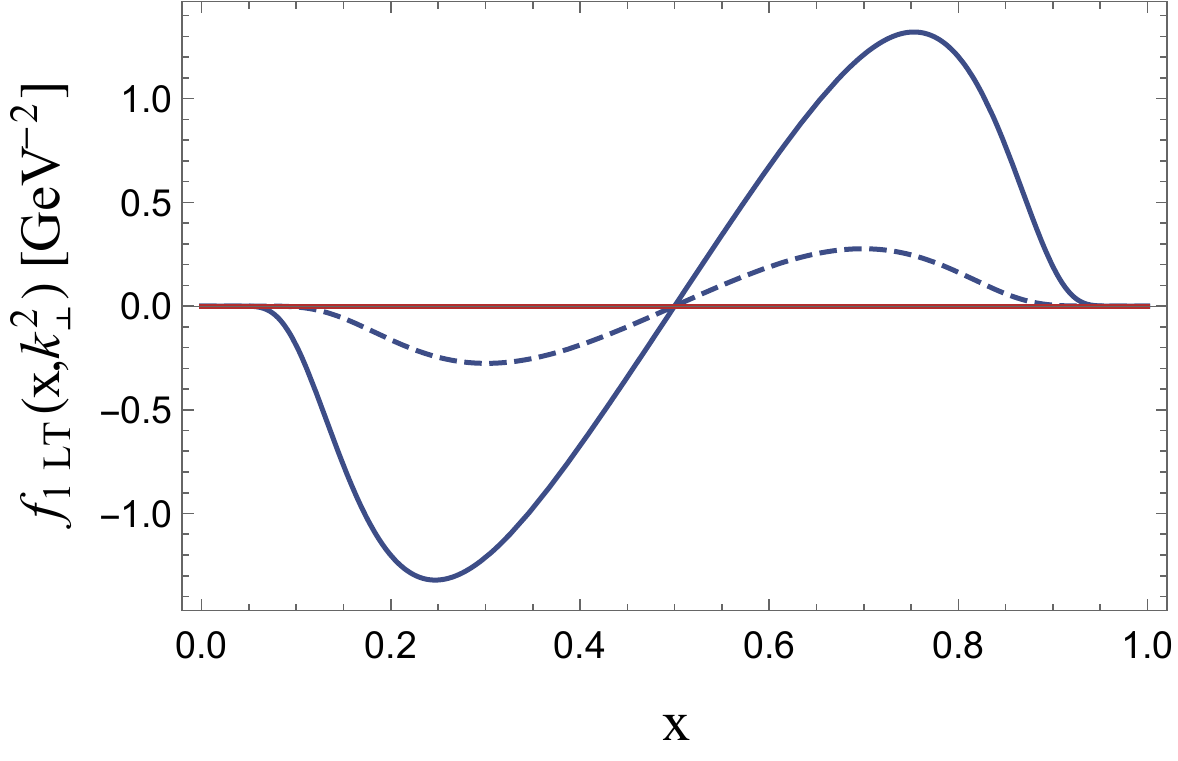}
(d)\includegraphics[width=.38\textwidth]{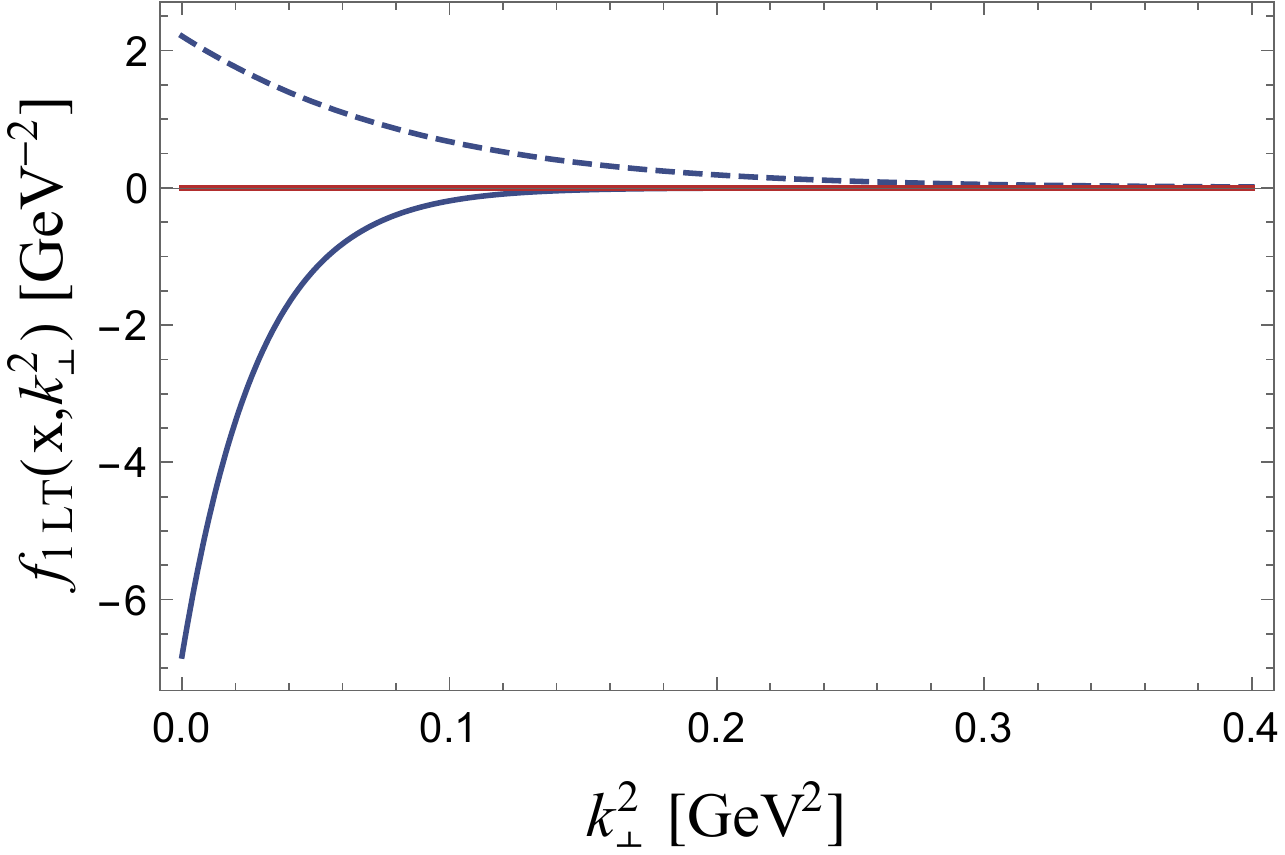}\end{center}
\end{minipage}
\begin{minipage}[c]{1\textwidth}\begin{center}
(e)\includegraphics[width=.38\textwidth]{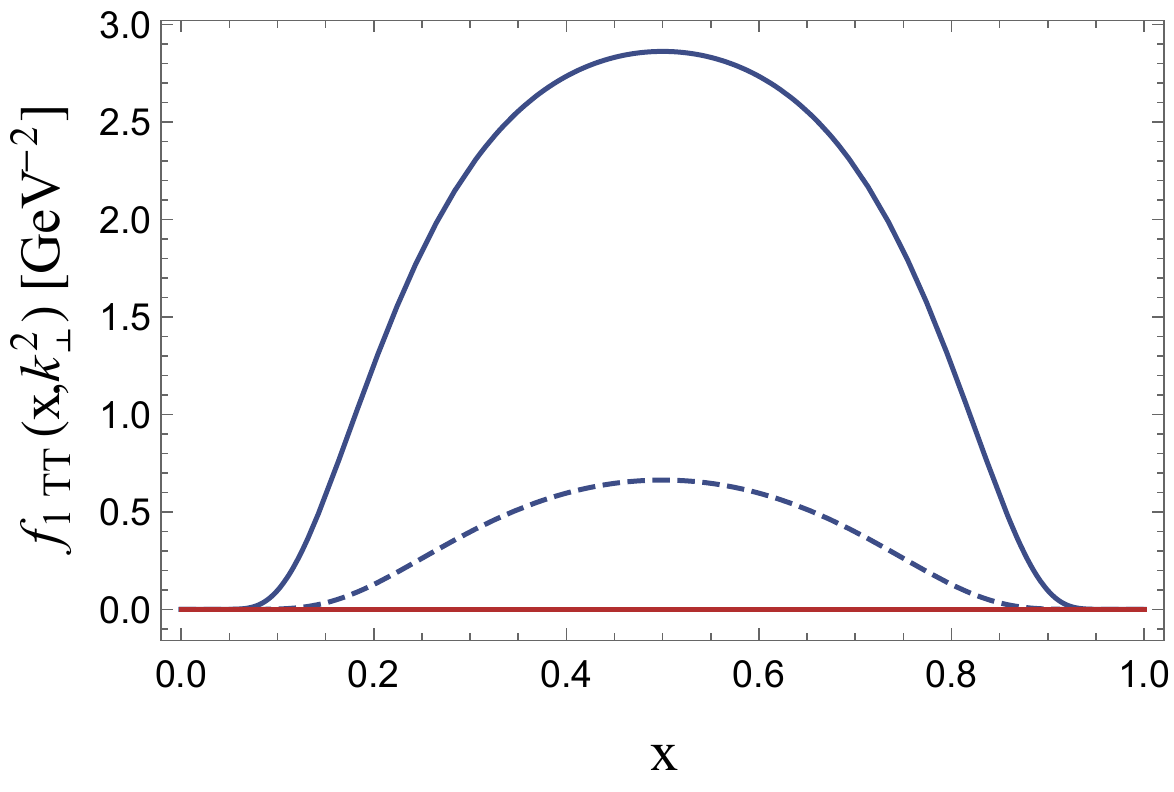}
(f)\includegraphics[width=.38\textwidth]{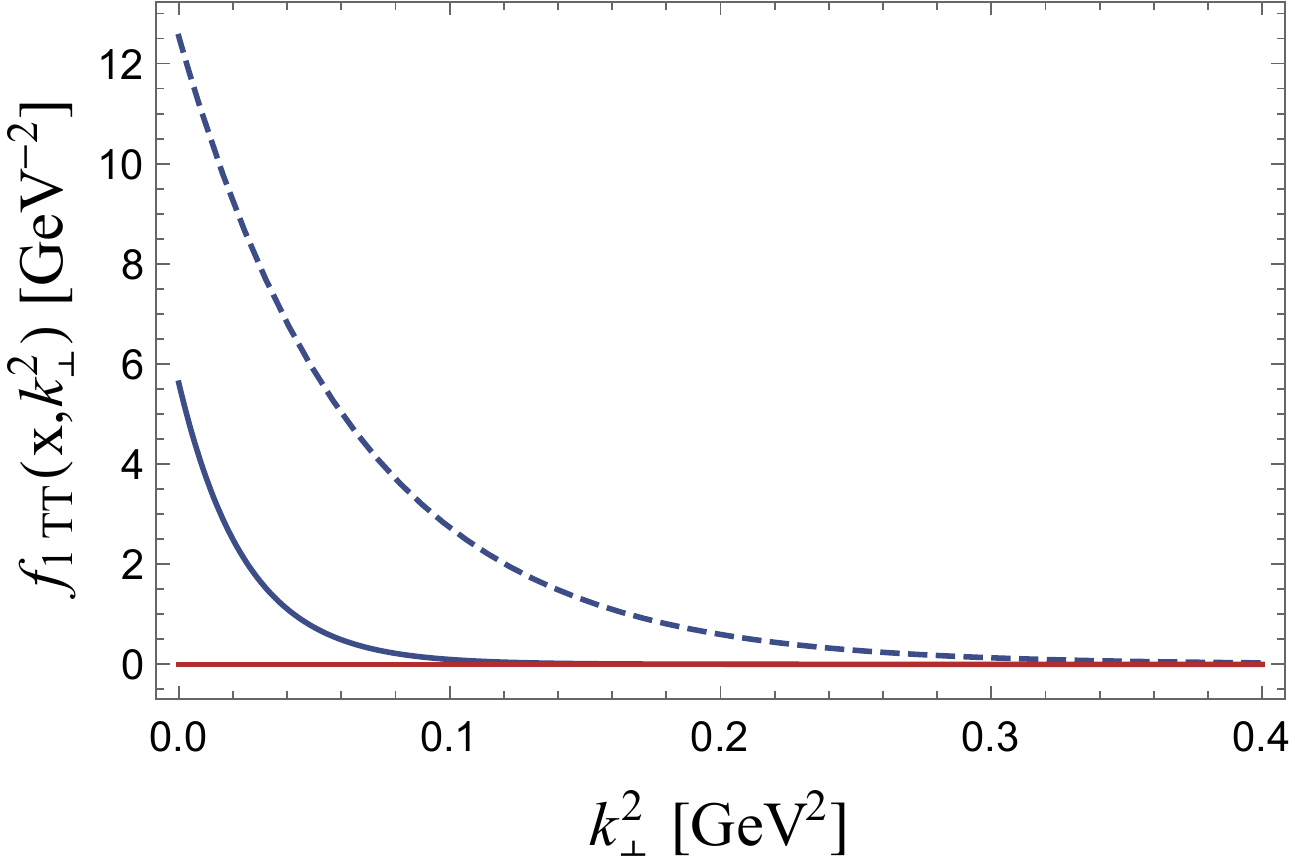}\end{center}
\end{minipage}
\caption{(Color online) The tensor-polarized TMDs $f_{1LL}(x,{\bf k}^2_\perp)$, $f_{1LT}(x,{\bf k}^2_\perp)$ and $f_{1TT}(x,{\bf k}^2_\perp)$ with respect to $x$ at different value of ${\bf k}^2_\perp$ (left panel), i.e., ${\bf k}^2_\perp=0.1 {\ \rm GeV}^2$ (solid curves) and ${\bf k}^2_\perp=0.2 {\ \rm GeV}^2$ (dashed curves). On the right panel, these TMDs are shown with respect to ${\bf k}^2_\perp$ at different values of $x$, i.e., $x=0.3$ (solid curves) and $x=0.6$ (dashed curves). The blue and red curves correspond to the light-front holographic model at the model scale $\mu_{\rm LFH}^2=0.20$ GeV$^2$ and the light-front quark model predictions at the model scale $\mu_{\rm LFQM}^2=0.19$ GeV$^2$, respectively.}
\label{tensor-tmds}
\end{figure}
\begin{figure}[hbt]
\centering
\begin{minipage}[c]{1\textwidth}
\begin{center}
(a)\includegraphics[width=.38\textwidth]{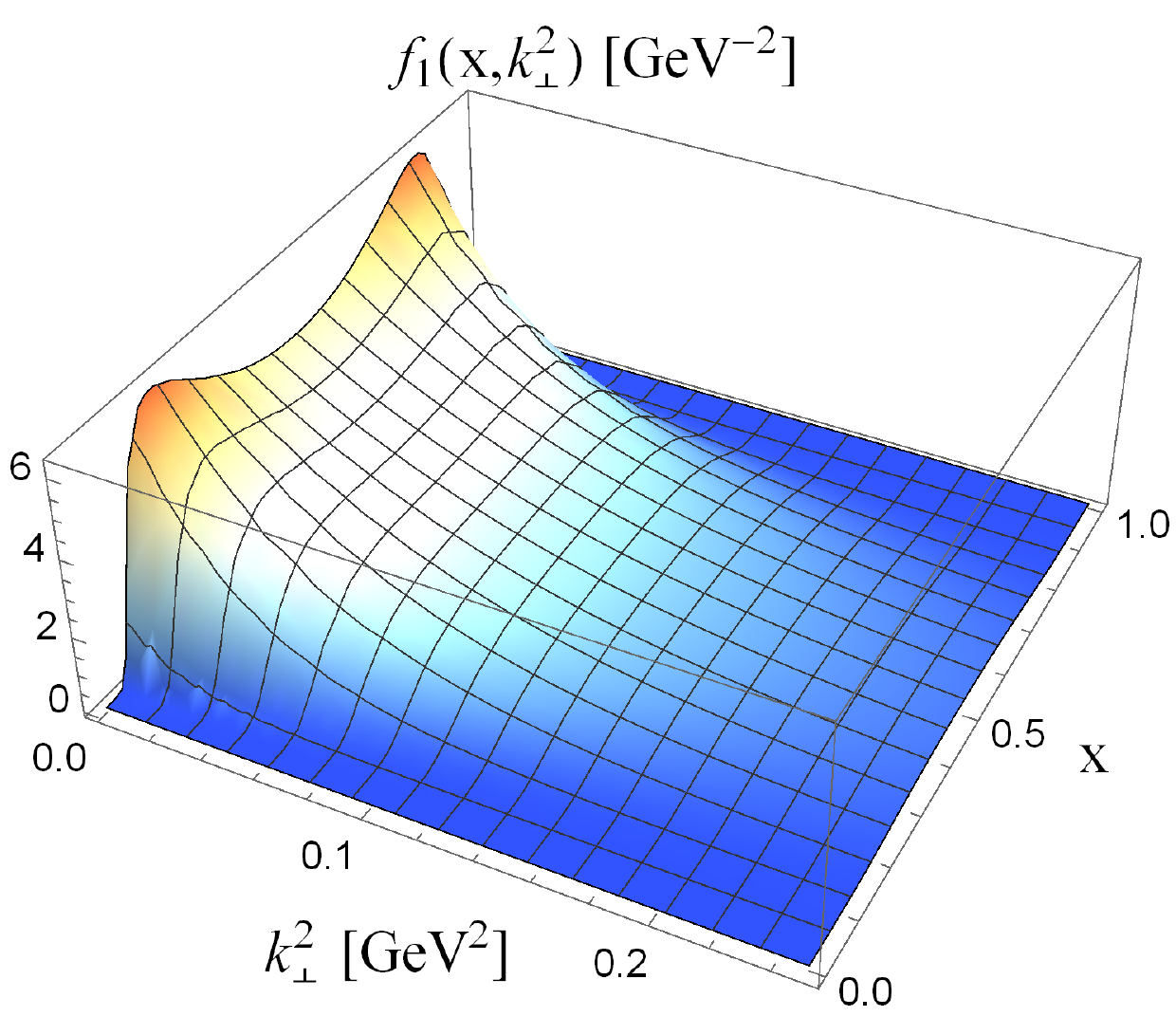}\hspace{1cm}
(b)\includegraphics[width=.38\textwidth]{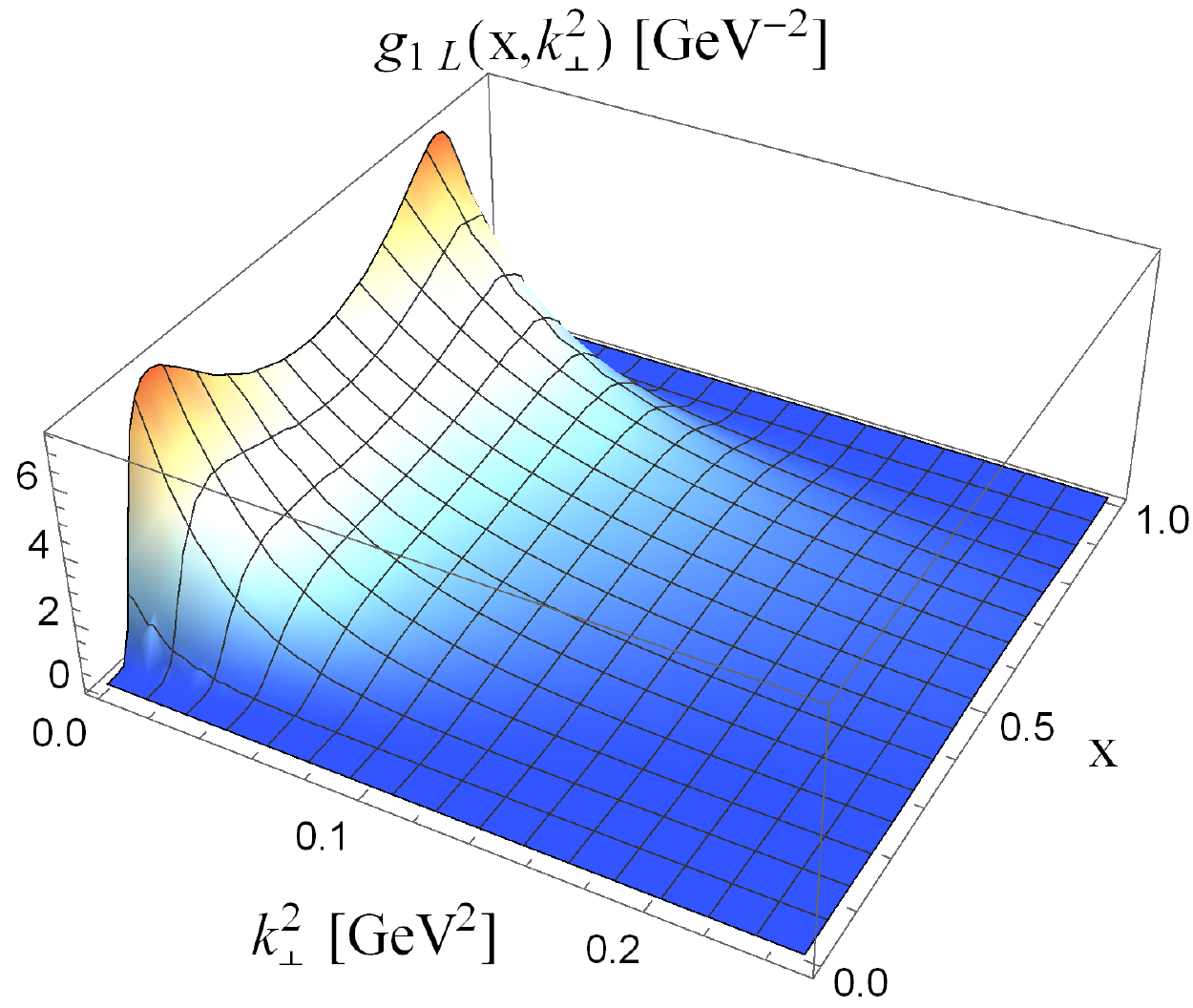}
\end{center}
\end{minipage}
\begin{minipage}[c]{1\textwidth}
\begin{center}
(c)\includegraphics[width=.38\textwidth]{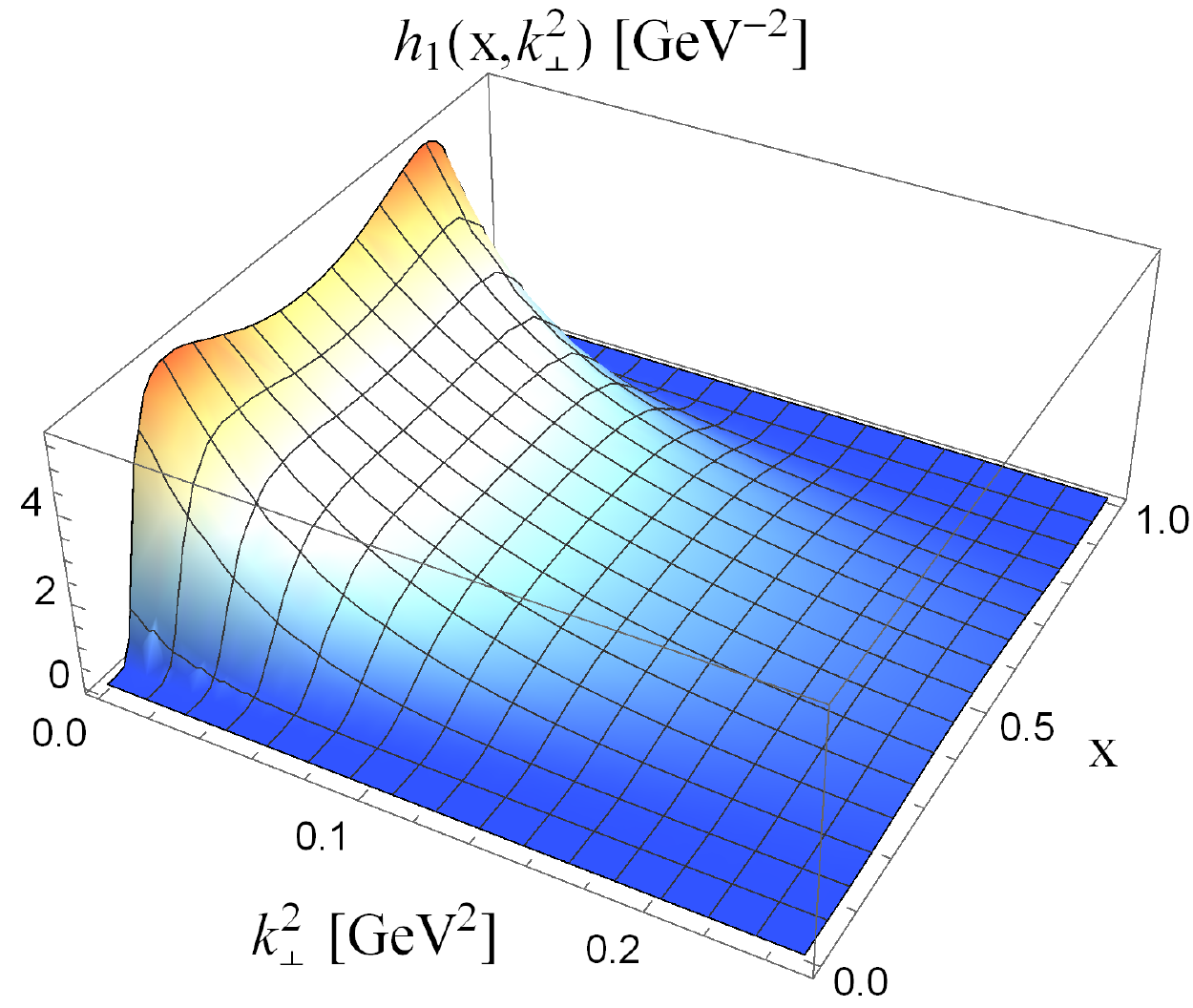}\hspace{1cm}
(d)\includegraphics[width=.38\textwidth]{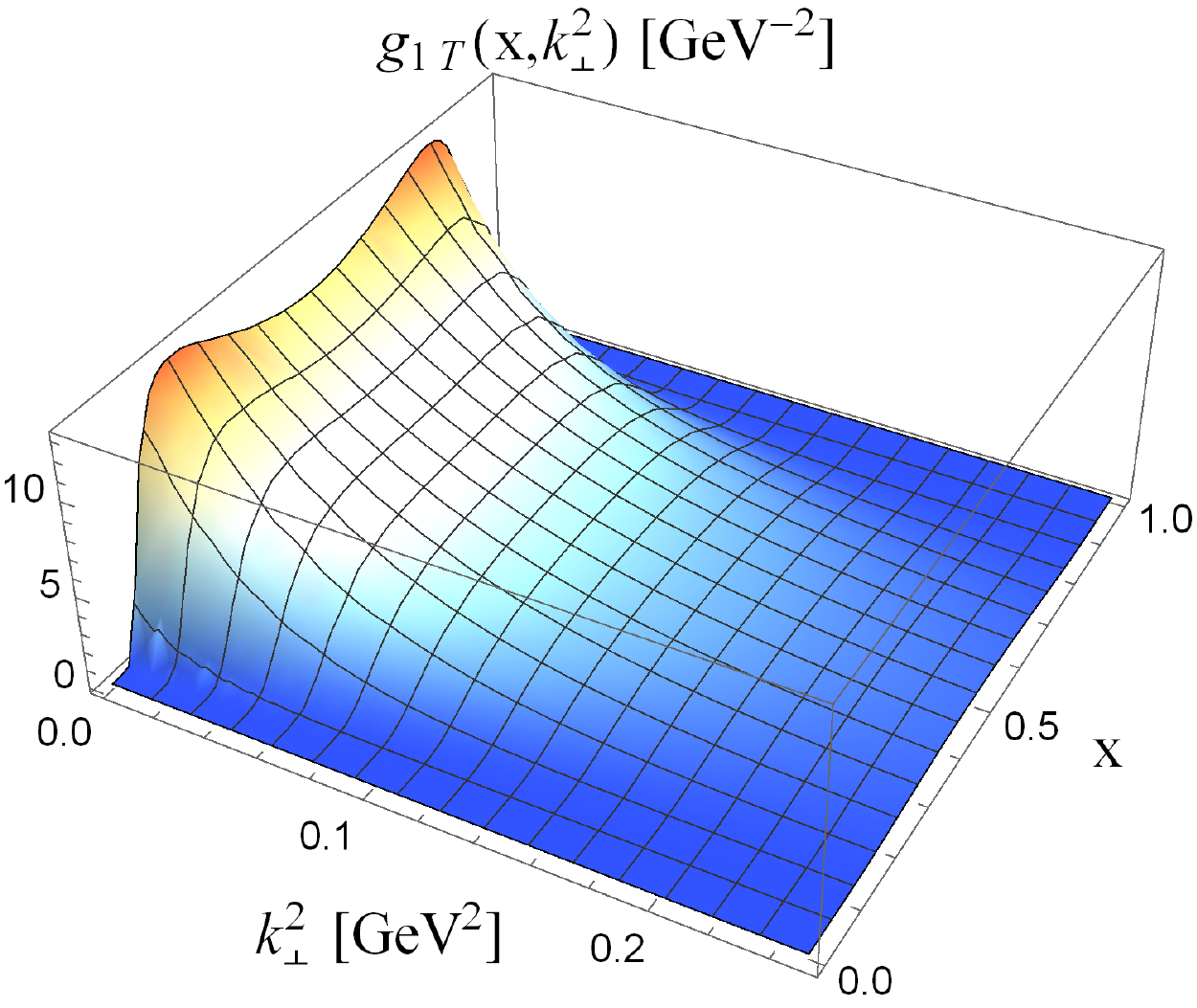}
\end{center}
\end{minipage}
\begin{minipage}[c]{1\textwidth}
\begin{center}
(e)\includegraphics[width=.38\textwidth]{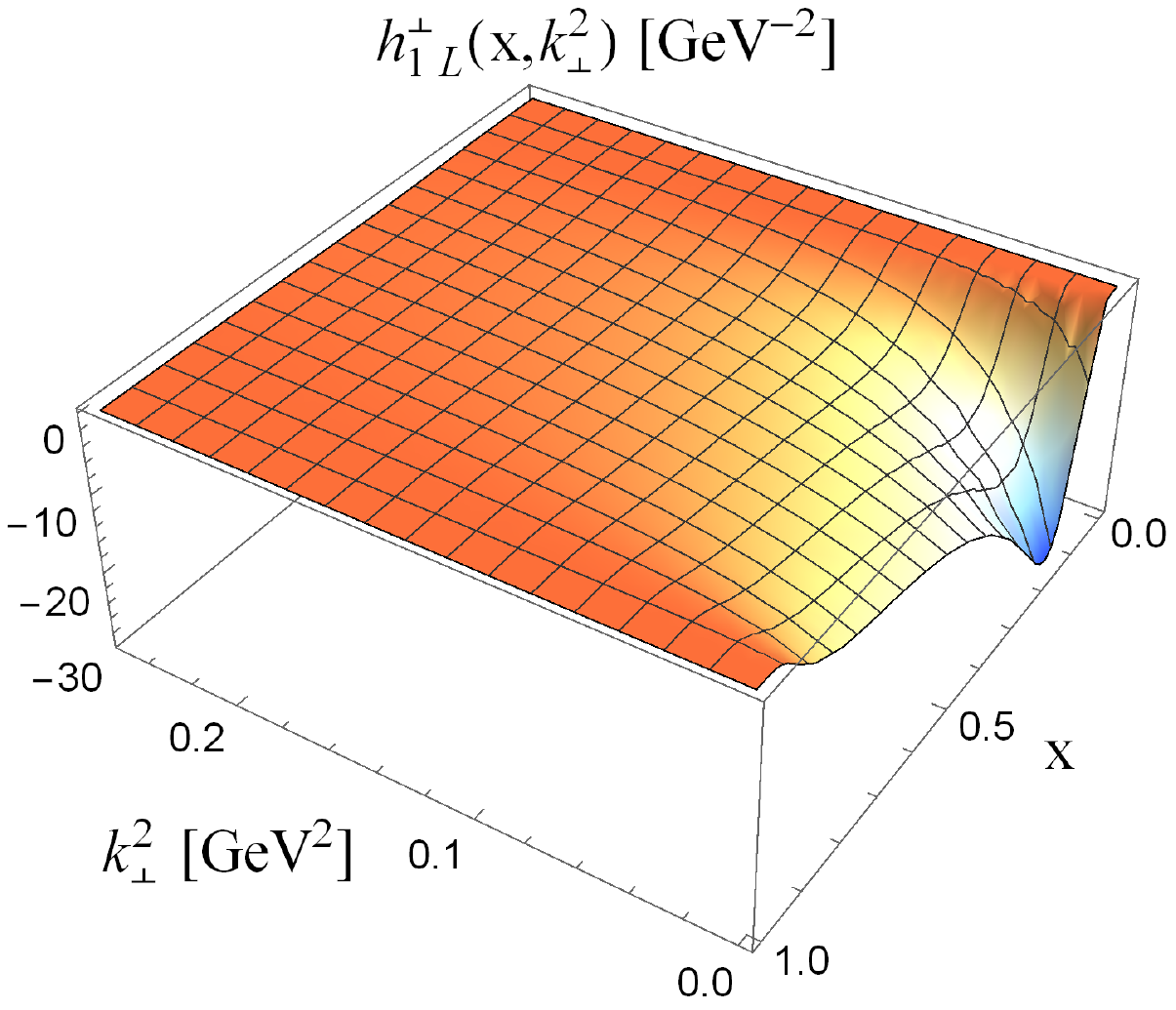}\hspace{1cm}
(f)\includegraphics[width=.38\textwidth]{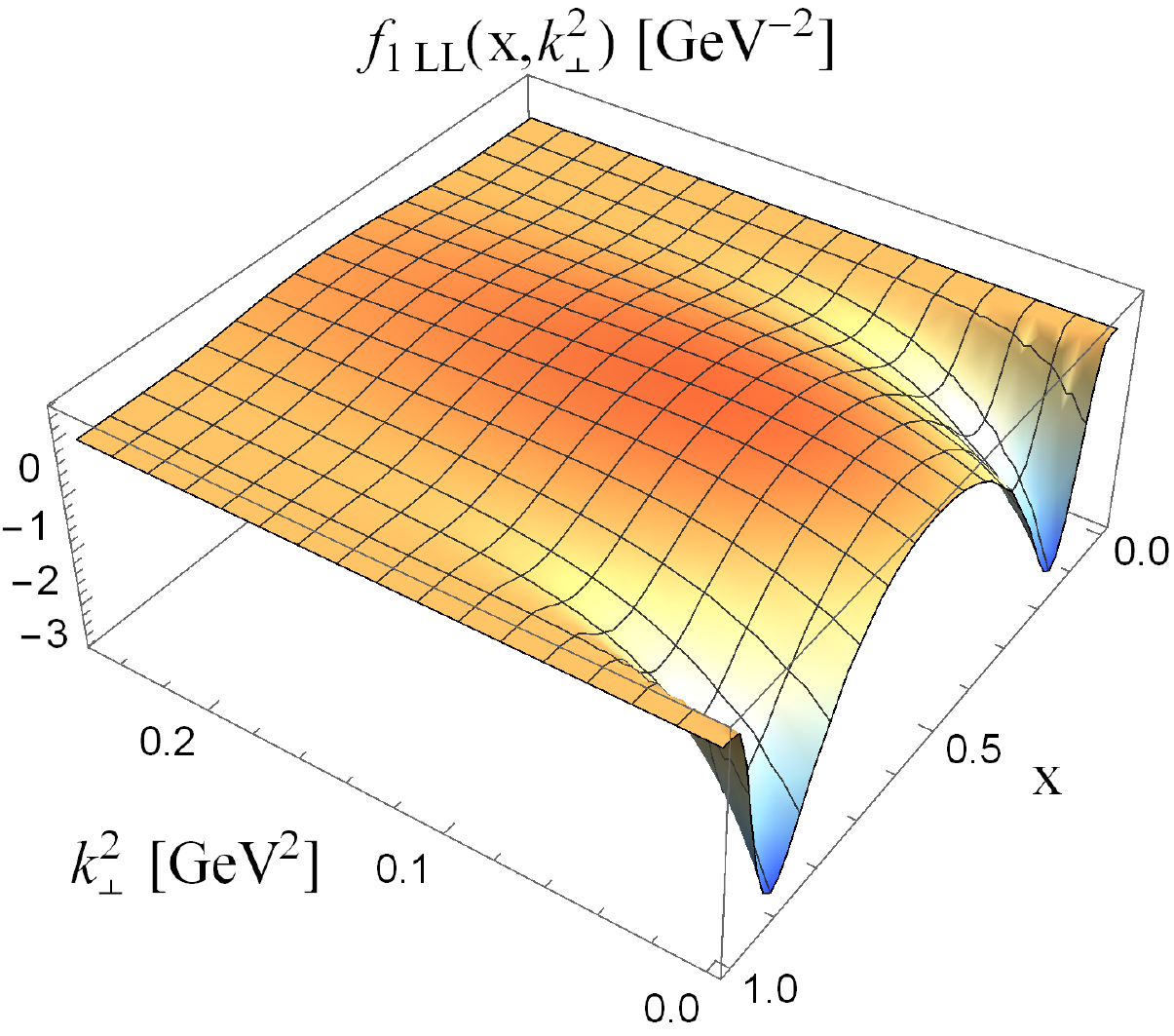}
\end{center}
\end{minipage}
\begin{minipage}[c]{1\textwidth}
\begin{center}
(g)\includegraphics[width=.38\textwidth]{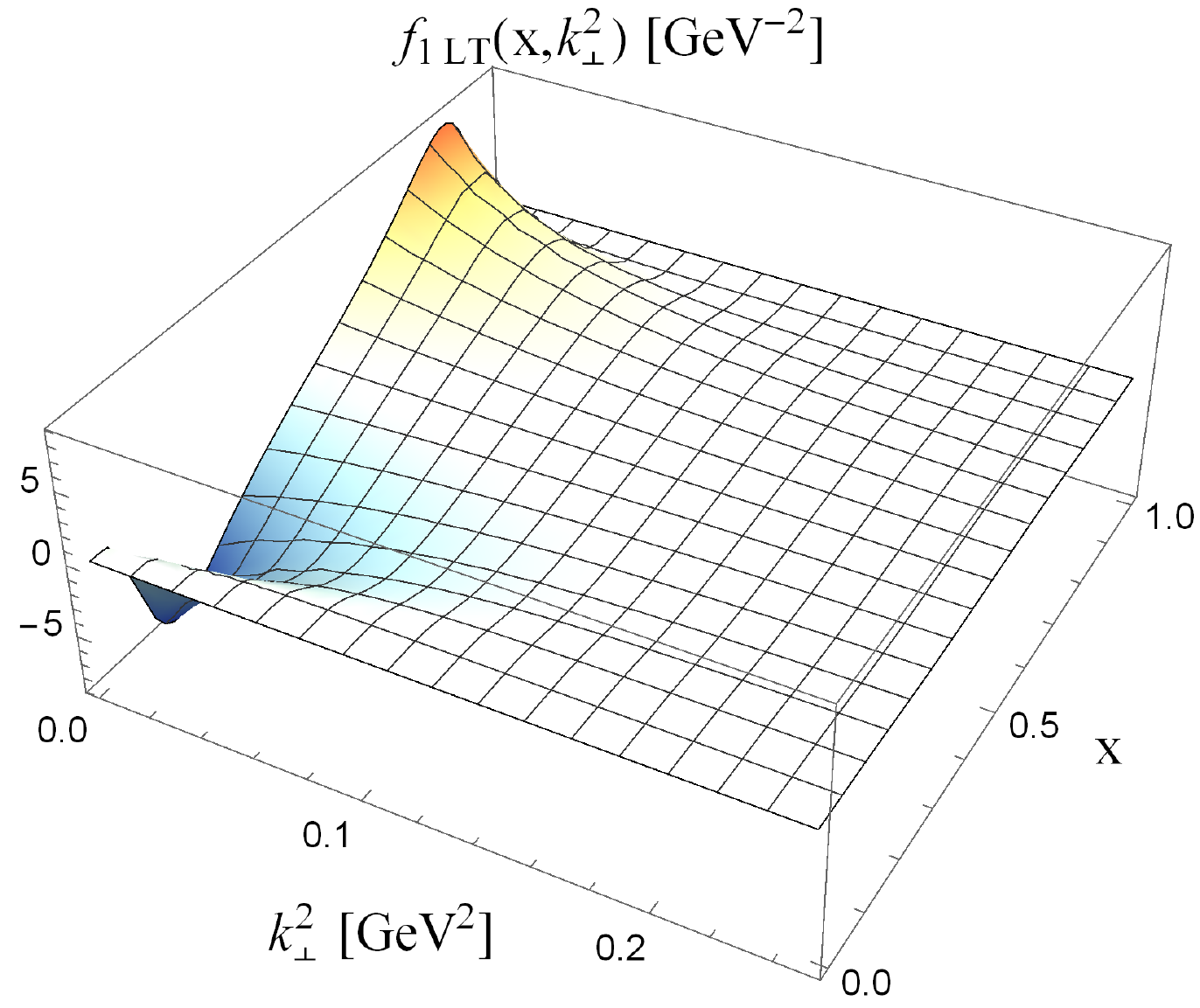}\hspace{1cm}
(h)\includegraphics[width=.38\textwidth]{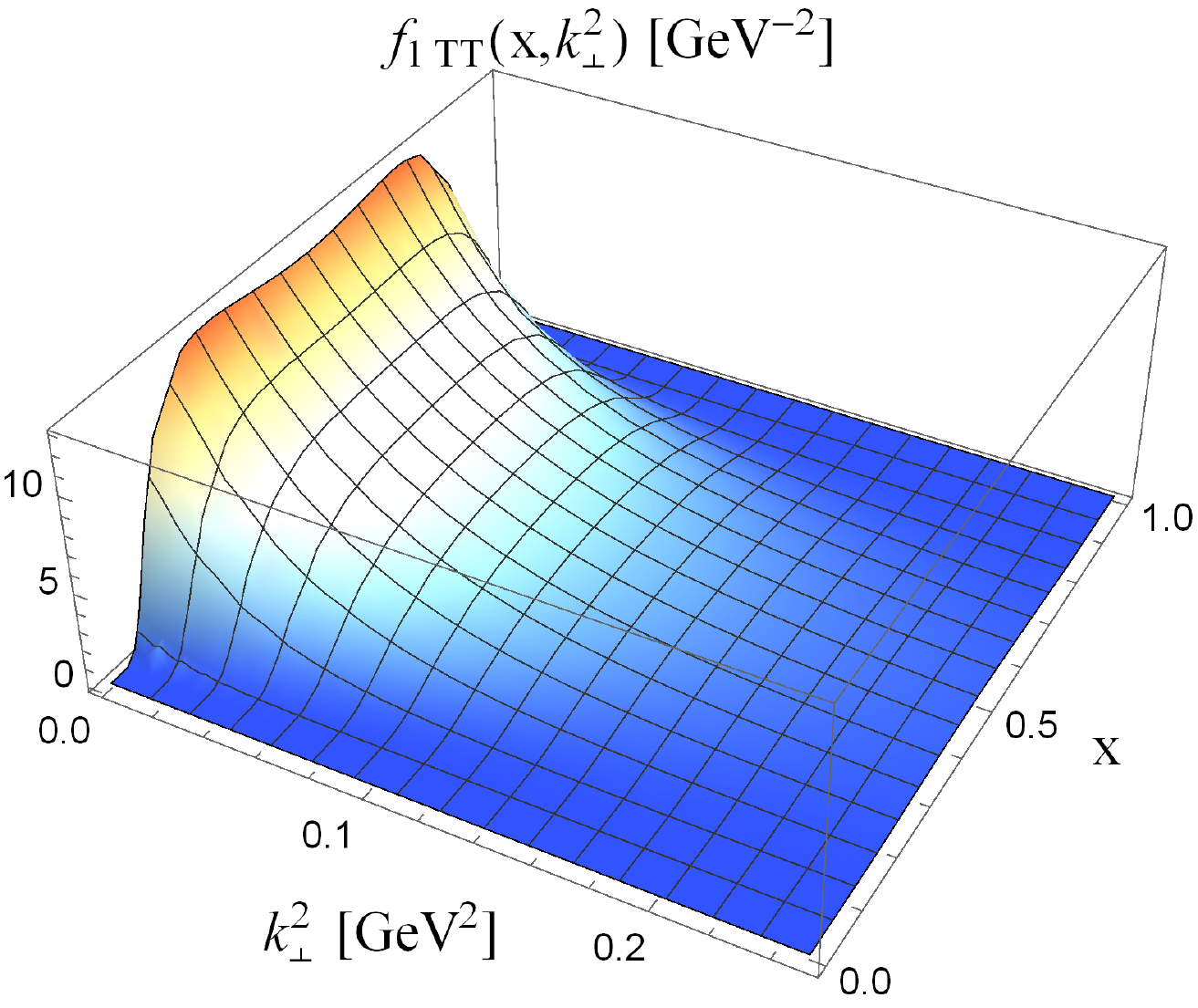}
\end{center}
\end{minipage}
\caption{(Color online) Light-front holographic  TMDs of the $\rho$-meson  as a function of $x$ and ${\bf k}^2_\perp$ at the model scale $\mu_{\rm LFH}^2=0.20$ GeV$^2$.}
\label{3d-tmds}
\end{figure}
\subsection{Numerical results}\label{results}
For the numerical predictions of the $\rho$-meson TMDs, we use the quark mass, $m_{u/d}=0.33$ GeV and the universal AdS/QCD scale, $\kappa=0.523$ GeV as in Ref.~\cite{Ahmady:2020mht,Ahmady:2018muv,Ahmady:2019yvo}.
%
%
In Figs.~\ref{tmds1}, \ref{tmds2}, and \ref{tensor-tmds}, we illustrate the $\rho$-meson TMDs in the LF holographic model at the model scale $\mu_{\rm LFH}^2=0.20$ GeV$^2$ and compare with the LF quark model predictions at $\mu_{\rm LFQM}^2=0.19$ GeV$^2$. On the left panels of these figures, the TMDs are shown as a function of $x$ when ${\bf k}_\perp$ is fixed, whereas, we show the TMDs as a function of ${\bf k}^2_\perp$ for fixed values of $x$ on the right panels. 
In Fig. $\ref{tmds1}$, we present the unpolarized quark TMD, $f_1(x,{\bf k}^2_\perp)$, as well as the longitudinally polarized quark TMDs: $g_{1L}(x,{\bf k}^2_\perp)$ and $g_{1T}(x,{\bf k}^2_\perp)$, while the  transversely polarized quark TMDs: $h_1(x,{\bf k}^2_\perp)$, $h_{1L}^\perp(x,{\bf k}^2_\perp)$, and $h_{1T}^\perp(x,{\bf k}^2_\perp)$ are displayed in Fig.~\ref{tmds2}. The qualitative behaviors of the TMDs $f_1$, $g_{1L}$, and $g_{1T}$ in the LF holographic model are found to be consistent with the LF quark model. We also observe the similar trend followed by the TMDs 
$h_1$ and $h_{1L}^\perp$. However, we find that  $h_{1T}^\perp$ is zero in the LF holographic model but it is nonzero in the LF quark model. Note that $h_{1T}^\perp$ has also been found to be zero in the NJL model~\cite{Ninomiya:2017ggn}.  The TMDs $f_1$, $g_{1T}$ and $h_1$ describe the momentum distributions of the unpolarized quark in the unpolarized meson, the longitudinally polarized quark in the transversely polarized meson, and the transversely polarized quark in the transversely polarized meson, respectively. It can be noticed that the TMDs $f_1$, $g_{1T}$ (known as ``worm gear 2" distribution) and $h_1$ (known as transversity distribution) exhibit symmetry under $x \leftrightarrow (1-x)$ in the LF holographic model. Similar behavior of these TMDs have been observed in the NJL model~\cite{Ninomiya:2017ggn}. However, in the LF quark model, only $f_1$ is symmetric under the transformation. The TMDs, which describe the momentum distributions of the longitudinally polarized quark and the transversely polarized quark in the longitudinally polarized meson, are defined as: the helicity TMD $g_{1L}$, and the ``worm gear 1" $h_{1L}^\perp$, respectively. Unlike $f_1$, $g_{1T}$ and $h_1$, the longitudinally polarized meson TMDs, $g_{1L}$ and $h_{1L}^\perp$, are asymmetric under $x\leftrightarrow (1-x)$ and $h_{1L}^\perp$ displays a negative distribution. It can be seen from Fig. \ref{tmds2}(c) that it needs less than half of the momentum fraction to be carried by the transversely polarized quark to get the distribution peak.
 The pretzelosity TMD $h_{1T}^\perp$ describes the momentum distribution when both the quark and the $\rho$-meson are transversely polarized and also, their polarizations are perpendicular to each other.
Therefore, $h_{1T}^\perp$ has different overlap contributions from $h_1$. 
 However, due to the different spin structures, $h_{1T}^\perp$ does not vanish in the LF quark model and found to be negative and asymmetrical with respect to $x$, shown in lower panel of Fig. \ref{tmds2}. 
 Neverthless, the different responses are observed in the LF quark model: (i) the lesser number of TMDs show symmetry as compared to the NJL model \cite{Ninomiya:2017ggn} and the LF holographic model, (ii) $h_{1T}^\perp\ne0$. The reason behind the difference in the observations lie in the spin structure of the $\rho$-meson. In other words, the presence of the P-wave component i.e. $L_z=\pm 1$ in the longitudinally polarized ($\Lambda=L$) and the D-wave component i.e. $L_z=\pm 2$ in the transversely polarized ($\Lambda=T$) $\rho$-meson wave functions in the LF quark model are responsible for the asymmetry in $g_{1T}$ and $h_1$ and the non-vanishing $h_{1T}^\perp$.  
Further, the dominance of TMDs on the quark longitudinal momentum fraction come to an end at the extended values of the quark transverse momentum and the quark TMDs get vanished.
  

Further, in Fig.~\ref{tensor-tmds}, we display the tensor-polarized TMDs designated for the unpolarized quark. As discussed before, we plot these TMDs with respect to $x$ at the fixed values of ${\bf k}^2_\perp$ (left panel) and vice versa (right panel). We observe that $f_{1LL}$ has a positive peak at $x=0.5$, and two negative peaks at lower $(<0.5)$ and higher $x$ $(>0.5)$ or  equivalently, it has two zero crossings over $x$. For $f_{1LL}$, there is no OAM transfer between the initial and final states as seen in the Eq. (\ref{f_1LL}).
Basically, the positive contribution from the S-wave and the negative contributions from the other wave compositions of the LFWFs cancel each other's effect which leads to the zero distribution at the crossing over points. The S-wave contribution dominates at the central region of $x$, whereas at lower and higher $x$ domains, the other contributions rule over. However, with increasing ${\bf k}_\perp$, the effect of cancellation decreases resulting in the small negative distribution peaks. $f_{1LL}$ shows symmetry under $x \leftrightarrow (1-x)$. Further, $f_{1LT}$ is shown in the middle panel of Fig.~\ref{tensor-tmds}. It vanishes at $x=0.5$ and exhibits the positive and the negative distributions at $x>0.5$  and $x<0.5$, respectively. The overlap of $f_{1LT}$, Eq. (\ref{f_1lt}), is observed to transfer one unit of OAM from the initial to the final states. In this case, the cancellation occurs due to $L_z= \pm 1$ contributions. $L_z=0$ component of the wave functions is always positive in nature, however, according to Eq. (\ref{f_1lt}), the difference in terms corresponding to $L_z=+1$ and $L_z=-1$ brings zero into the picture. $f_{1LT}$ is anti-symmetric under $x \leftrightarrow (1-x)$. The left over tensor-polarized TMD $f_{1TT}$, shown in the lower panel of Fig.~\ref{tensor-tmds}, shows the sum of the overlaps providing the two units of OAM transfer from the initial to the final states. Again, because of the different spin structure, the tensor-polarized TMDs do not survive in the LF quark model. We also remark that in addition to the different spin structures of two light-front models, the overall factor $1/\sqrt{x(1-x)}$ in the holographic wave function enhances the distributions, which are diagonal in OAM in the overlap representation, i.e., $f_1$, $g_{1L}$, $h_1$ and $f_{1LL}$ at ${\bf k}^2_\perp\to 0$ compared to those distributions in the LF quark model. The enhancement of the distributions in the holographic model can be seen in the Figs. $\ref{tmds1}$(b), $\ref{tmds1}$(d), $\ref{tmds2}$(b), and \ref{tensor-tmds}(b). Note that as expected all the TMDs in both the LF models vanish at the end points $x\to\{0,1\}$ for any value of ${\bf k}_\perp$. Surprisingly, a distinct feature of the TMDs has been observed in NJL model~\cite{Ninomiya:2017ggn}, where the TMDs are non-zero at the end points for low ${\bf k}_\perp$.
To understand $x$ and ${\bf k}^2_\perp$ dependence together, the three-dimensional structure of the eight non-zero TMDs in the LF holographic model is shown in Fig. \ref{3d-tmds}. Overall features of all the TMDs have also been observed in the NJL model~~\cite{Ninomiya:2017ggn}.


Further, to compare our results with the available theoretical predictions, we compute the ${\bf k}_\perp$ moments for several TMDs~\cite{Ninomiya:2017ggn}
\begin{eqnarray}
\langle k_{\perp}^a \rangle_{\rm TMD} \equiv \frac{{\int} {\rm d}x ~{\rm d}^2{\bf k}_\perp | {\bf k}_\perp|^a  {\rm TMD}(x,{\bf k}^2_\perp)}{\int {\rm d}x ~{\rm d}^2{\bf k}_\perp {\rm TMD}(x,{\bf k}^2_\perp)},
\label{moments}
\end{eqnarray}
 where $a$ represents the order of the moment.
 %
In Table \ref{table-moment}, we compare our predictions for the first and the second order ${\bf k}_\perp$ moments in the LF holographic model and the LF quark model with the only available theoretical results from the NJL model~\cite{Ninomiya:2017ggn}. We observe that except for $g_{1L}$, our predictions are under estimated and they are in more or less accord with the NJL model~\cite{Ninomiya:2017ggn}, however, the results in the LF quark model are closer to the results predicted in the NJL model compared to the LF holographic model. Our predictions for the moments of $g_{1L}$ differ significantly from the  NJL model. $f_{1LL}$ and $f_{1LT}$ are not shown in table, because in both the LF models, the denominator of the ${\bf k}_\perp$ moment is evaluated to be zero for both $f_{1LL}$ and $f_{1LT}$ TMDs. Similar observation has been made in NJL model. Also, the moments corresponding to $f_{1TT}$ in the LF quark model are not filled up because the denominator of Eq. (\ref{moments}) vanishes.
 
\begin{table}
\centering
\begin{tabular}{|c|c|c|c|c|c|c|}
\hline \hline
 & \multicolumn{2}{c|}{LF holographic model} &\multicolumn{2}{c|}{LF quark model} & \multicolumn{2}{c|}{NJL model}\\
 TMDs & \multicolumn{2}{c|}{(This work)} &\multicolumn{2}{c|}{(This work)} & \multicolumn{2}{c|}{\cite{Ninomiya:2017ggn}} \\
\cline{2-7}
& $\langle{{k}_\perp}\rangle$ & $\langle{{k}^2_\perp}\rangle$ & $\langle{{k}_\perp}\rangle$ & $\langle{{k}^2_\perp}\rangle$ & $\langle{{k}_\perp}\rangle$ &  $\langle{{k}^2_\perp}\rangle$ \\
\hline
$f_1$ & $0.238 \pm 0.011$ & $0.073\pm 0.007$ & $0.328 \pm 0.016$ & $0.140 \pm 0.013$ & 0.32 & 0.13 \\
$g_{1L}$ & $0.204 \pm 0.008$ & $0.054 \pm 0.004$ & $0.269 \pm 0.012$ & $0.098 \pm 0.009$ & 0.08 & -0.11 \\
$g_{1T}$ & $0.229 \pm 0.010$  & $0.077 \pm 0.006$ & $0.269 \pm 0.012$ & $0.098 \pm 0.009$ & 0.34 & 0.16 \\
$h_1$ & $0.229 \pm 0.010$ & $0.077 \pm 0.006$ & $0.307 \pm 0.014$ & $0.124 \pm 0.011$ & 0.34 & 0.16 \\
$h^\perp_{1L}$ & $0.204 \pm 0.008$ & $0.054 \pm 0.004$ & $0.269 \pm 0.012$ & $0.098 \pm 0.009$ & 0.33 & 0.15 \\
$h^\perp_{1T}$ & - & - & $0.237 \pm 0.011$ & $0.077 \pm 0.007$ & - & - \\
$f_{1LL}$ & - & - & - & - & - & -\\
$f_{1LT}$ & - & - & - & - & - & -\\
$f_{1TT}$ & $0.211 \pm 0.009$ & $0.067 \pm 0.005$ & - & - & 0.32 & 0.14 \\
\hline
\end{tabular}
\caption{The first moment $\langle { k}_\perp \rangle $ [in GeV] and the second moment $\langle { k}^2_\perp \rangle $ [in GeV$^2$] predictions corresponding to several $\rho$-meson TMDs are compared with NJL model results \cite{Ninomiya:2017ggn}. 
The theory uncertainties result from the uncertainties in the constituent quark mass $m_q=0.33 \pm 0.03$ GeV and the AdS/QCD scale $\kappa=0.523 \pm 0.024$ GeV in the LF holographic model whereas in the LF quark model, the values of the parameters are $m_q=0.20 \pm 0.02$ GeV and $\beta=0.41 \pm 0.02$ GeV.}
\label{table-moment}
\end{table}
%
\begin{figure}[hbt!]
\centering
\includegraphics[width=1\textwidth]{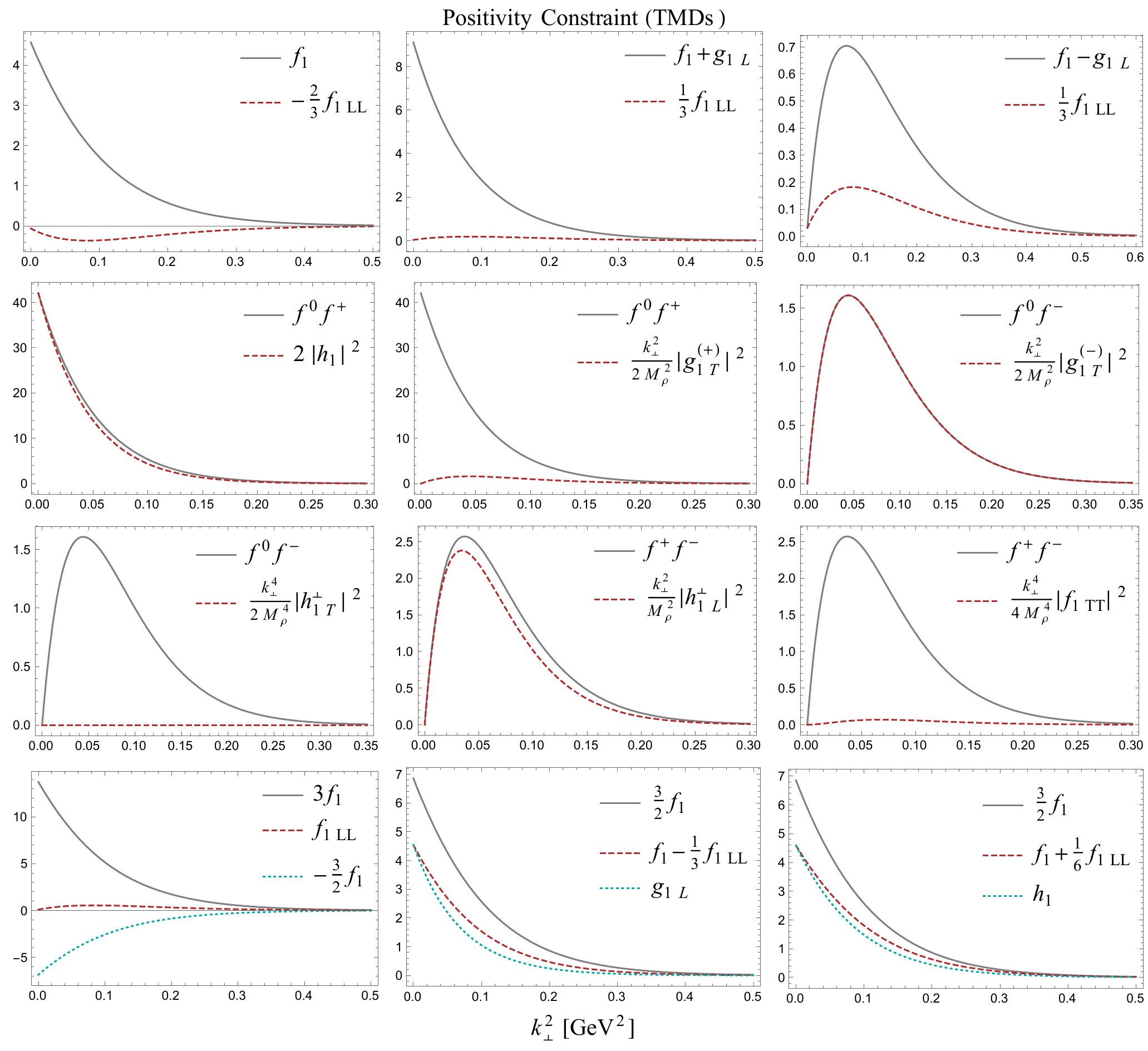}
\caption{The positivity constraints on the holographic TMDs, plotted with respect to ${\bf k}^2_\perp$ at fixed $x=0.5$.}
\label{positivity-constraints}
\end{figure}
\begin{figure}[hbt!]
\centering
\includegraphics[width=1\textwidth]{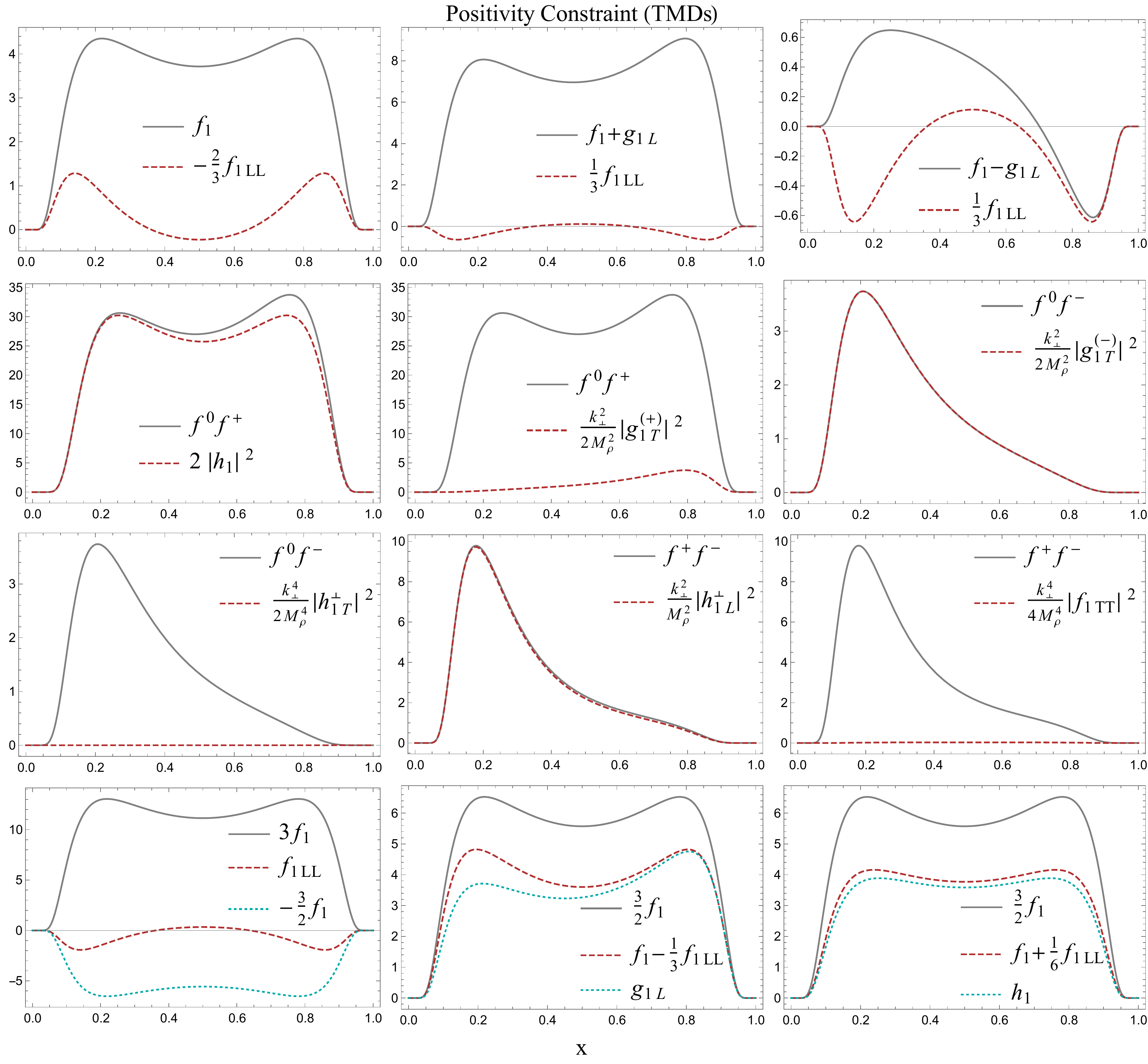}
\caption{The positivity constraints on the holographic TMDs, plotted with respect to $x$ at fixed ${\bf k}_\perp=0.15$ GeV.}
\label{positivity-constraints-x}
\end{figure}
\subsection{Positivity constraints}
Let us now check the positivity constraints of our holographic TMDs for the $\rho$-meson.  For the spin-1 hadron, the TMDs satisfy the following relations \cite{Bacchetta:2001rb, Ninomiya:2017ggn}:
\begin{equation}
f_1(x,{\bf k}^2_\perp) \geq 0 \,,
\label{pc1}
\end{equation}
\begin{equation}
f^0(x,{\bf k}^2_\perp) \geq 0 \, \Rightarrow f_1(x,{\bf k}^2_\perp) \geq -\frac{2}{3}f_{1LL}(x,{\bf k}^2_\perp)\,,
\label{pc2} 
\end{equation}
\begin{equation}
f^+(x,{\bf k}^2_\perp) \geq 0 \, \Rightarrow f_{1}(x,{\bf k}^2_\perp)+g_{1L}(x,{\bf k}^2_\perp) \geq \frac{1}{3}f_{1LL}(x,{\bf k}^2_\perp)\,,
\end{equation}
\begin{equation}
f^-(x,{\bf k}^2_\perp) \geq 0 \, \Rightarrow f_{1}(x,{\bf k}^2_\perp)-g_{1L}(x,{\bf k}^2_\perp) \geq \frac{1}{3}f_{1LL}(x,{\bf k}^2_\perp)\,,
\end{equation}
\begin{equation}
f^0(x,{\bf k}^2_\perp) f^+(x,{\bf k}^2_\perp) \geq 2 | h_1(x,{\bf k}^2_\perp)|^2 \,,
\end{equation}
\begin{equation}
 f^0(x,{\bf k}^2_\perp) f^+(x,{\bf k}^2_\perp) \geq \frac{{\bf k}^2_\perp}{2 M^2_\rho} | g^{(+)}_{1T}(x,{\bf k}^2_\perp)|^2 \,,
\end{equation}
\begin{equation}
f^0(x,{\bf k}^2_\perp) f^-(x,{\bf k}^2_\perp) \geq \frac{{\bf k}^2_\perp}{2 M^2_\rho} | g^{(-)}_{1T}(x,{\bf k}^2_\perp)|^2 \,,
\end{equation}
\begin{equation}
f^0(x,{\bf k}^2_\perp) f^-(x,{\bf k}^2_\perp) \geq \frac{{\bf k}^4_\perp}{2 M^4_\rho} | h^\perp_{1T}(x,{\bf k}^2_\perp) |^2 \,,
\end{equation}
\begin{equation}
f^+(x,{\bf k}^2_\perp) f^-(x,{\bf k}^2_\perp) \geq \frac{{\bf k}^2_\perp}{M^2_\rho} | h^\perp_{1L}(x,{\bf k}^2_\perp) |^2 \,, 
\end{equation}
\begin{equation}
f^+(x,{\bf k}^2_\perp) f^-(x,{\bf k}^2_\perp) \geq \frac{{\bf k}^4_\perp}{4 M^4_\rho} | f_{1TT}(x,{\bf k}^2_\perp) |^2 \,,
\label{pc13}
\end{equation}
\begin{equation}
3 f_1(x,{\bf k}^2_\perp) \geq f_{1LL}(x,{\bf k}^2_\perp) \geq -\frac{3}{2} f_1(x,{\bf k}^2_\perp)\,, 
\end{equation}
\begin{equation}
\frac{3}{2} f_1(x,{\bf k}^2_\perp) \geq f_1(x,{\bf k}^2_\perp) -\frac{1}{3} f_{1LL}(x,{\bf k}^2_\perp) \geq g_{1L}(x,{\bf k}^2_\perp) \,, 
\end{equation}
\begin{equation}
\frac{3}{2} f_1(x,{\bf k}^2_\perp) \geq f_1(x,{\bf k}^2_\perp)+\frac{1}{6} f_{1LL}(x,{\bf k}^2_\perp) \geq h_1(x,{\bf k}^2_\perp) \,. 
\end{equation}


Figs.~\ref{positivity-constraints} and \ref{positivity-constraints-x} confirm that the different positivity constraints, defined in Eqs. (\ref{pc2})-(\ref{pc13}), are satisfied by our holographic TMDs for the $\rho$-meson. In Fig.~\ref{positivity-constraints}, the corresponding TMDs in each constraint equation are shown as a function of ${\bf k}^2_\perp$ at fixed $x=0.5$, while Fig.~\ref{positivity-constraints-x} displays the constraint equations of TMDs as a function of $x$ for fixed ${\bf k}_\perp=0.15$ GeV. 
\section{Parton distribution functions}\label{pdfs_rho}
The PDFs encode the distribution of the longitudinal momentum and the polarization carried by the partons  with no information on the parton intrinsic transverse momentum ${\bf k}_\perp$.
Therefore, one can retrieve the PDFs by integrating Eqs.~(\ref{form1})-(\ref{form3}) over ${\bf k}_\perp$ \cite{Bacchetta:2000jk,Hino:1999qi}:
\begin{eqnarray}
\langle \gamma^+ \rangle_{\boldsymbol{ \mathcal{S}}}^{(\Lambda)} (x) 
&\equiv f_1(x) + \mathcal{S}_{LL} \,
f_{1LL}(x) \,,
\label{form1-pdf} 
\end{eqnarray}
\begin{eqnarray}
\langle\gamma^{+} \gamma_{5}\rangle^{(\Lambda)}_{\boldsymbol{ \mathcal{S}}} (x) 
&\equiv  \mathcal{S}_L\,g_{1}(x) \,,
\label{form2-pdf} 
\end{eqnarray}
\begin{eqnarray}
\langle\gamma^+\gamma^i\gamma_{5}\rangle^{(\Lambda)}_{\boldsymbol{ \mathcal{S}}} (x) 
& \equiv \mathcal{S}_\perp^i h_1(x)\,.
\label{form3-pdf}
\end{eqnarray}
After integrating over the quark transverse momenta, the $6 \times 6$ light-front helicity amplitudes matrix, Eq.~(\ref{matrix-helicity}), can then be parameterized by the leading-twist PDFs, defined as \cite{Bacchetta:2001rb}:
\begin{eqnarray}
&&\Phi(x)= \nonumber\\
&&\left(
\begin{array}{cccccc}
f_1+g_1-\frac{f_{1LL}}{3} & 0 & 0 & 0 & \sqrt{2}h_1 & 0\\
0 & f_1+\frac{2 f_{1LL}}{3} & 0 & 0 & 0 & \sqrt{2}h_1\\
0 & 0 & f_1-g_1-\frac{f_{1LL}}{3} & 0 & 0 & 0\\
0 & 0 & 0 & f_1-g_1-\frac{f_{1LL}}{3} & 0 & 0\\
\sqrt{2}h_1 & 0 & 0 & 0 & f_1+\frac{2 f_{1LL}}{3} & 0\\
0 & \sqrt{2}h_1 & 0 & 0 & 0 & f_1+g_1-\frac{f_{1LL}}{3}
\end{array}
\right)\,,\nonumber\\
\end{eqnarray}
where the positivity constraints on the PDFs are generated as
\begin{equation}
f_1(x)\geq 0 \,, 
\label{pos_f1}
\end{equation}
\begin{equation}
3f_1(x) \geq f_{1LL}(x) \geq -\frac{3}{2}f_1(x)\,,
\end{equation}
\begin{equation}
\frac{3}{2} f_1(x)\geq f_1(x)-\frac{1}{3} f_{1LL}(x)\geq | g_1(x) |\,,
\end{equation}
\begin{equation}
\left(f_1(x)+\frac{2}{3}f_{1LL}(x)\right)\left(f_1(x)+g_1(x)-\frac{1}{3}f_{1LL}(x)\right) \geq 2 | h_1(x)|^2\,.
\label{pos_f1LL}
\end{equation}
\begin{figure}[tbh]
\begin{minipage}[c]{1\textwidth}\begin{center}
(a)\includegraphics[width=.38\textwidth]{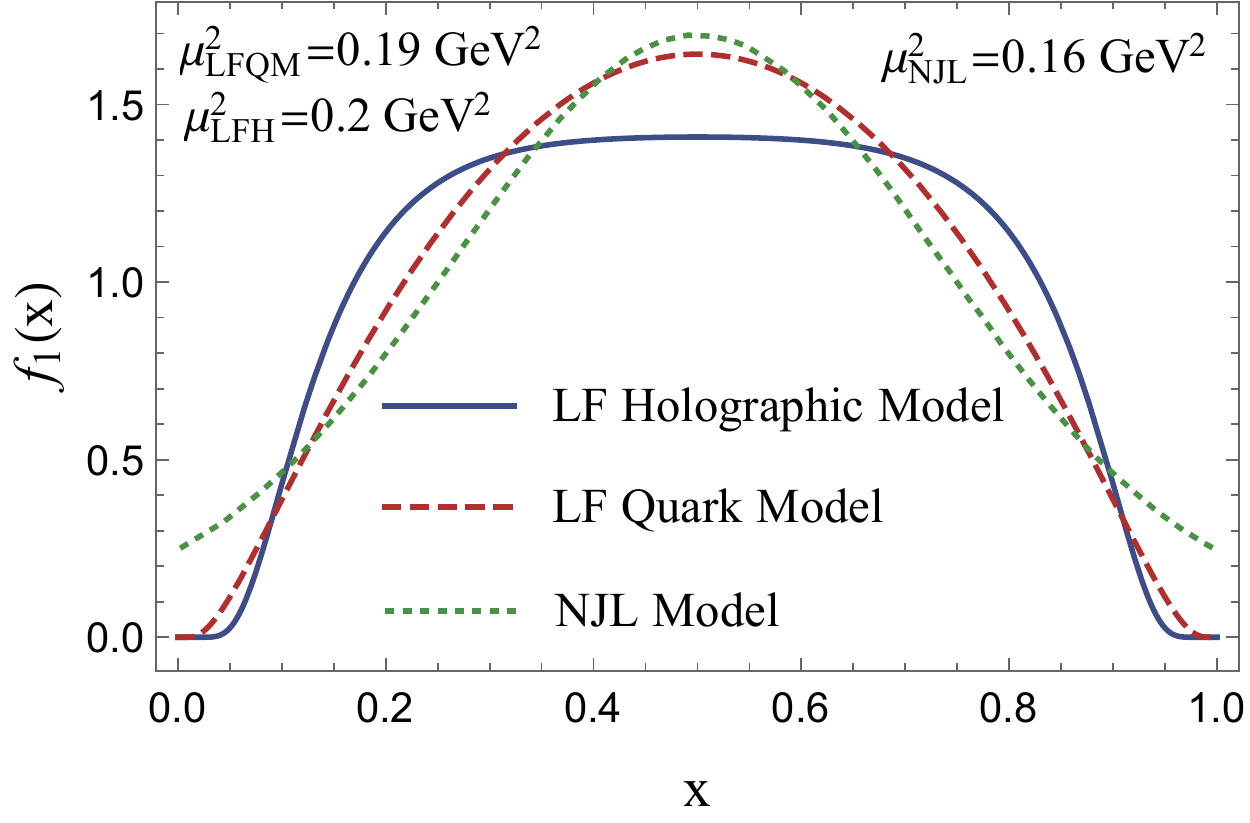}
(b)\includegraphics[width=.38\textwidth]{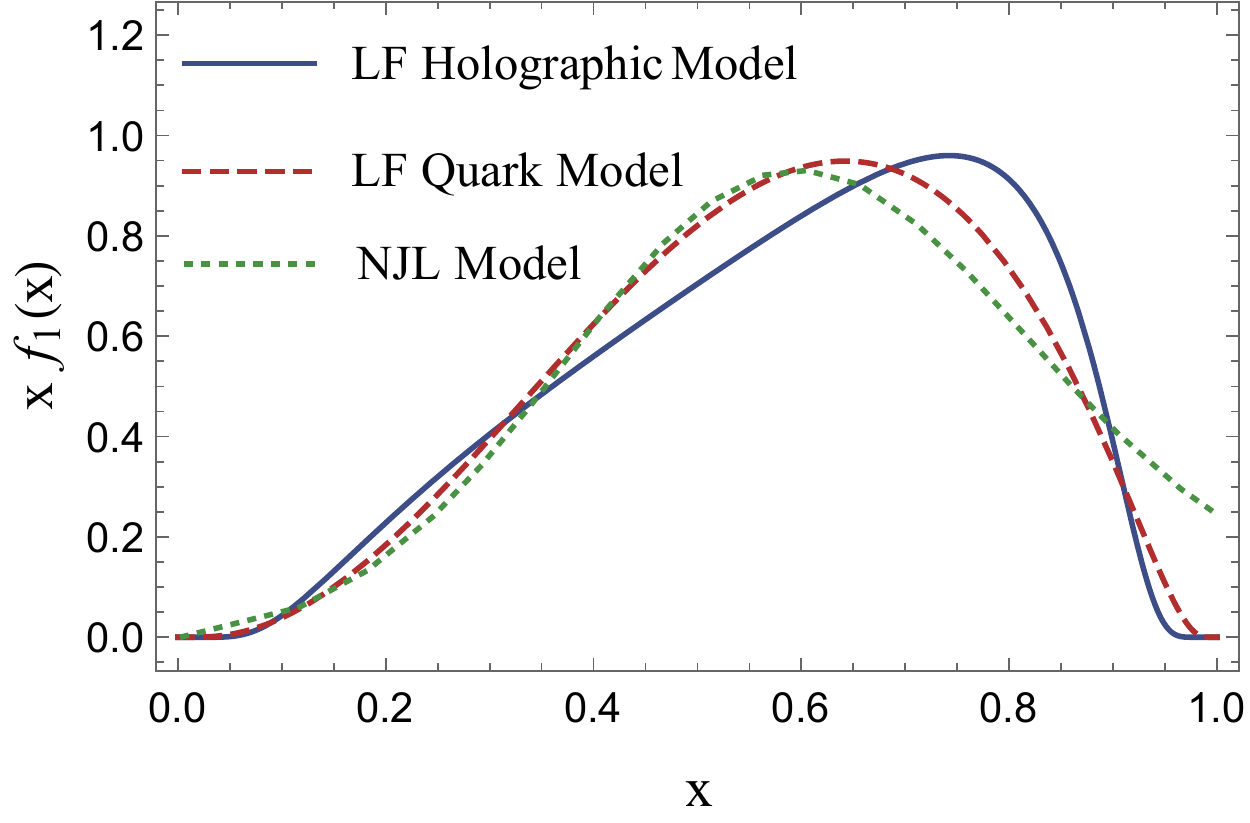}\end{center}
\end{minipage}
\begin{minipage}[c]{1\textwidth}\begin{center}
(c)\includegraphics[width=.38\textwidth]{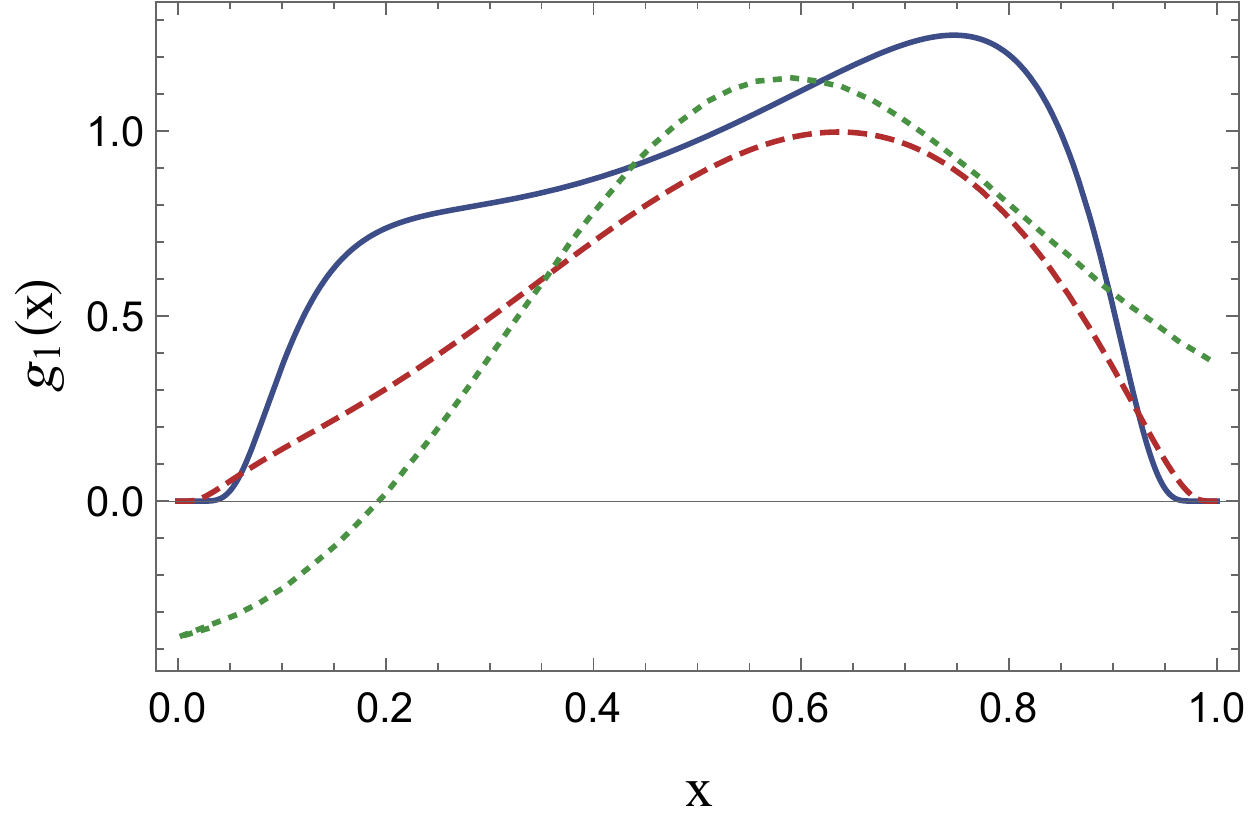}
(d)\includegraphics[width=.38\textwidth]{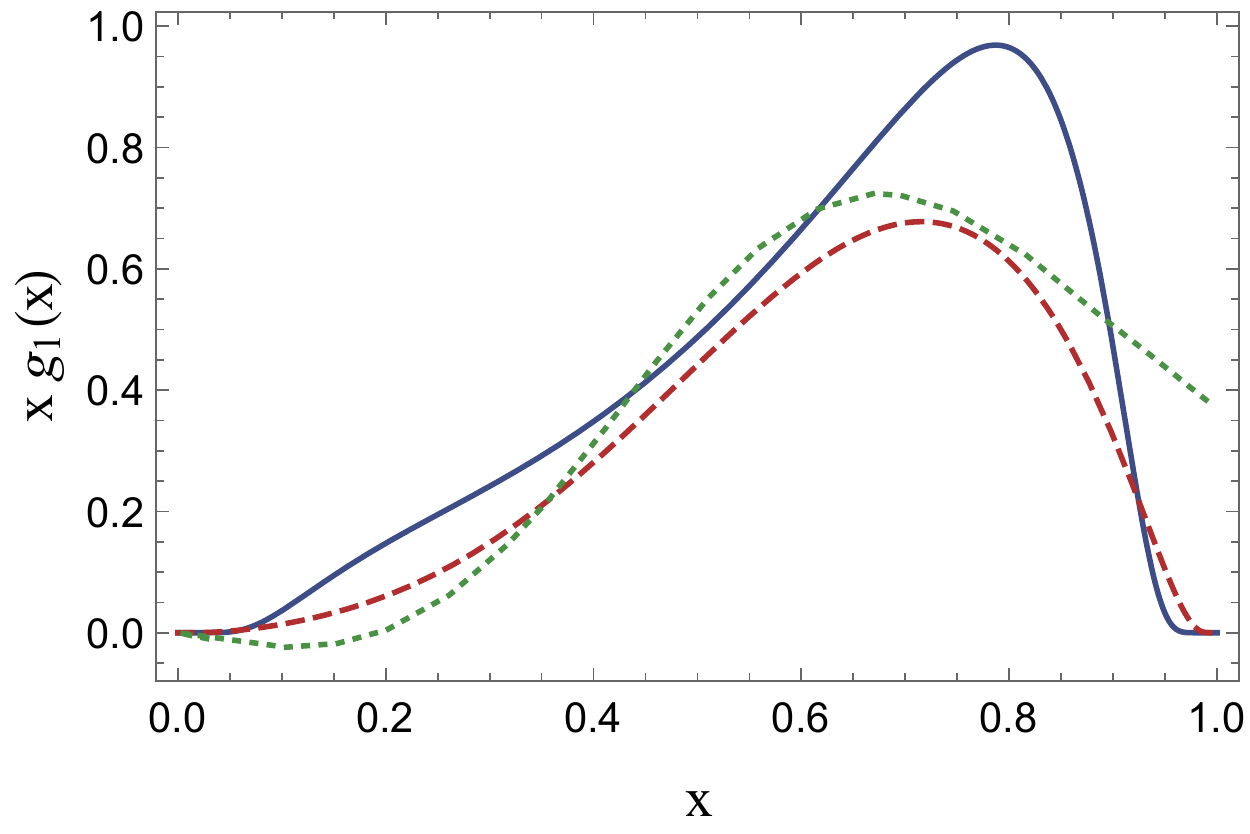}\end{center}
\end{minipage}
\begin{minipage}[c]{1\textwidth}\begin{center}
(e)\includegraphics[width=.38\textwidth]{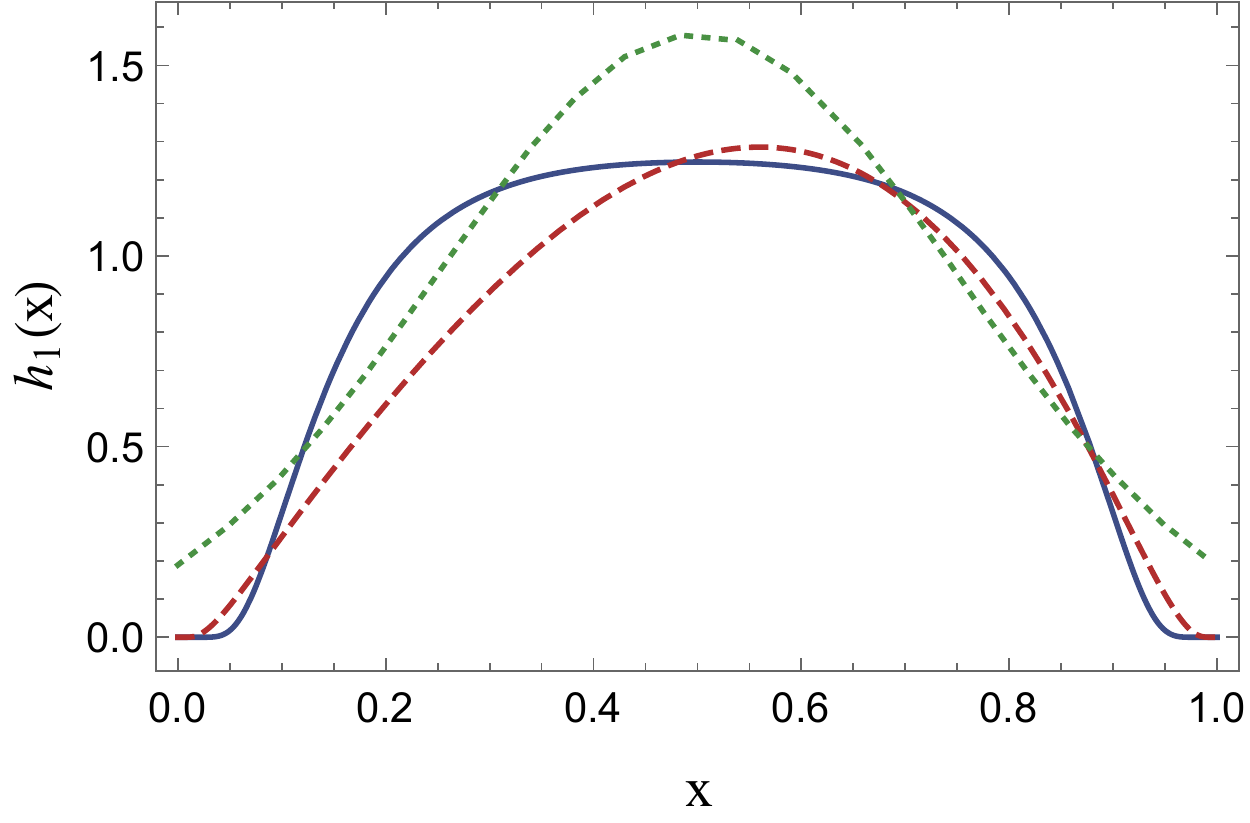}
(f)\includegraphics[width=.38\textwidth]{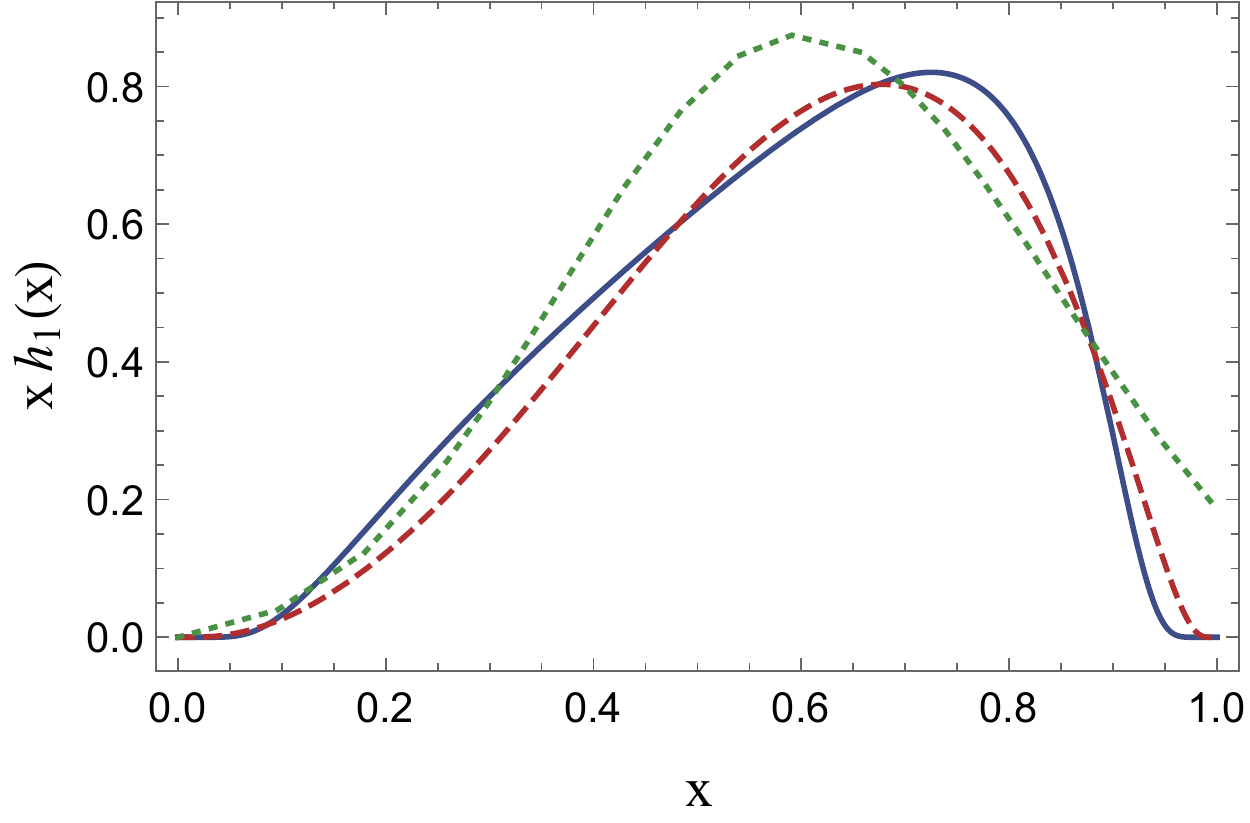}\end{center}
\end{minipage}
\begin{minipage}[c]{1\textwidth}\begin{center}
(g)\includegraphics[width=.38\textwidth]{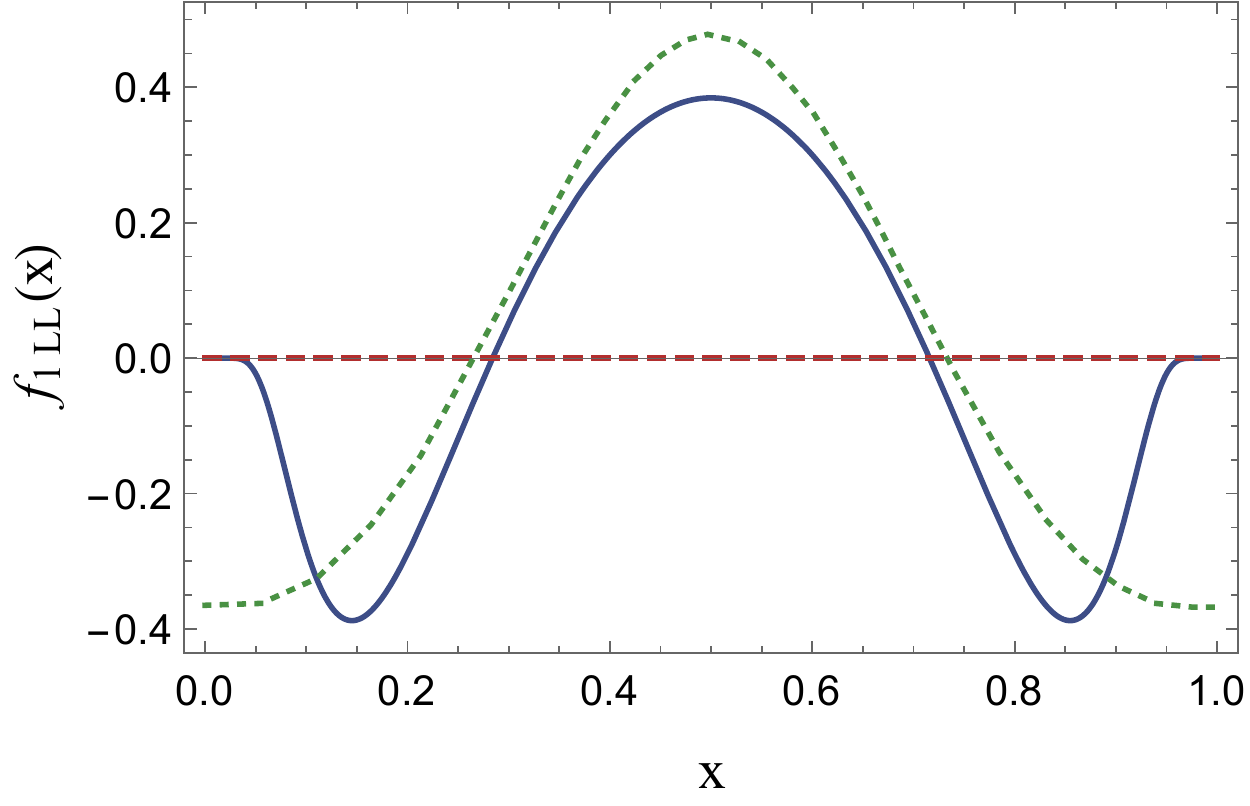}
(h)\includegraphics[width=.38\textwidth]{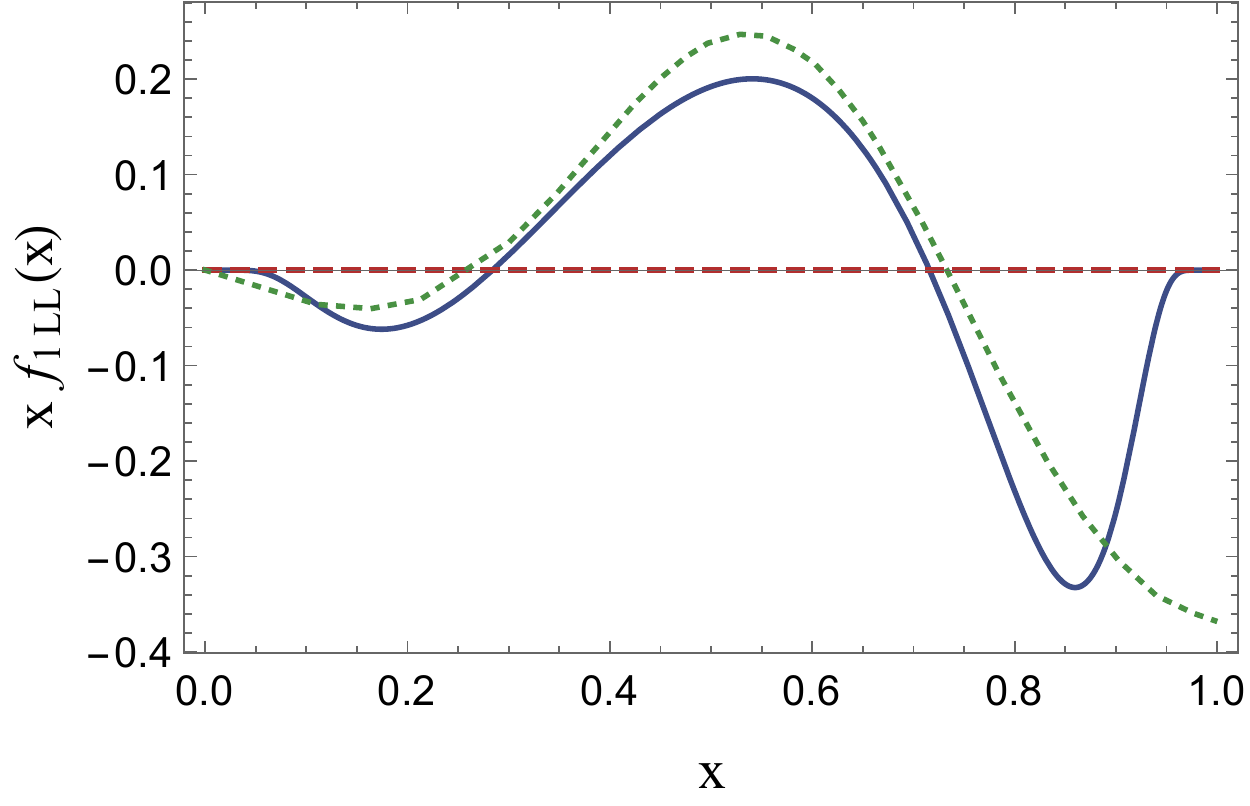}\end{center}
\end{minipage}
\caption{Left panel: The PDFs: $f_1(x),~ g_{1L}(x), ~h_1(x)$ and $f_{1LL}(x)$ as functions of $x$. Right panel:  The PDFs multiplied by $x$, i.e., $x f_1(x),~ x g_{1L}(x),~ x h_1(x)$ and $x f_{1LL}(x)$ are plotted with respect to $x$. The solid-blue and dashed-red curves represent the results evaluated in the LF holographic model and the LF quark model, respectively. Our results are compared with the predictions of the NJL model (dotted-green curves)~\cite{Ninomiya:2017ggn}}
\label{pdfs}
\end{figure}

\begin{figure}[tbh]
\includegraphics[width=1\textwidth]{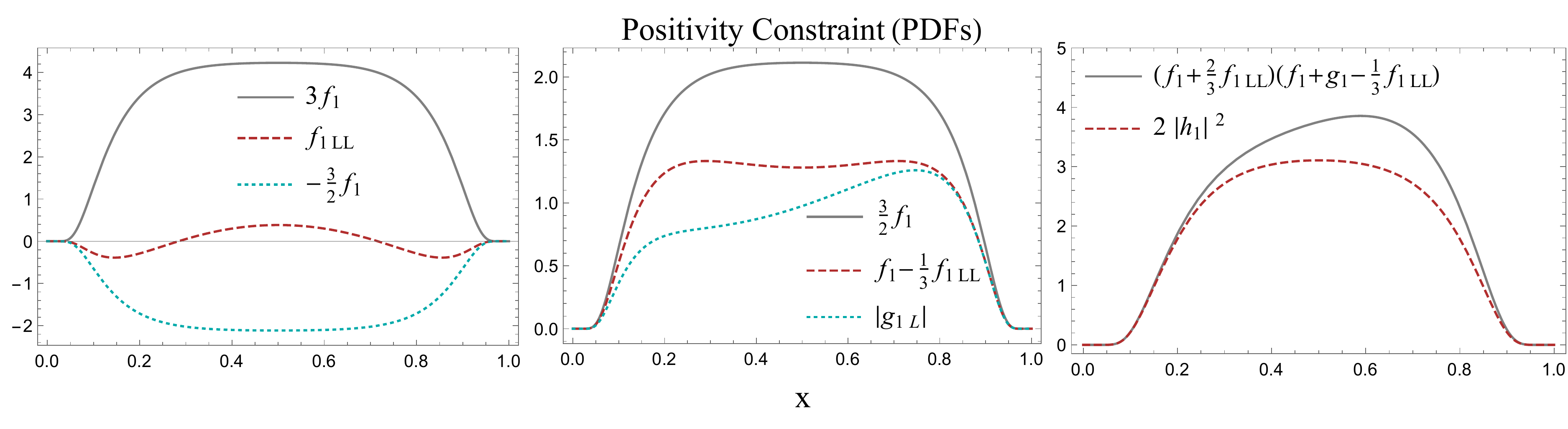}
\caption{Several positivity constraints on PDFs plotted with respect to $x$.}
\label{positivity-constraints-pdfs}
\end{figure}

In Fig. \ref{pdfs}, we show the behavior of the leading-twist PDFs of $\rho$-meson namely, the unpolarized distribution $f_1(x)$, the helicity distribution $g_1(x)$, the transversity distributon $h_1(x)$, and the tensor polarized distribution $f_{1LL}(x)$ at the model scales with respect to the longitudinal momentum fraction carried by the quark. We compare our results in the LF holographic model as well as in the LF quark model with the predictions of NJL model~\cite{Ninomiya:2017ggn}. We obtain the holographic PDFs by integrating out ${\bf k}_\perp$ of the TMDs $f_{1}(x,{\bf k}^2_\perp)$, $g_{1L}(x,{\bf k}^2_\perp)$, $h_{1}(x,{\bf k}^2_\perp)$ and $f_{1LL}(x,{\bf k}^2_\perp)$ given in Eqs. (\ref{f1-LFH}), (\ref{g1l-LFH}), (\ref{h1-LFH}) and (\ref{f1ll-LFH}) respectively, while in the LF quark model, the corresponding TMDs are evaluated in Eqs.~(\ref{f1_LF}), (\ref{g1_LF}), and (\ref{f1LL_LF}). Overall, the qualitative behavior of the holographic PDFs and ones in the LF quark model are consistent with the predictions in the NJL model~\cite{Ninomiya:2017ggn}. The tensor-polarized PDF $f_{1LL}$ being an important quantity, related to $b_1$ structure function~\cite{Hoodbhoy:1988am}, vanishes in the LF quark model, whereas the holographic $f_{1LL}$ is in more or less agreement with the one in the NJL model~\cite{Ninomiya:2017ggn}, within the range $0.1<x<0.9$. However, it differs significantly when $x\to \{0,~1\}$, as shown in \ref{pdfs} (left panel). Surprisingly, the PDFs in the NJL model do not vanish at the end points, however, as expected, appear to be zero at $x\to\{0,1\}$ in both the LF models. $f_{1LL}$ has been measured by HERA for the deutron, spin-1 target~\cite{Airapetian:2005cb}. In Fig.~\ref{positivity-constraints-pdfs}, we illustrate that our holographic PDFs also satisfy  the positivity constraints mentioned in Eqs. (\ref{pos_f1})-(\ref{pos_f1LL}).

Further, at the model scale, the following sum rules are satisfied by our PDFs,
\begin{equation}
\int_0^1 {\rm d}x~f_1(x)=1 \,,
\end{equation}
\begin{equation}
\int_0^1 {\rm d}x~x\,f_1(x) + \int_0^1 {\rm d}x~(1-x)\,f_1(x)=1\,,
\end{equation}
\begin{equation}
\int_0^1 {\rm d}x~f_{1LL}(x)=0\ \ \ ; \ \ \ \int_0^1  {\rm d}x~x\,f_{1LL}(x)=0\,. 
\end{equation}
\begin{figure}[tbh]
\begin{minipage}[c]{1\textwidth}\begin{center}
(a)\includegraphics[width=.38\textwidth]{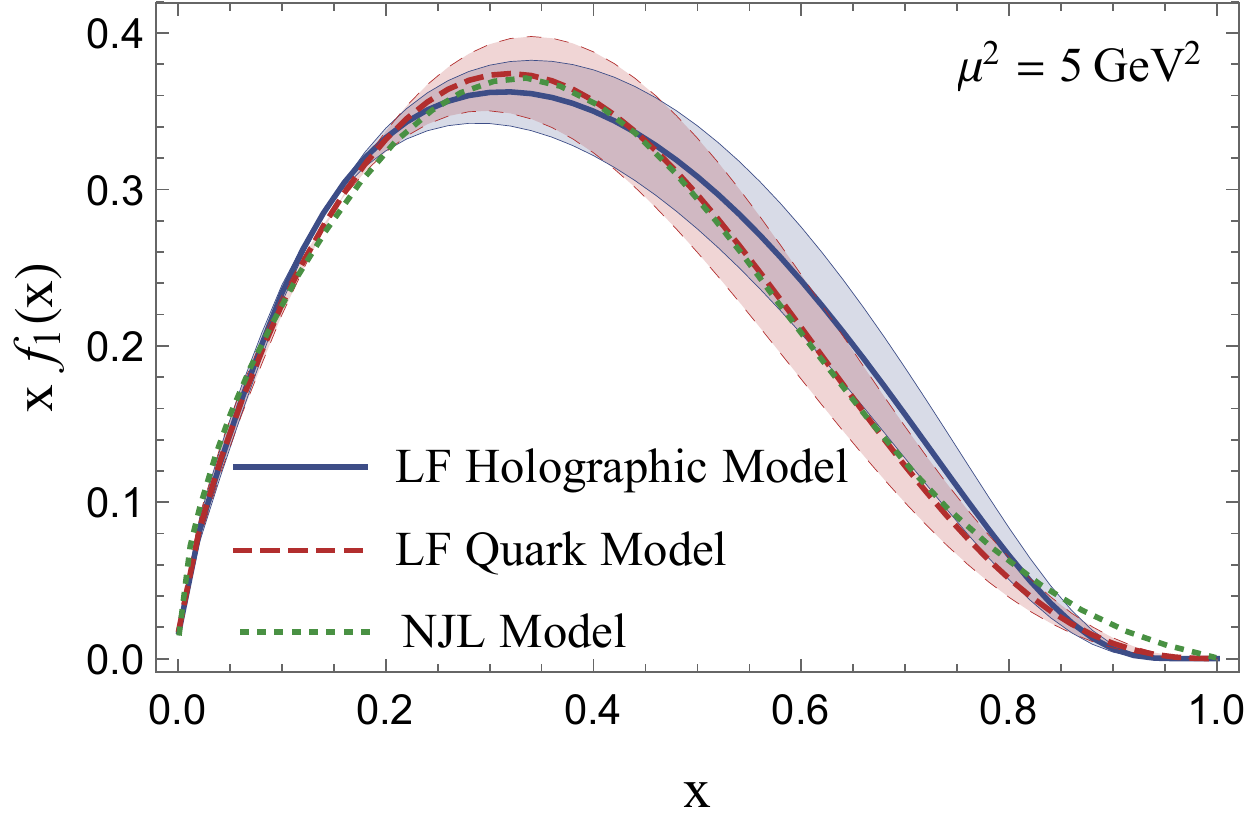}
(b)\includegraphics[width=.38\textwidth]{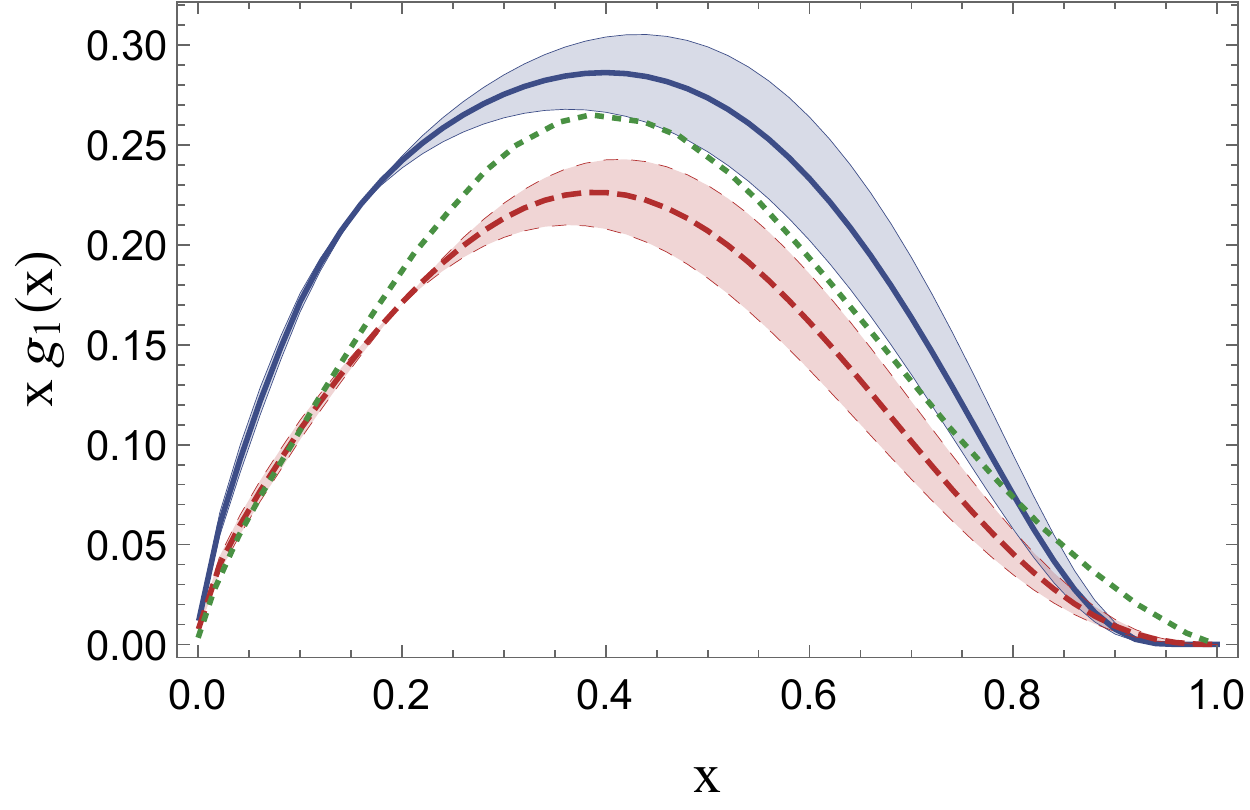}\end{center}
\end{minipage}
\begin{minipage}[c]{1\textwidth}\begin{center}
(c)\includegraphics[width=.38\textwidth]{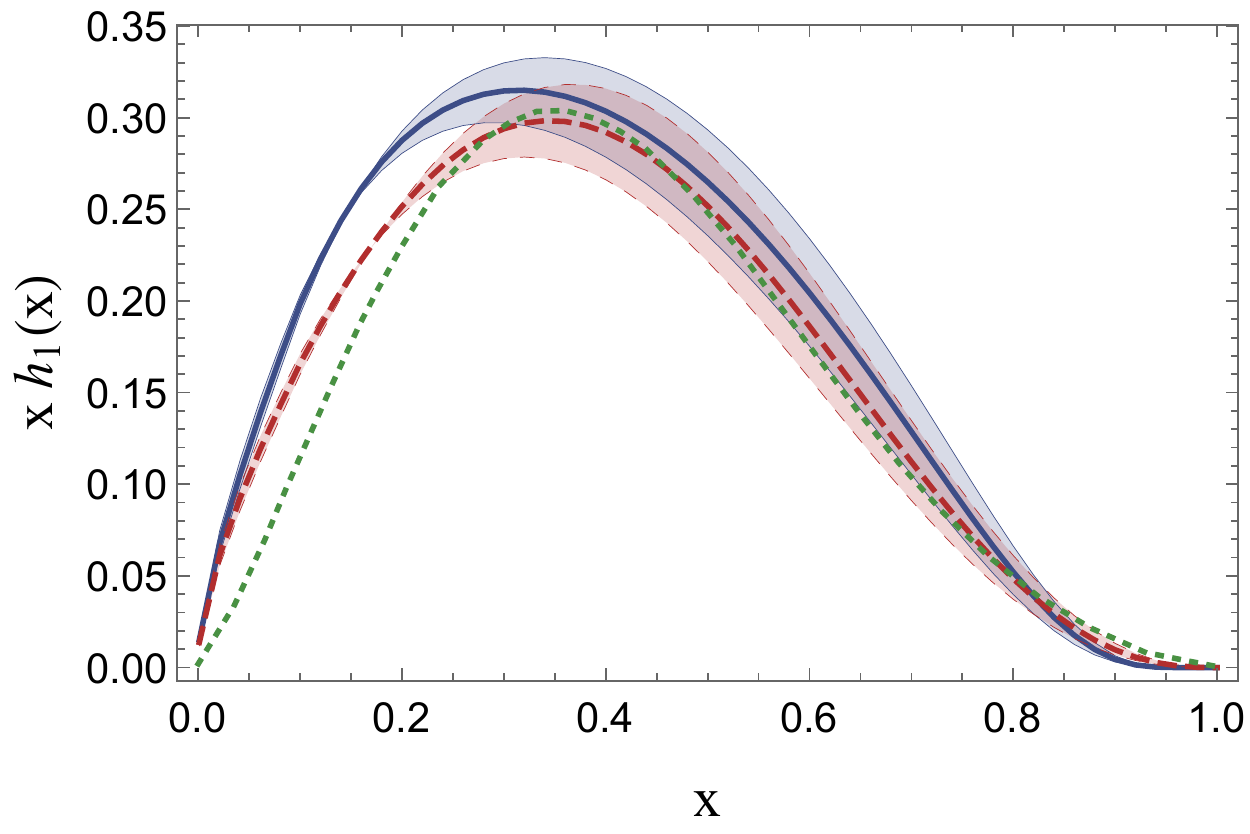}
(d)\includegraphics[width=.38\textwidth]{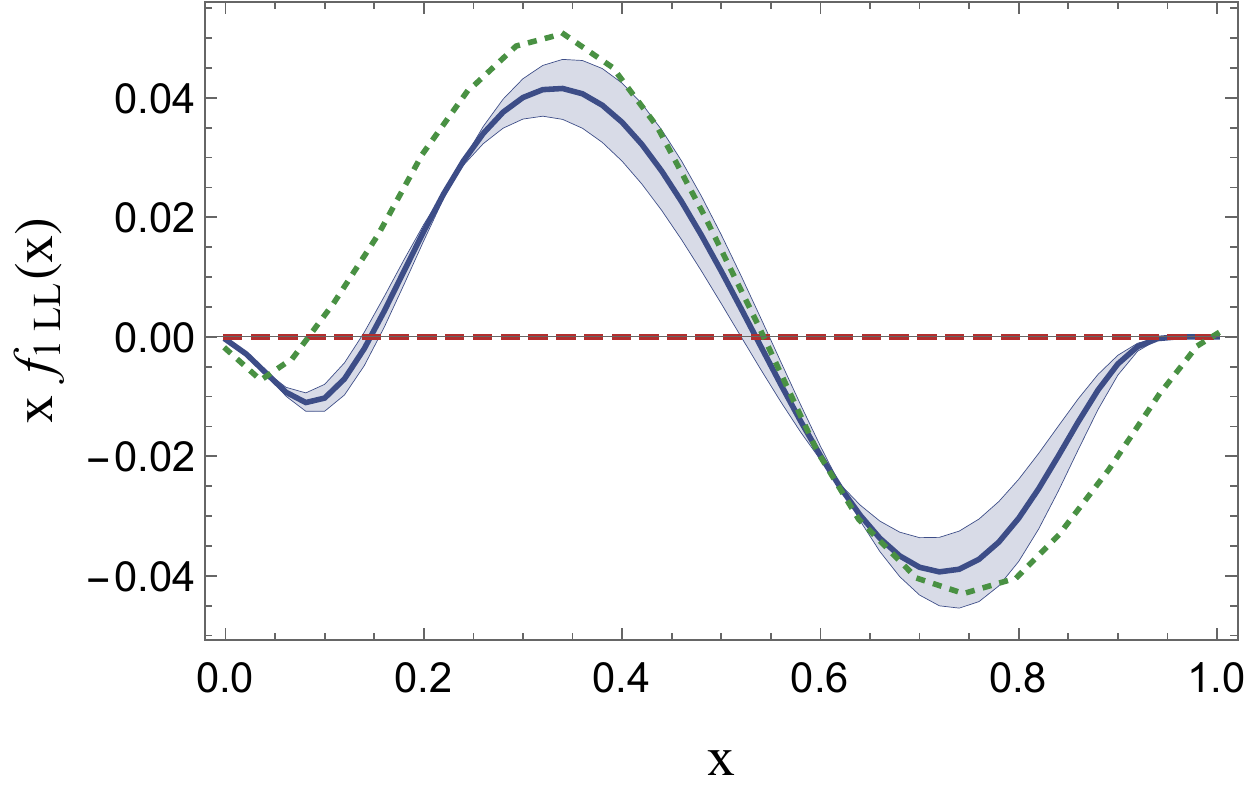}\end{center}
\end{minipage}
\caption{The evolved PDFs for $\rho$ meson: (a) $x f_1(x)$, (b) $x g_{1L}(x)$, (c) $x h_1(x)$ and (d) $x f_{1LL}(x)$ as functions of $x$. The blue bands are LF holographic model results evolved from the initial scale $\mu^2_{\rm LFH}=0.20 \pm 0.02$ GeV$^2$ using the NNLO DGLAP equations to the scale $\mu^2=5$ GeV$^2$. The red bands are LF quark model results evolved from the initial scale $\mu^2_{\rm LFQM}= 0.19 \pm 0.02$ GeV$^2$ to the same final scale. 
The green dotted lines represent the results from the NJL model~\cite{Ninomiya:2017ggn}.}
\label{pdfs-evolved}
\end{figure}
 
\subsection{QCD evolution for $\rho$ meson PDFs}
By performing the QCD evolution, the valence quark distributions at high $\mu^2$ scale can be obtained with the initial input PDFs. We use the  next-to-next-to-leading order (NNLO) Dokshitzer-Gribov-Lipatov-Altarelli-Parisi (DGLAP) equations~\cite{Dokshitzer:1977sg,Gribov:1972ri,Altarelli:1977zs} of QCD, to evolve our PDFs from our model scales to higher scale $\mu^2$. 
We explicitly evolve our initial $\rho$ meson PDFs in both the LF holographic and the LF quark model to the scale ${\mu^2=5~\mathrm{GeV}^2}$ utilizing the Higher Order Perturbative Parton Evolution toolkit to numerically solve the NNLO DGLAP equations~\cite{Salam:2008qg}. 

We adopt $\mu_{\rm LFH}^2=0.20\pm 0.02$ GeV$^2$ and $\mu_{\rm LFQM}^2=0.19\pm 0.02$ GeV$^2$ for the initial scales of the LF holographic and the LF quark models, respectively, which we determine by matching the valence quarks moment: $\langle x\rangle=\int_0^1 \,dx\, x\, f_1(x)$ at $\mu^2=5$ GeV$^2$, with the result from the NJL model,  after performing the QCD evolution of the valence quark PDF.
 Since the experimental data for $\rho$-meson PDFs are not available, we consider the results of the NJL model as the testimonial~\cite{Ninomiya:2017ggn} and set a $10\%$ uncertainty in the initial scales. We notice that at $\mu^2=5$ GeV$^2$, the valence quark and valence antiquark together carry $\sim 41\%$ of the total moment of $\rho$-meson. Note that the obtained initial scales of the LF holographic and the LF quark models are close to that of the NJL model, where $\mu^2_{\rm NJL}=0.16$ GeV$^2$.  We interpret the initial scales associated with the LF models as effective scales where the structures of the $\rho$ meson is described by the motion of the valence quarks only.

We show our result for the valence quark PDFs of the $\rho$ meson at $\mu^2=5$ GeV$^2$ after QCD evolution in Fig.~\ref{pdfs-evolved}, where we compare the valence quark distributions with the results from the NJL model~\cite{Ninomiya:2017ggn}.  We again notice that except $f_{1LL}$ in LF quark model, our results agree with the NJL model predictions. The error band in the valence quark distributions is due to the spread in the initial scales $\mu_{\rm LFH}^2=0.20\pm 0.02$ GeV$^2$ and $\mu_{\rm LFQM}^2=0.19\pm 0.02$ GeV$^2$ propagated by the QCD evolution.


\section{Spin densities in the momentum space}\label{density}
\begin{figure}[hbt]
\centering
\begin{minipage}[c]{1\textwidth}\begin{center}
(a)\includegraphics[width=.41\textwidth]{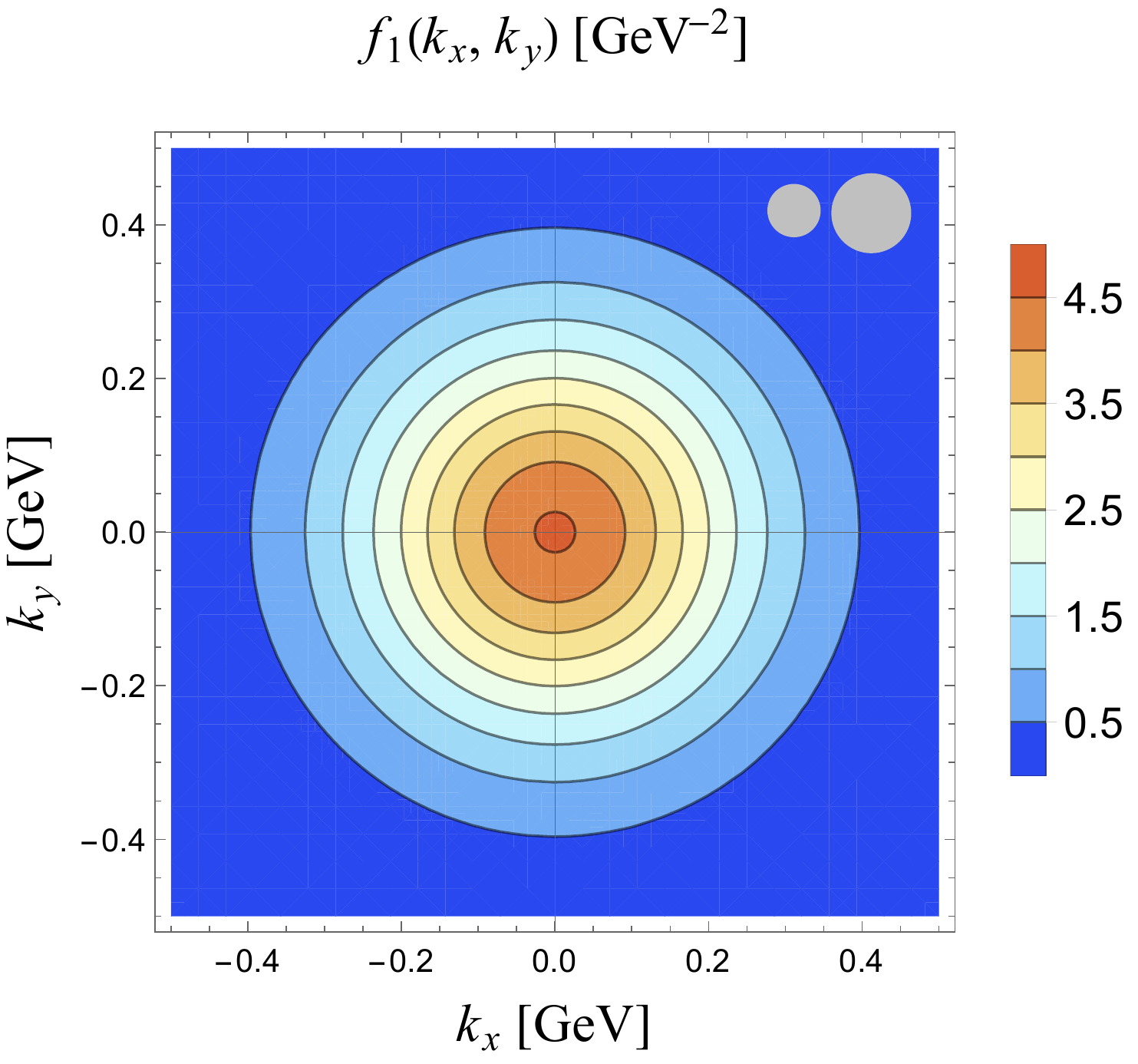} \end{center}
\end{minipage}
\begin{minipage}[c]{1\textwidth}
\begin{center}
(b)\includegraphics[width=.40\textwidth]{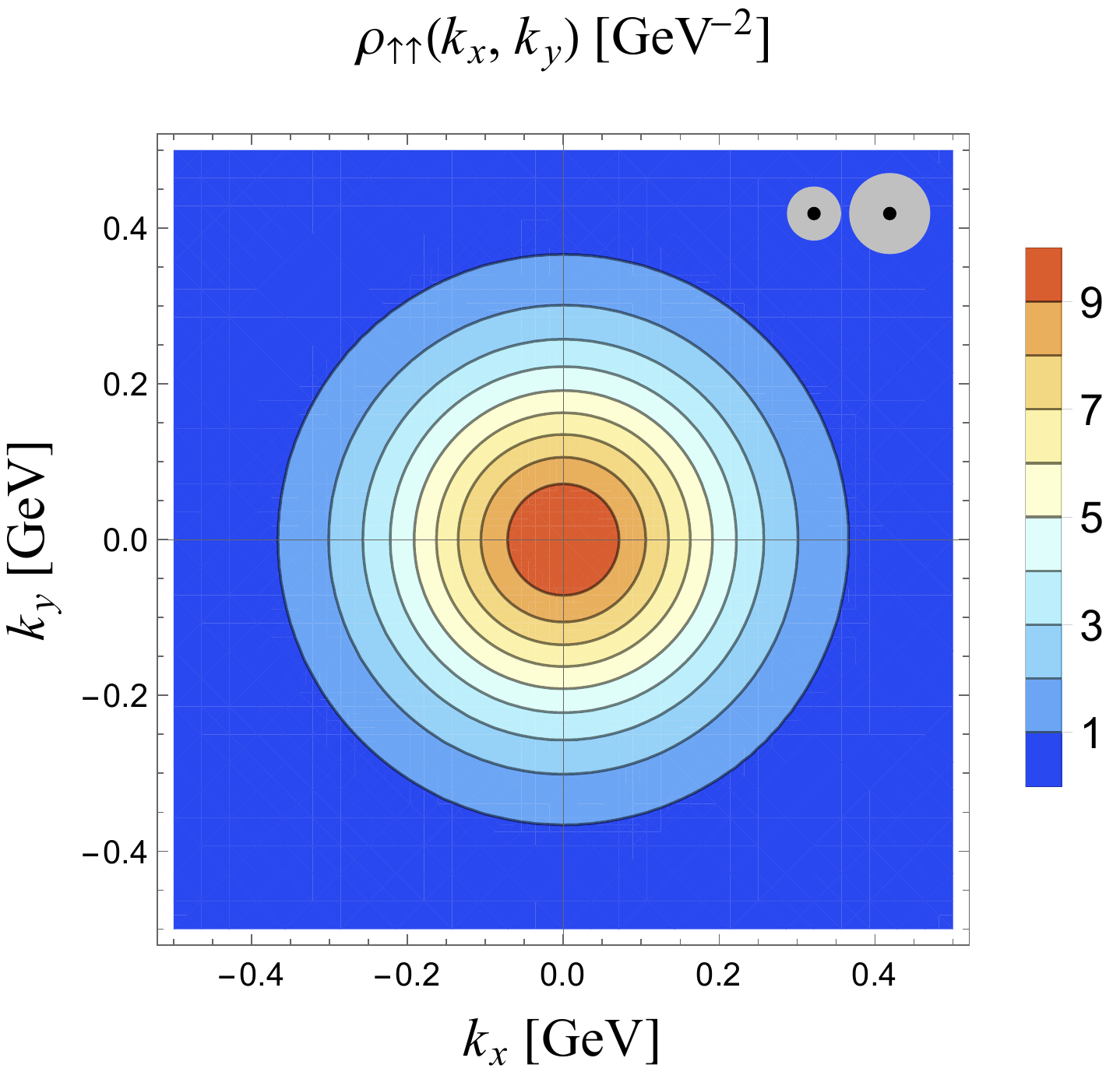}
(c)\includegraphics[width=.41\textwidth]{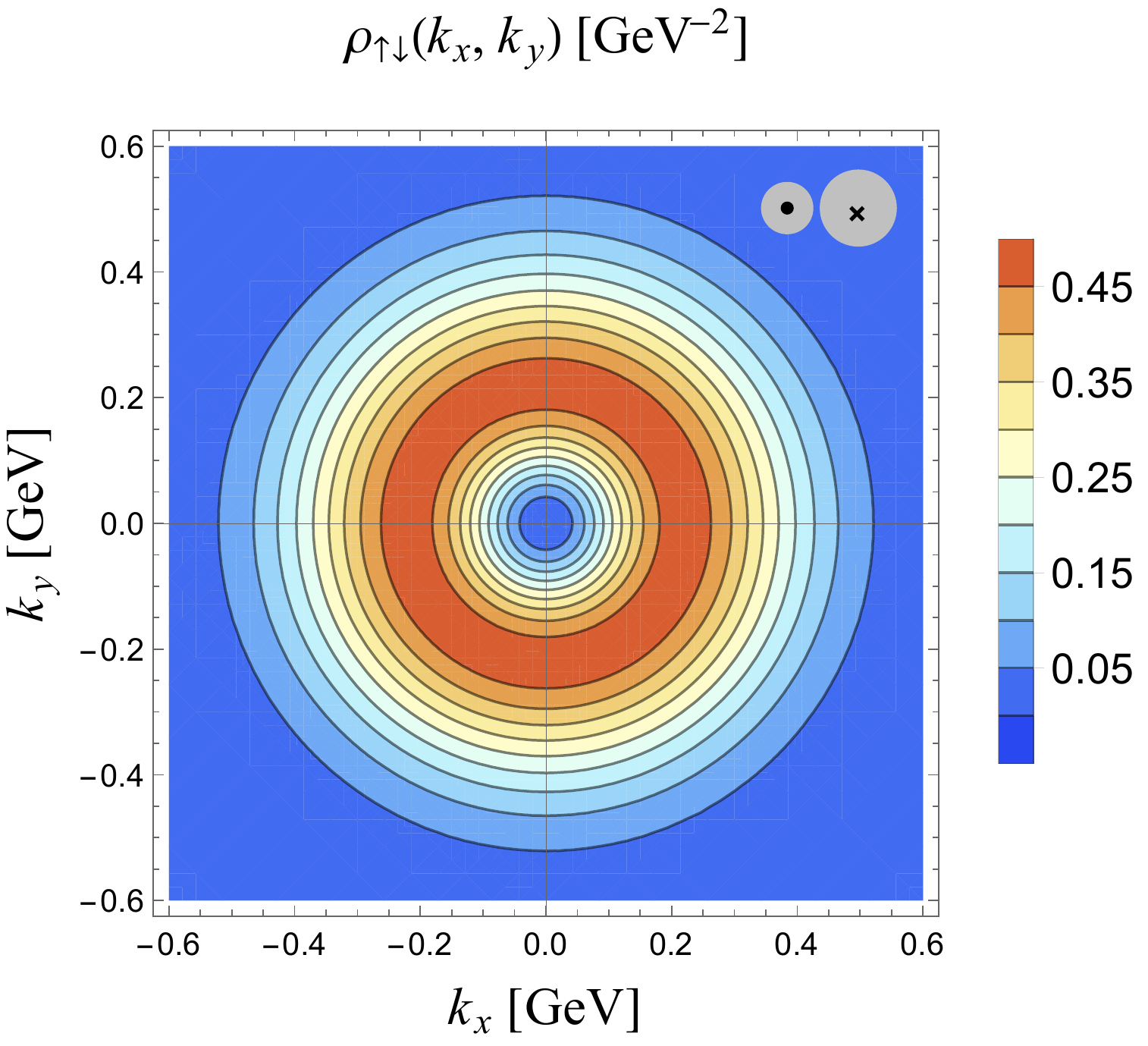}
\end{center}
\end{minipage}
\begin{minipage}[c]{1\textwidth}
\begin{center}
(d)\includegraphics[width=.40\textwidth]{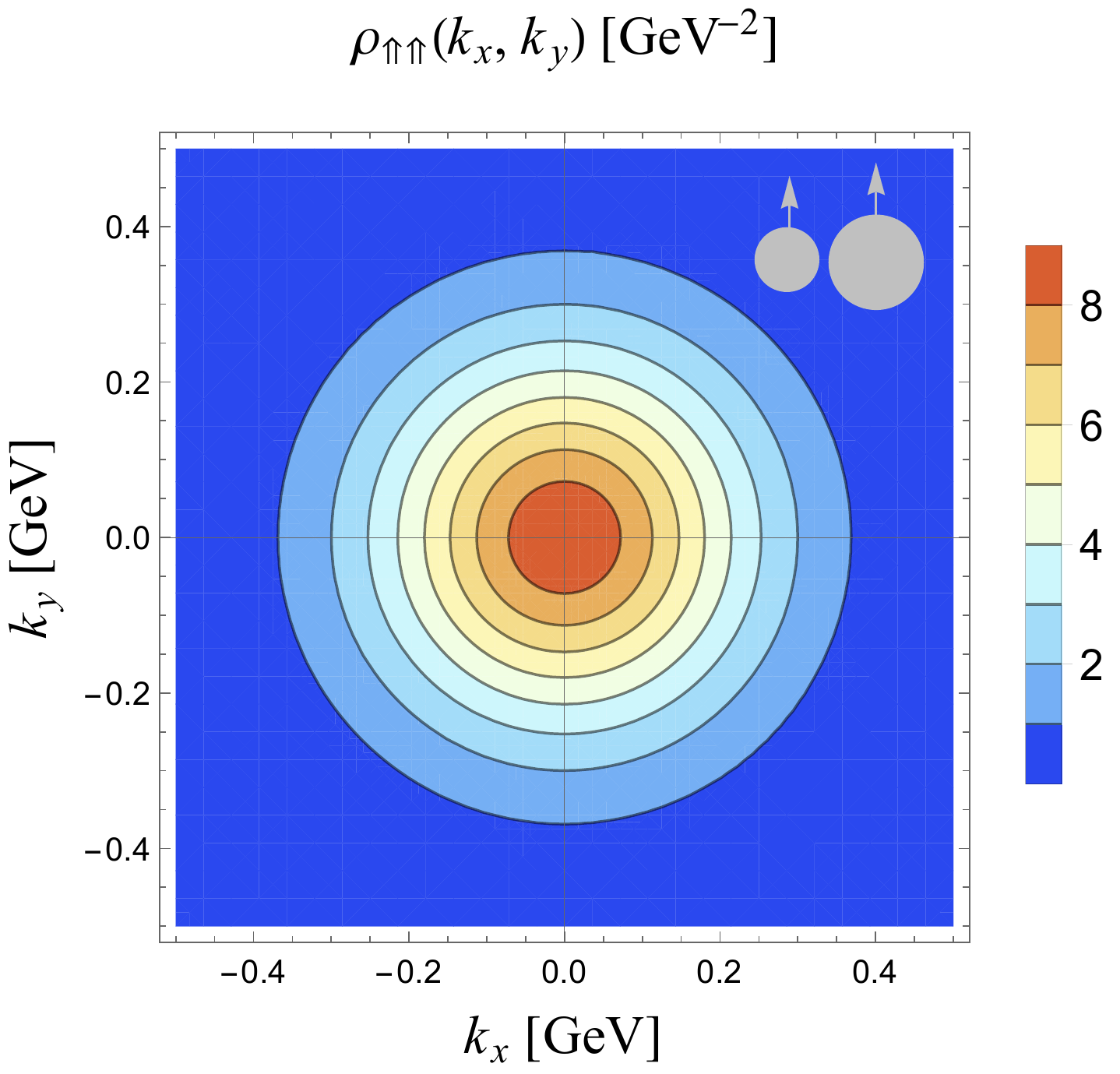}
(e)\includegraphics[width=.41\textwidth]{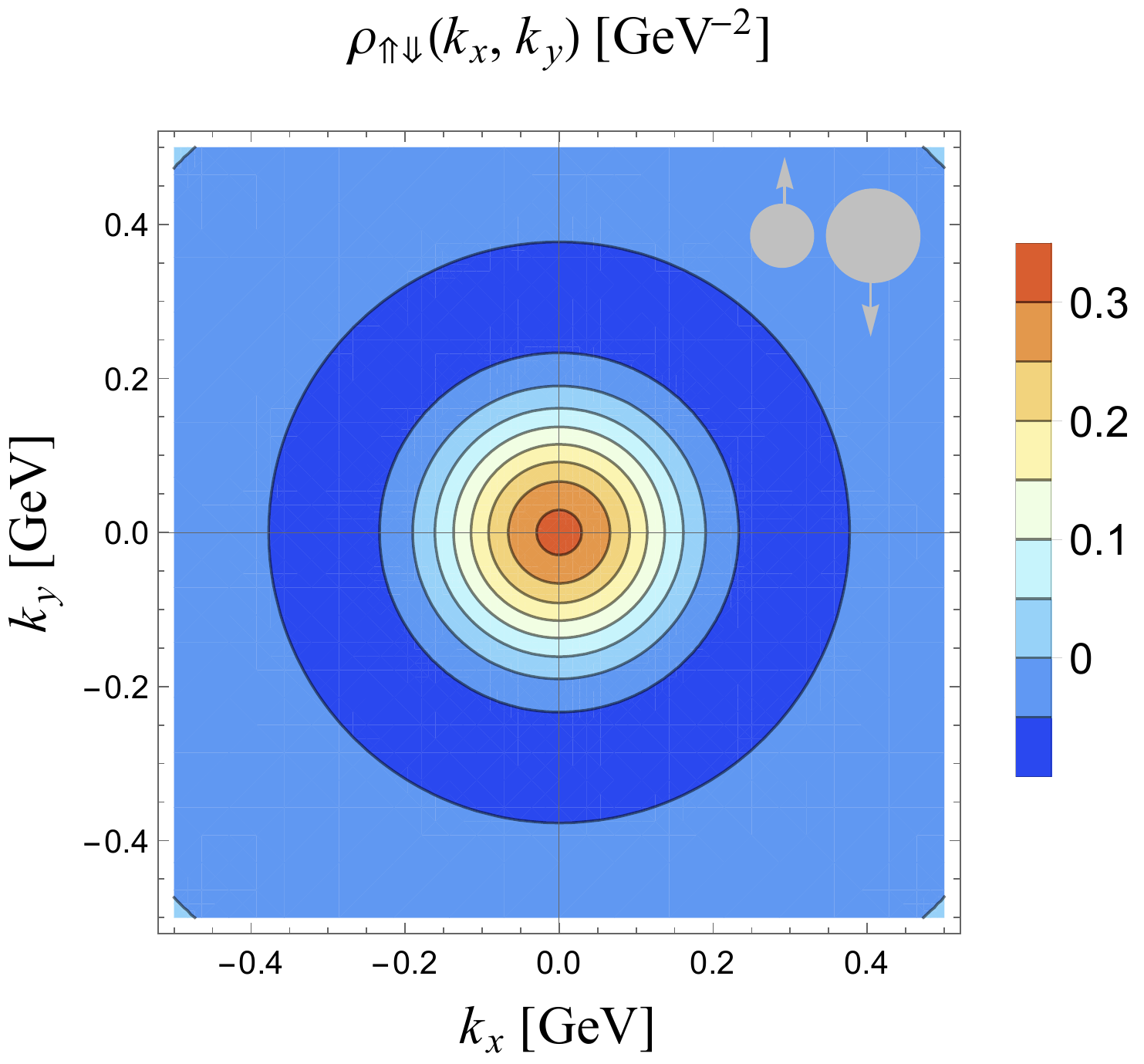}
\end{center}
\end{minipage}
\caption{(Color online) Quark density plots for $f_1(k_x,k_y)$(upper panel); $\rho_{\uparrow \uparrow}(k_x,k_y)$, $\rho_{\uparrow \downarrow}(k_x,k_y)$  (middle panel) and $\rho_{\Uparrow \Uparrow}(k_x,k_y)$, $\rho_{\Uparrow \Downarrow}(k_x,k_y)$ (lower panel) in the momentum plane. The gray colored vacant small and large circles (upper right corner) corresponds to both quark and the $\rho$-meson being unpolarized. The dot and cross inside the circles denote the longitudinal polarization in same and opposite directions respectively. The arrow along upward and downward directions symbolize the transverse polarization of the quark and the $\rho$-meson.}
\label{spin-densities1}
\end{figure}
\begin{figure}[hbt]
\begin{minipage}[c]{1\textwidth}\begin{center}
(a)\includegraphics[width=.41\textwidth]{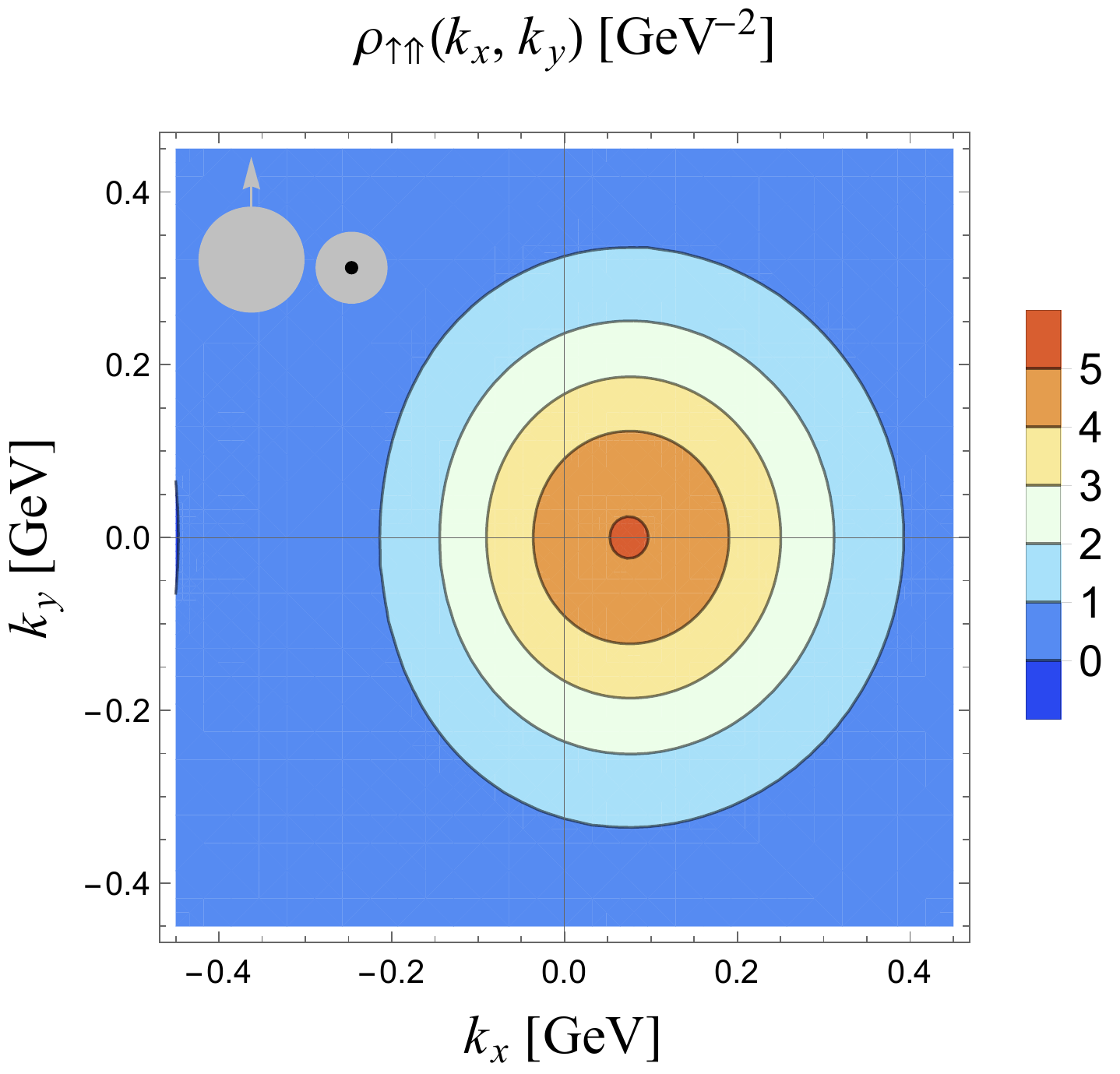}
(b)\includegraphics[width=.41\textwidth]{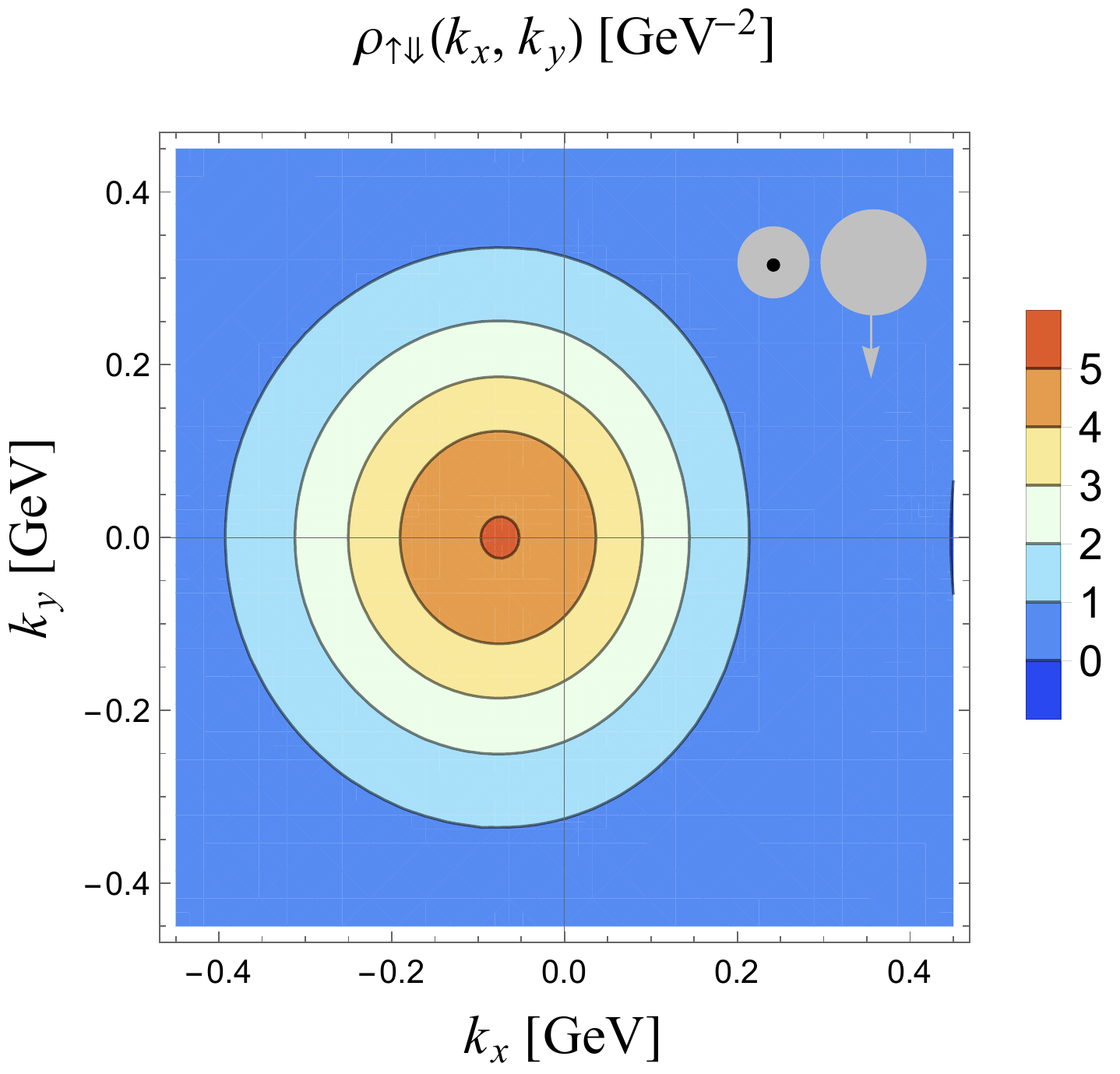}\end{center}
\end{minipage}
\begin{minipage}[c]{1\textwidth}\begin{center}
(c)\includegraphics[width=.41\textwidth]{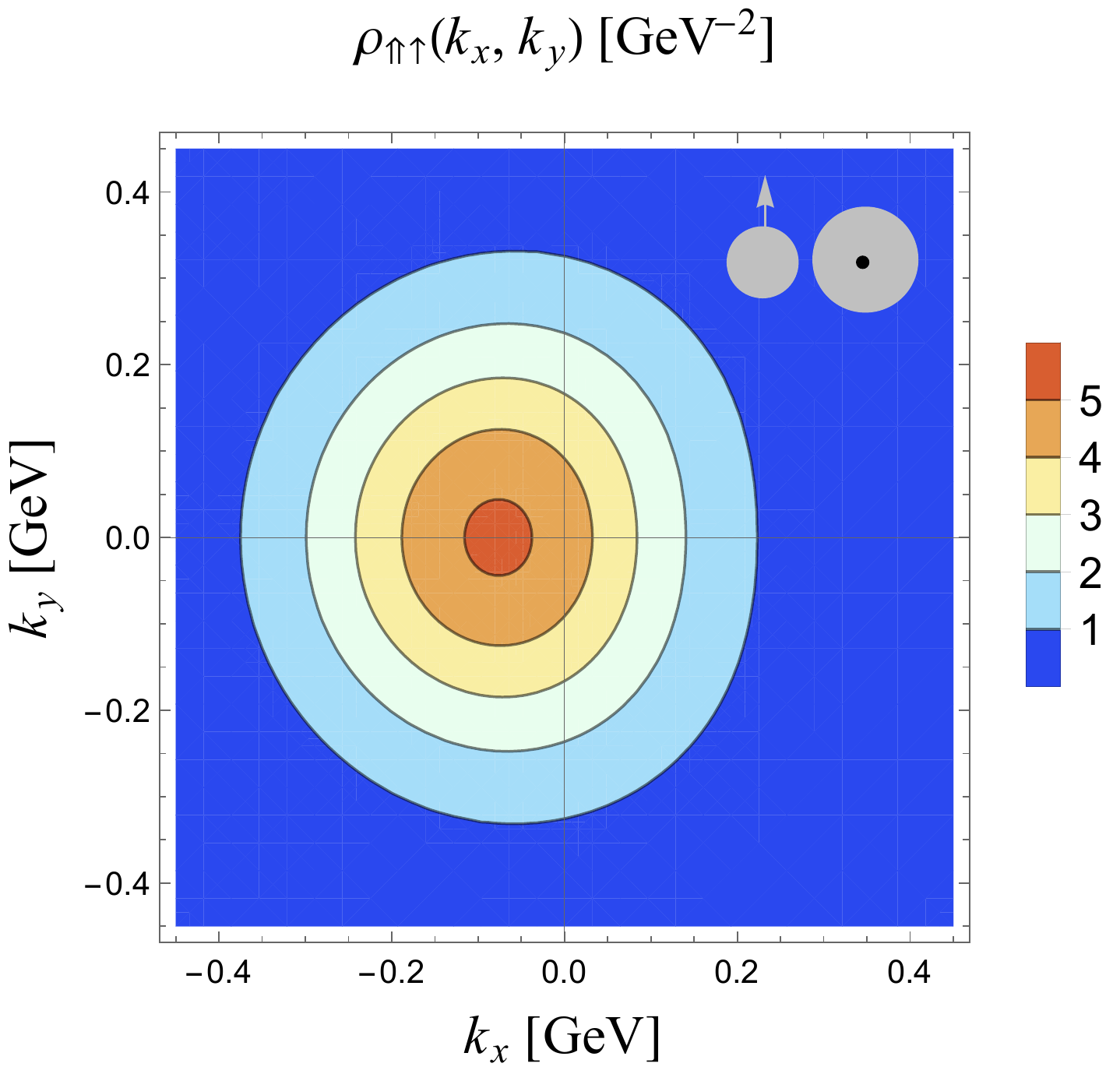}
(d)\includegraphics[width=.41\textwidth]{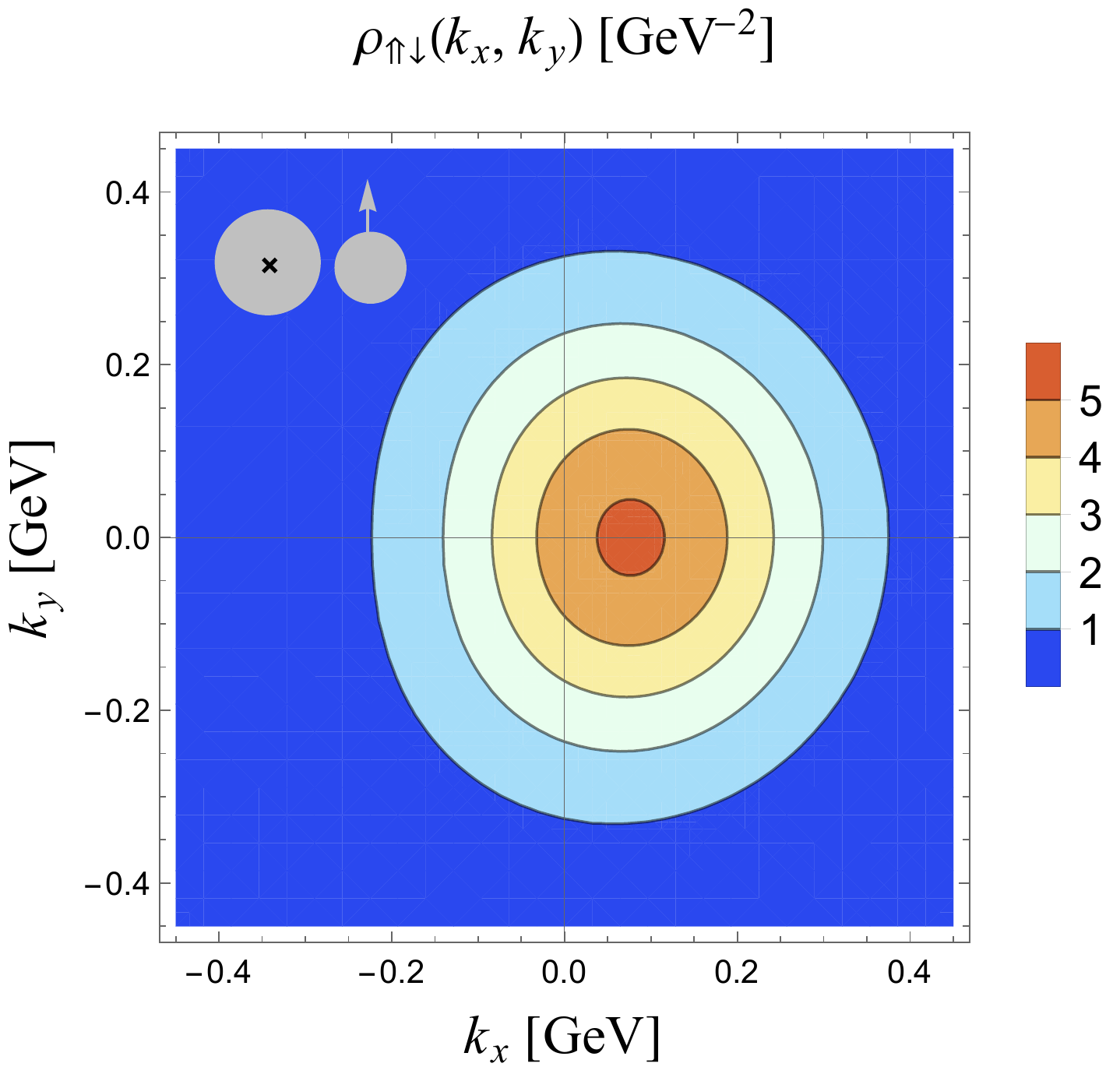}\end{center}
\end{minipage}
\caption{(Color online) Quark density plots for $\rho_{\uparrow \Uparrow}(k_x,k_y)$, $\rho_{\uparrow \Downarrow}(k_x,k_y)$(upper panel) and $\rho_{\Uparrow \uparrow}(k_x,k_y)$, $\rho_{\Uparrow \Downarrow}(k_x,k_y)$  (lower panel) in the momentum plane. The gray colored small and large circles (upper right or left corner) with dot and cross inside, denote the longitudinal polarization in same and opposite directions respectively. The arrow along upward and downward directions symbolize the transverse polarization of the quark and the $\rho$-meson.}
\label{spin-densities2}
\end{figure}
The TMDs can be interpreted as the quark densities inside the hadron. One can define the quark momentum distributions inside the target with the different polarization combinations via
TMDs. The spin densities describe the correlation between the quark and the target spins. Following Eqs. (\ref{form1})-(\ref{form3}), we define the quark spin densities in the momentum space for the spin-1 target as,
\begin{eqnarray}
\rho\left(x, k_x, k_y, (\lambda,\boldsymbol{\lambda}_\perp),(\Lambda, \boldsymbol{\Lambda}_\perp)\right)&=&
f_1 + \lambda \Lambda g_{1L}+\lambda \Lambda_\perp^i\frac{k_\perp^i}{M_\rho}g_{1T}+\lambda_\perp^i \Lambda_\perp^i h_1 +\lambda_\perp^i \Lambda \frac{k_\perp^i}{M_\rho}h_{1L}^\perp \nonumber\\
&&+(3 \lambda^2-2)\Bigg(\left(\frac{1}{6}-\frac{1}{2}\Lambda^2\right)f_{1LL}+\Lambda \Lambda_\perp^i \frac{k_\perp^i}{M_\rho}f_{1LT} \nonumber\\
&&+ \left(\Lambda_\perp^i \Lambda_\perp^j-\frac{1}{2}\Lambda_\perp^2 \delta^{ij}\right)\frac{k_\perp^i k_\perp^j}{M_\rho^2}f_{1TT}\Bigg).
\label{spin-densities}
\end{eqnarray}
Here $\lambda$ and $\Lambda$ correspond to the quark and the target spins in the longitudinal direction. Note that in this Section, $\lambda$ and $\Lambda$ denote different quantities from the previous sections. The configurations for these two can be $\lambda= \uparrow, \downarrow$ (or +1,-1) and $\Lambda=\uparrow, \downarrow$ (or +1,-1). $\boldsymbol{\lambda}_\perp=\Uparrow, \Downarrow$ (or +1,-1) and $\boldsymbol{\Lambda}_\perp=\Uparrow, \Downarrow$ (or +1,-1) symbolize the transverse spins of the quark and the target $\rho$-meson, respectively. Here, we consider the transverse polarization to be along $x$-direction. Depending on the different spin directions of the quark and the $\rho$-meson, we predict the various spin correlations, which are discussed below.

We integrate out the longitudinal momentum fraction $x$ to get all the spin densities in the transverse momentum plane.
In Fig. \ref{spin-densities1}, we show the spin densities in the transverse momentum plane by considering the different polarization configurations of the quark and the $\rho$-meson in the longitudinal direction. $\rho_{\uparrow \uparrow}=f_1-\frac{1}{3} f_{1LL}+ g_{1L}$ and $\rho_{\uparrow \downarrow}=f_1-\frac{1}{3}f_{1LL}-g_{1L}$ designated to $\lambda=\Lambda=\uparrow$ and $\lambda=\uparrow, \Lambda=\downarrow$ are explained in Figs.~\ref{spin-densities1}(b) and \ref{spin-densities1}(c), respectively. In view of $\rho_{\uparrow \uparrow}$, one can observe the probability of finding the quark in the $\rho$-meson with the spin aligned to the spin of the composite system, while $\rho_{\uparrow \downarrow}$ explains the probability when both spins are anti-aligned. $\rho_{\uparrow \uparrow}(k_x,k_y)$ and $\rho_{\uparrow \downarrow}(k_x,k_y)$, allow only those overlap configurations of LFWFs, which display the effect of only two wave contributions out of three. $\rho_{\uparrow \uparrow}$ has the contributions from the squared of the wave functions which describe the S-wave and the P-wave separation, while $\rho_{\uparrow \downarrow}$ can be obtained from the squared of the P-wave and the D-wave components. In other words, no OAM transfer occur between the initial and the final states in these cases. Also, both the densities $\rho_{\uparrow \uparrow}$ and $\rho_{\uparrow \downarrow}$ are axially symmetric. However, due to constructive interference between $f_{1}$ and $g_{1L}$ in $\rho_{\uparrow \uparrow}$, much larger magnitude has been observed as compared to $\rho_{\uparrow \downarrow}$, where $f_{1}$ and $g_{1L}$ appear with the opposite signs. To shed light on the transverse spin densities, let us consider $\boldsymbol{\lambda}_\perp=\boldsymbol{\Lambda}_\perp=\Uparrow$ and $\boldsymbol{\lambda}_\perp=\Uparrow, \boldsymbol{\Lambda}_\perp=\Downarrow$ shown in Figs.~\ref{spin-densities1}(d) and \ref{spin-densities1}(e) respectively, indicated by $\rho_{\Uparrow \Uparrow}=f_1-\frac{{\bf k}_\perp^2}{M_\rho^2}f_{1TT}+h_1$ and $\rho_{\Uparrow \Downarrow}=f_1-\frac{{\bf k}_\perp^2}{M_\rho^2}f_{1TT}-h_1$. These spin densities are the mixture of zero and two units of OAM transfer overlap terms. These are also axially symmetric. Due to a similar reason mentioned in the case of the longitudinal spin densities, $\rho_{\Uparrow \Uparrow}$ dominates over $\rho_{\Uparrow \Downarrow}$.

The distorting effects are observed in $\rho_{\uparrow \Uparrow(\uparrow \Downarrow)}$ and $\rho_{\Uparrow \uparrow(\Uparrow \downarrow)}$ as shown in Fig. \ref{spin-densities2}. $\rho_{\uparrow \Uparrow(\uparrow \Downarrow)}$ spin densities come into the picture when the spin directions of the quark and the $\rho$-meson are longitudinal and transverse respectively, i.e. $\lambda=\uparrow, \boldsymbol{\Lambda}_\perp=\Uparrow( \Downarrow)$. To describe $\rho_{\Uparrow \uparrow(\Uparrow \downarrow)}$, the spin directions $\boldsymbol{\lambda}_\perp=\Uparrow, \Lambda=\uparrow( \downarrow)$ are considered. We observe that the distortion effect takes place with these considerations because of the terms 
$\frac{k_x}{M_\rho} g_{1T}$ in $\rho_{\uparrow \Uparrow(\uparrow \Downarrow)}(=f_1\pm \frac{k_x}{M_\rho} g_{1T})$ and $\frac{k_x}{M_\rho} h_{1L}^\perp$ in $\rho_{\Uparrow \uparrow(\Uparrow \downarrow)}(=f_1\pm \frac{k_x}{M_\rho} h_{1L}^\perp)$.
In these cases, the densities feature a significant dipole deformation along $x$-direction arising
due to the terms mentioned above, while
the $f_1$ is stick to the monopole effect. These terms lead to the distortion in the plots when implemented together. One can also notice that the distortion of the longitudinally-polarized quark in the transversely-polarized target is opposite to that of the transversely-polarized
quark in the longitudinally polarized target. The reason is that $g_{1T}$ is positive, while 
$h_{1L}^\perp$ is
negative for $\rho$-meson.
\section{Conclusions}\label{summary}
In this work, we have presented the leading-twist TMDs for the $\rho$-meson in the LF holographic model where the dynamical spin effect has been taken into account in the wave functions. The TMDs have been analyzed by practising the overlap representation of the light-front wave functions in the constituent valence quark Fock space via the model independent overlap forms of the different light-front amplitudes.
 We have compared the holographic predictions with the distributions evaluated in the LF quark model, which contains a different spin structure from that in the LF holographic model.
 
 We have observed that the TMDs $f_1$, $g_{1L}$, $g_{1T}$, $h_1$, and $h_{1L}^{\perp}$ show a quite similar behavior in both the light-front models. However, the holographic $h_{1T}^{\perp}$ TMD vanishes and shows the negative distribution in the LF quark model. On the other hand, the tensor polarized TMDs, $f_{1LL}$, $f_{1LT}$ and $f_{1TT}$ in the LF quark model appear to be zero but nonzero in the LF holographic model, where $f_{1LL}$ and $f_{1TT}$ exhibit symmetry and $f_{1LT}$ shows anti-symmetry under $x\leftrightarrow(1-x)$. 
Nevertheless, our holographic predictions on all the TMDs have been found consistent with the previous finding in the NJL model~\cite{Ninomiya:2017ggn}. All the TMDs satisfy the necessary positivity constraints~\cite{Bacchetta:2001rb, Ninomiya:2017ggn}. We have also presented first two moments, $\langle k_\perp\rangle$ and  $\langle k_\perp^2\rangle$, of various TMDs, which are comparable with the available predictions of the NJL model. 

Next, we have evaluated the four survived PDFs for the $\rho$-meson, namely, the unpolarized $f_1$, the helicity $g_1$, the transversity $h_1$ and the tensor $f_{1LL}$ PDFs. We have again compared our results in the LF holographic and the LF quark models with the available NJL model predictions and observed the qualitative agreement between our results (except $f_{1LL}$ in the LF quark model) and the NJL model predictions. The PDF $f_{1LL}$, which requires the tensor polarization of the meson, is an important measurable quantity, vanishes in the LF quark model. The valence quark PDFs in the holographic model, after QCD evolution, are consistent with the NJL model results.

We have also studied the spin densities in the transverse momentum plane of the quark inside the $\rho$-meson with the different polarization configurations. The distributions for both the quark and the target polarized in the longitudinal or in the transverse directions have been found to be axially symmetric. Meanwhile, we have observed the dipolar distortions on top of the unpolarized symmetric distribution when the quark is longitudinally polarized and the target is transversely polarized, or vice-versa. The distortions have been found to be opposite for the longitudinal-transverse and the transverse-longitudinal polarization configurations of the quark and the $\rho$-meson.

For further investigation, the future developments should focus on the inclusion of the nontrivial gauge link that will provide a prediction of the various T-odd $\rho$-meson TMDs.
The presented results in this study together with other theoretical predictions on the
TMDs and the PDFs may help the experimental groups to measure these distributions for the $\rho$-meson. Any experimental data on these distributions and the comparison with the theoretical predictions can help one to gain the valuable knowledge on the internal structure of the $\rho$-meson.

\acknowledgments
We thank  Ruben
Sandapen and Mohammad Ahmady for many useful discussions. C.M. is supported by the National Natural Science Foundation of China (NSFC) under the Grants No.
11850410436 and No. 11950410753. C.M is also
supported by new faculty start up funding by the Institute of Modern Physics, Chinese Academy of Sciences and the Strategic Priority Research Program of Chinese Academy of Sciences, Grant No. XDB34000000. H.D. would like to thank the Department of Science and Technology (Ref No. EMR/2017/001549) Government of India for financial support.
\appendix
\section{Light-front quark model}\label{LFmodel}
The complete light-front wave function is accomplished by appraising the spin and the momentum wave functions i.e. $\chi$ and $\psi$ depending upon the spin projections of the $\rho$-meson, $\Lambda$, at the scale $\mu_{\rm LFQM}^2=0.19$ GeV$^2$ \cite{Yu:2007hp,Qian:2008px}:
\begin{eqnarray}
\Psi^\Lambda_{h_q,h_{\bar{q}}}(x,{\bf k}_\perp)=\chi_{h_q,h_{\bar{q}}}^\Lambda (x,{\bf k}_\perp) \psi(x,{\bf k}^2_\perp),
\end{eqnarray}
with
\begin{eqnarray}
\sum_{h_q,h_{\bar{q}}}  \chi_{h_q,h_{\bar{q}}}^{\Lambda *}(x,{\bf k}_\perp)\chi_{h_q,h_{\bar{q}}}^\Lambda(x,{\bf k}_\perp) =1. 
\end{eqnarray}
According to the Brodsky-Huang-Lepage (BHL) prescription, the momentum wave function is written as 
\begin{eqnarray}
 \psi(x,{\bf k}^2_\perp)= \mathcal{N} \,{ \rm exp} \left[-\frac{{\bf k}^2_\perp+m_q^2}{8 \beta^2 x(1-x)} \right].
 \label{momentum-space-wf}
\end{eqnarray}
The spin part of the wave function is provided by relating the spin states transforming from the instant form to the light-front form by using the Melosh-Wigner method. For $\Lambda=T (+)$ with the quark and the antiquark helicities being $h_q$ and $h_{\bar{q}}$, we have 
\begin{eqnarray}
\chi^{T(+)}_{+,+}(x,{\bf k}_\perp)&=&\frac{m_q(\mathcal{M}+2 m)+{\bf k}^2_\perp}{\left(\mathcal{M}+2 m_q \right)\sqrt{{\bf k}^2_\perp+m_q^2}}\, ,\\
\chi^{T(+)}_{+,-}(x,{\bf k}_\perp)&=&\frac{\left(x \mathcal{M}+m_q\right) k_R}{\left(\mathcal{M}+2 m_q \right)\sqrt{{\bf k}^2_\perp+m_q^2}}\, ,\\
\chi^{T(+)}_{-,+}(x,{\bf k}_\perp)&=&-\frac{\left(\left(1-x\right) \mathcal{M}+m_q \right) k_R}{\left(\mathcal{M}+2 m_q \right)\sqrt{{\bf k}^2_\perp+m_q^2}}\, ,\\
\chi^{T(+)}_{-,-}(x,{\bf k}_\perp)&=& -\frac{k_R^2}{\left(\mathcal{M}+2 m_q \right)\sqrt{{\bf k}^2_\perp+m_q^2}}\, ,
\end{eqnarray} 
for $\Lambda=L$,
\begin{eqnarray}
\chi^{L}_{+,+}(x,{\bf k}_\perp)&=&\frac{\left(1-2x \right) \mathcal{M} k_L}{\left(\mathcal{M}+2 m_q \right)\sqrt{2\left({\bf k}^2_\perp+m_q^2\right)}} \, ,\\
\chi^{L}_{+,-}(x,{\bf k}_\perp)&=&\frac{m_q(\mathcal{M}+2 m_q)+2{\bf k}^2_\perp}{\left(\mathcal{M}+2 m_q \right)\sqrt{2\left({\bf k}^2_\perp+m_q^2\right)}}\, ,\\
\chi^{L}_{-,+}(x,{\bf k}_\perp)&=&\frac{m_q(\mathcal{M}+2 m_q)+2{\bf k}^2_\perp}{\left(\mathcal{M}+2 m_q \right)\sqrt{2\left({\bf k}^2_\perp+m^2\right)}}\, ,\\
\chi^{L}_{-,-}(x,{\bf k}_\perp)&=&-\frac{\left(1-2x \right) \mathcal{M} k_R}{\left(\mathcal{M}+2 m_q \right)\sqrt{2\left({\bf k}^2_\perp+m_q^2\right)}} \, ,
\end{eqnarray}
for $\Lambda=T (-)$
\begin{eqnarray}
\chi^{T(-)}_{+,+}(x,{\bf k}_\perp)&=& -\frac{k_L^2}{\left(\mathcal{M}+2 m_q \right)\sqrt{{\bf k}^2_\perp+m_q^2}}\, ,\\
\chi^{T(-)}_{+,-}(x,{\bf k}_\perp)&=& \frac{\left(\left(1-x\right) \mathcal{M}+m_q \right) k_L}{\left(\mathcal{M}+2 m_q \right)\sqrt{{\bf k}^2_\perp+m_q^2}}\, ,\\
\chi^{T(-)}_{-,+}(x,{\bf k}_\perp)&=&-\frac{\left(x \mathcal{M}+m_q\right) k_L}{\left(\mathcal{M}+2 m_q \right)\sqrt{{\bf k}^2_\perp+m_q^2}}\, ,\\
\chi^{T(-)}_{-,-}(x,{\bf k}_\perp)&=&\frac{m_q(\mathcal{M}+2 m_q)+{\bf k}^2_\perp}{\left(\mathcal{M}+2 m_q \right)\sqrt{{\bf k}^2_\perp+m_q^2}}\, ,
\end{eqnarray}
where 
\begin{eqnarray}
\mathcal{M}=\sqrt{\frac{{\bf k}^2_\perp+m_q^2 }{x(1-x)}} \, .
\label{factor-M}
\end{eqnarray}
Following Eqs.~(\ref{f_1})-(\ref{f_1tt}), the explicit expressions of TMDs in LF quark model are given by
\begin{eqnarray}
f_1(x,{\bf k}^2_\perp)&=&\frac{1}{3(2 \pi)^3}\bigg(\frac{1}{2}\left(3\left(m_q\left(\mathcal{M}+2m_q\right)\right)^2+(1-2x)^2\mathcal{M}^2{\bf k}^2_\perp \right)\nonumber\\
&&+4 {\bf k}^2_\perp\left( m_q(\mathcal{M}+2m_q)+{\bf k}^2_\perp\right)\nonumber\\
&&+{\bf k}^2_\perp\left(2m_q(\mathcal{M}+m_q)+(1-2x+2x^2)\mathcal{M}^2\right)\bigg)\frac{| \psi(x,{\bf k}^2_\perp)|^2}{\omega^2}\, ,
\label{f1_LF}
\end{eqnarray}
\begin{eqnarray}
g_{1L}(x,{\bf k}^2_\perp)&=&\frac{1}{2(2 \pi)^3}\big(m_q(\mathcal{M}+2m_q)\left(m_q(\mathcal{M}+2m_q)+2 {\bf k}^2_\perp\right)\nonumber\\
&&-\mathcal{M}(\mathcal{M}+2m_q)(1-2x)\big)\frac{| \psi(x,{\bf k}^2_\perp)|^2}{\omega^2}\, ,
\label{g1_LF}
\end{eqnarray}
\begin{eqnarray}
g_{1T}(x,{\bf k}^2_\perp)&=&\frac{M_\rho}{2 (2 \pi)^3}\left(\mathcal{M}+2m_q\right)\left(m_q \mathcal{M}(1-2x)+\left(m_q(\mathcal{M}+2m_q)+2 {\bf k}^2_\perp\right)\right)\nonumber\\
&&\times\frac{| \psi(x,{\bf k}^2_\perp)|^2}{\omega^2}\, ,
\end{eqnarray}
\begin{eqnarray}
h_1(x,{\bf k}^2_\perp)&=&\frac{1}{2 (2 \pi)^3}\big(\left(m_q(\mathcal{M}+2m_q)+2 {\bf k}^2_\perp\right)\left(m_q(\mathcal{M}+2m_q)+ {\bf k}^2_\perp\right)\nonumber\\
&&-\mathcal{M}(x\mathcal{M}+m_q)(1-2x){\bf k}^2_\perp\big)\frac{| \psi(x,{\bf k}^2_\perp)|^2}{\omega^2}\, ,
\label{h1_LF}
\end{eqnarray}
\begin{eqnarray}
h^\perp_{1L}(x,{\bf k}^2_\perp)&=&-\frac{M_\rho}{(2 \pi)^3}(\mathcal{M}+2m_q)\left(m_q\left((1-x)\mathcal{M}+m_q\right)+{\bf k}^2_\perp\right)\frac{| \psi(x,{\bf k}^2_\perp)|^2}{\omega^2}\, ,\nonumber\\
\end{eqnarray}
\begin{eqnarray}
h^\perp_{1T}(x,{\bf k}^2_\perp)&=&-\frac{M_\rho^2}{(2 \pi)^3}\left(\mathcal{M}\left((1-x)\mathcal{M}+m_q\right)(1-2x)+\left(m_q(\mathcal{M}+2m_q)+2 {\bf k}^2_\perp\right)\right)\nonumber\\
&&\times\frac{| \psi(x,{\bf k}^2_\perp)|^2}{\omega^2}\, ,
\end{eqnarray}
\begin{eqnarray}
f_{1LL}(x,{\bf k}^2_\perp)&=& 0 \, ,
\label{f1LL_LF}
\end{eqnarray}
\begin{eqnarray}
f_{1LT}(x,{\bf k}^2_\perp)&=& 0 \, ,
\end{eqnarray}
\begin{eqnarray}
f_{1TT}(x,{\bf k}^2_\perp) &=& \frac{M_\rho^2}{(2 \pi)^3}\left(\mathcal{M}^2x(1-x)-({\bf k}^2_\perp+m_q^2)\right )\frac{| \psi(x,{\bf k}^2_\perp)|^2}{\omega^2}\,,
\end{eqnarray}
with
\begin{eqnarray}
\omega=(\mathcal{M}+2m_q)\sqrt{{\bf k}^2_\perp+m_q^2}\, ,
\end{eqnarray}
where $\mathcal{M}$ and $\psi(x,{\bf k}^2_\perp)$ are defined in Eqs. (\ref{factor-M}) and (\ref{momentum-space-wf}) respectively. To find the numerical results, we use the quark mass and $\beta$ parameter as: $m_q=0.2$ GeV and $\beta=0.41$ GeV respectively \cite{Yu:2007hp}. 
\section{Density matrix description of spin-1 hadron}\label{densitymartix}
The spin density matrix $\rho (\boldsymbol{\mathcal{S}})$ of spin-$J$ is totally related to the tensor matrices of 2$J$ rank. The spin-1 matrices correspond to three cartesian and six traceless and symmetric tensor matrices denoted by $\Sigma^i$ and $\Sigma^{ij}\left(=\frac{1}{2}\left(\Sigma^i \Sigma^j + \Sigma^j \Sigma^i\right)-\frac{2}{3} \delta^{ij} {\bf I}\right)$ respectively, which are given by
\begin{eqnarray}
\Sigma^x=\frac{1}{\sqrt{2}}\left(
\begin{array}{c c c}
0~~ & 1~~ & 0\\
1~~ & 0~~ & 1\\
0~~ & 1~~ & 0\\
\end{array}
\right) \, , 
\Sigma^y=\frac{1}{\sqrt{2}}\left(
\begin{array}{c c c}
0~ & -\iota & 0\\
\iota~ & 0 & -\iota\\
0~ & \iota & 0\\
\end{array}
\right) \, ,\Sigma^z=\left( \begin{array}{c c c}
1~~ & 0~ & 0\\
0~~ & 0~ & 0\\
0~~ & 0~ & -1\\
\end{array}
\right)\,,
\end{eqnarray}
\begin{eqnarray}
\Sigma^{xx}=\frac{1}{6}\left(
\begin{array}{c c c}
-1~ & 0~ & 3\\
0~ & 2~ & 0\\
3~ & 0~ & -1\\
\end{array}
\right) \, , 
\Sigma^{xy}=\frac{1}{2}\left(
\begin{array}{c c c}
0~~ & 0~ & -\iota\\
0~~ & 0~ & 0\\
\iota~~ & 0~ & 0\\
\end{array}
\right) \, ,\Sigma^{xz}=\frac{1}{2 \sqrt{2}}\left( \begin{array}{c c c}
0~ & ~1 & ~~0\\
1~ & ~0 & -1\\
0 & -1 & ~~0\\
\end{array}
\right)\,,
\end{eqnarray} 
\begin{eqnarray}
\Sigma^{yy}=\frac{1}{6}\left(
\begin{array}{c c c}
-1~~ & 0~ & -3\\
~~0~~ & 2~ & ~~0\\
-3~~ & 0~ & -1\\
\end{array}
\right) \, , 
\Sigma^{yz}=\frac{1}{2\sqrt{2}}\left(
\begin{array}{c c c}
0 & -\iota~ & 0\\
\iota~ & 0~ & \iota\\
0 & -\iota~ & 0\\
\end{array}
\right) \, ,
\Sigma^{zz}=\frac{1}{3}\left( \begin{array}{c c c}
1 & ~~0 & ~~0\\
0 & -2 & ~~0\\
0 & ~~0 & ~~1\\
\end{array}
\right)\,.
\end{eqnarray}
Now, the spin density matrix $\rho (\boldsymbol{\mathcal{S}})$ is expressed as
\begin{eqnarray}
\rho (\boldsymbol{\mathcal{S}})=\frac{1}{3} \left(1+\frac{3}{2}\, \Sigma^i \mathcal{S}^i+3\, \Sigma^{ij} T^{ij}\right)\,,
\label{density-matrix}
\end{eqnarray}  
with 
\begin{eqnarray}
\boldsymbol{\mathcal{S}}= \left(\mathcal{S}^x_T,\mathcal{S}^y_{T}, \mathcal{S}_L\right),
\end{eqnarray}
and
\begin{eqnarray}
T^{ij}=\frac{1}{2}\left(
\begin{array}{ccc}
\mathcal{S}^{xx}_{TT}+\mathcal{S}_{LL} & \mathcal{S}^{xy}_{TT} & \mathcal{S}^x_{LT}\\ \\
\mathcal{S}^{yx}_{TT} & \mathcal{S}^{yy}_{TT}+\mathcal{S}_{LL} & \mathcal{S}^y_{LT}\\ \\
\mathcal{S}^x_{LT} & \mathcal{S}^y_{LT} & -2 \mathcal{S}_{LL}
\end{array}
\right)\,,
\end{eqnarray}
where $\mathcal{S}^{yy}_{TT}=-\mathcal{S}^{xx}_{TT}$ and $\mathcal{S}^{xy}_{LT}=\mathcal{S}^{yx}_{LT}$.
Therefore, from Eq. (\ref{density-matrix})
\begin{eqnarray}
\rho(\boldsymbol{\mathcal{S}}) = \left(
\begin{array}{ccc}
\frac{1}{3}-\frac{\mathcal{S}_{LL}}{2}+\frac{\mathcal{S}_L}{2}  & \frac{\mathcal{S}_{LT}^x -\iota \mathcal{S}_{LT}^y}{2\sqrt{2}}+\frac{\mathcal{S}_{T}^x -\iota \mathcal{S}_{T}^y}{2\sqrt{2}} & \frac{\mathcal{S}_{TT}^{xx}-\mathcal{S}_{TT}^{yy} -2 \iota \mathcal{S}_{TT}^{xy}}{4}\\ \\
\frac{\mathcal{S}_{LT}^x +\iota \mathcal{S}_{LT}^y}{2\sqrt{2}}+\frac{\mathcal{S}_{T}^x + \iota \mathcal{S}_{T}^y}{2\sqrt{2}} & \frac{1}{3}+ \mathcal{S}_{LL} & -\frac{\mathcal{S}_{LT}^x -\iota \mathcal{S}_{LT}^y}{2\sqrt{2}}+\frac{\mathcal{S}_{T}^x -\iota \mathcal{S}_{T}^y}{2\sqrt{2}}\\ \\
\frac{\mathcal{S}_{TT}^{xx}-\mathcal{S}_{TT}^{yy} + 2 \iota \mathcal{S}_{TT}^{xy}}{4} &  -\frac{\mathcal{S}_{LT}^x + \iota \mathcal{S}_{LT}^y}{2\sqrt{2}}+\frac{\mathcal{S}_{T}^x + \iota \mathcal{S}_{T}^y}{2\sqrt{2}} & \frac{1}{3}-\frac{\mathcal{S}_{LL}}{2}-\frac{\mathcal{S}_L}{2}
\end{array}
\right)
\end{eqnarray}
with
\begin{eqnarray}
-1 \leq \mathcal{S}_L \leq 1, ~~~~~   -\frac{1}{3} \leq \mathcal{S}_{LL} \leq \frac{2}{3}.
\end{eqnarray}
\bibliographystyle{JHEP} 
\bibliography{ref-v2.bib}
\end{document}